\documentclass[twocolumn,prd, aps, notitlepage, nofootinbib, showpacs, superscriptaddress,longbibliography]{revtex4-1}
\usepackage[utf8]{inputenc}
\usepackage{graphicx,mathtools,xcolor,latexsym,rotating,soul}
\usepackage[T1]{fontenc}
\usepackage{tikz}
\usepackage{comment}
\usepackage{amsmath, amssymb, amsthm, amsfonts,breqn}
\usepackage{multirow}
\usetikzlibrary{decorations.pathmorphing}
\usepackage[colorlinks=true,citecolor=green,urlcolor=blue,linkcolor=blue]{hyperref}
\usepackage{pifont}
\usepackage{gensymb}
\usepackage{enumitem}
\usepackage{mathrsfs}
\usepackage[scr=boondox,  
            cal=esstix]   
           {mathalpha}

\newtheoremstyle{named}{}{}{\itshape}{}{\bfseries}{.}{.5em}{\thmnote{#3 }#1}
\theoremstyle{named}

\newcommand*{\rom}[1]{\expandafter\@slowromancap\romannumeral #1@}

\makeatletter
\let\cat@comma@active\@empty
\makeatother
\allowdisplaybreaks
\begin{document}
\title{A relativistic treatment of accretion disk torques on \\ extreme mass-ratio inspirals around non-spinning black holes}
\author{Abhishek Hegade K. R.}
\email{ah4278@princeton.edu}
\affiliation{Illinois Center for Advanced Studies of the Universe, Department of Physics, University of Illinois Urbana-Champaign, Urbana, IL 61801, USA}
\affiliation{Princeton Gravity Initiative, Princeton University, Princeton, NJ 08544, USA}

\author{Charles F. Gammie}
\email{gammie@illinois.edu}
\affiliation{Department of Astronomy, Department of Physics, ICASU, and NCSA, University of Illinois Urbana-Champaign}

\author{Nicol\'as Yunes}
\email{nyunes@illinois.edu}
\affiliation{Illinois Center for Advanced Studies of the Universe, Department of Physics, University of Illinois Urbana-Champaign, Urbana, IL 61801, USA}

\begin{abstract}
We develop a relativistically accurate formalism to model the interaction between stellar mass compact objects embedded in thin accretion disks around a non-spinning supermassive black hole, using tools from self-force theory and Hamiltonian perturbation theory. 
We then apply this formalism to analyze the evolution of a compact object on a nearly circular and equatorial orbit interacting with a thin equatorial disk.
We provide analytic and relativistically-accurate expressions for the rates of energy and angular momentum exchanged during interactions due to Lindblad and corotation resonances. 
Our results show that relativistic corrections can enhance the magnitude of the torque by 1--2 orders of magnitude compared to purely Newtonian expressions when the orbit of the compact object is smaller than $10$ Schwarzschild radii of the supermassive black hole. 
We also demonstrate that strong relativistic shifts the inner Lindblad resonances closer to the compact object than the outer Lindblad resonances when the compact object is closer than 4 Schwarzschild radii to the supermassive black hole, potentially leading to a reversal in the direction of the torque acting on the compact object. 
Finally, we provide a dephasing estimate and show that using the relativistic torque formula is crucial to obtain reliable estimates for extreme mass ratio inspirals in orbits closer than 5 Schwarzschild radii to the supermassive black hole.
Our results highlight the importance of using relativistically-accurate models of environmental interactions in extreme mass-ratio inspirals close to a supermassive black hole.
\end{abstract}
\maketitle
\section{Introduction}
The Laser Interferometer Space Antenna (LISA) is a space-based gravitational wave detector that will measure gravitational-wave signals in the range of 0.1 mHz to 1.0 Hz from sources such as supermassive black hole mergers, extreme mass-ratio inspirals (EMRIs), stochastic gravitational-wave background, and galactic white dwarf binaries~\cite{colpi2024lisadefinitionstudyreport,amaroseoane2017laserinterferometerspaceantenna}. The detection of gravitational waves from these sources has the potential to test general relativity, reveal the properties of supermassive black hole mergers, test for the existence of intermediate mass-ratio black holes, and shed light on the stochastic gravitational-wave background~\cite{Berti:2018vdi,LISA:2022yao,amaroseoane2017laserinterferometerspaceantenna,Babak:2017tow,Caprini:2018mtu}. The accurate modeling of these systems, therefore, is crucial to avoid systematic biases and fulfill the promise of space-based gravitational-wave astrophysics~\cite{LISA:2022yao}.

In this paper, we study EMRIs in which a small, stellar-mass compact object $(10 M_{\odot} - 100 M_{\odot})$ slowly spirals into a supermassive black hole $(10^5 M_{\odot} - 10^8 M_{\odot})$ by emitting gravitational waves. 
The large mass-ratio makes EMRIs excellent probes of the spacetime geometry around the supermassive black hole. EMRIs can accumulate $\gtrsim \mathcal{O}(10^5)$ wave-cycles in the LISA band, and thus, they can be used to probe the mass and spin of the supermassive black hole with extreme accuracy, to test general relativity, and to understand the nature of the environment close to the supermassive black hole~\cite{Babak:2017tow,Kocsis_2011,Yunes:2011ws,Speri_2023,duque2025constrainingaccretionphysicsgravitational}.

Many important questions need to be addressed before the gravitational-waves emitted by EMRIs can be used to achieve the scientific goals described above.
From an astrophysical perspective, the origin of EMRIs remains uncertain. Several mechanisms, such as two-body relaxation in active galactic nuclei, tidal disruption of binaries around a massive black hole, and formation in active galactic nuclei, have been proposed but have highly uncertain rates  (see e.g.~\cite{LISA:2022yao,Amaro_Seoane_2018,C_rdenas_Avenda_o_2024} and references therein). 
Moreover, using LISA to understand EMRIs presents interesting data analysis challenges due to galactic binaries that create confusion noise~\cite{LISA:2022yao,2017CQGra..34x4002R,Bender:1997hs,katz2024efficientgpuacceleratedmultisourceglobal,strub2024globalanalysislisadata}.
Finally, the enormous number of wave cycles spent in the LISA band by EMRI systems presents a modeling challenge from a theoretical viewpoint, as these models must be accurate and precise.  

Before we can discuss the theoretical modeling problem for EMRIs that are immersed in an accretion disk, we first must review how EMRIs are modeled in vacuum.
In the absence of any \textit{matter} or \textit{external source}, the motion of the small compact object (SCO) can be described using vacuum general relativity. In this setting, the EMRI follows an approximately geodesic path around the supermassive black hole, and the gravitational back-reaction onto the system causes a self-force on the SCO forcing it to inspiral and emit gravitational waves~\cite{Poisson_2011,Sasaki_2003,Pound_2021}. 
The vacuum dynamics of EMRIs can be described using self-force theory and black hole perturbation theory~\cite{Poisson_2011,Sasaki_2003,Pound_2021}. 
Significant advances in self-force theory and black hole perturbation theory in recent years have paved the way for accurately modeling the motion of EMRIs in (vacuum) Kerr spacetimes~\cite{Poisson_2011,Albertini:2022rfe,Warburton:2024xnr,LISAConsortiumWaveformWorkingGroup:2023arg,Wardell:2021fyy,Wardell_2023,Bourg:2024vre}. 

The \textit{vacuum} or \textit{no-external-source} assumption on the supermassive black hole background may not be a valid assumption if the supermassive black hole is actively accreting matter. 
Several studies have explored the impact of the environment of the supermassive black hole on the motion and gravitational-wave emission from EMRIs
~\cite{Kocsis_2011,Yunes:2011ws,Derdzinski:2018qzv,Derdzinski:2020wlw}. 
Environmental effects can include the interaction between the SCO and the accretion disk of the supermassive black hole~\cite{2000ApJ...536..663N,Kocsis_2011,Yunes:2011ws,Speri_2023,duque2025constrainingaccretionphysicsgravitational,Derdzinski:2018qzv,Derdzinski:2020wlw}, mass accretion on to the supermassive black hole and the SCO~\cite{Kocsis_2011,Yunes:2011ws}, viscous drag on the SCO due to gas surrounding the SCO~\cite{Barausse:2007dy}, disk-SCO collisions~\cite{Franchini_2023}, and perturbations due to the presence of a third body~\cite{Yunes:2010sm,Naoz_2013,Silva:2022blb,Silva:2025lkl}.
In addition, there could be exotic scenarios where a SCO interacts with bosonic clouds and dark matter particles~\cite{Baumann_2022,Maselli_2020,Zhang_2020,Arvanitaki_2011,Vicente_2022}.

The energy and angular momentum exchanged between a SCO and an accretion disk surrounding the supermassive black hole can be the dominant source of measurable departures from vacuum evolution in the LISA band~\cite{Kocsis_2011,Yunes:2011ws,Speri_2023,duque2025constrainingaccretionphysicsgravitational}.
Hydrodynamic interactions between a small secondary object and a disk orbiting a large primary object have a rich phenomenology and have been extensively studied in the context of proto-planetary disks and planetary rings~\cite{GT-disc-satellite-interaction,GT-Uranus-Ring,Baruteau_2014,Paardekooper:2023,Murray-Dermott-Book,Ward-1986,2002ApJ...565.1257T}. In proto-planetary disks, a small secondary object (usually a planet) and the disk interact via resonant energy exchange~\cite{GT-disc-satellite-interaction}, which impacts the motion of the secondary object and leads to spiral density waves in the disk~\cite{1978ApJ...222..850G,GT-disc-satellite-interaction,1979ApJ...233..857G}. This interaction can cause slow migration of the secondary, forcing the orbit to inspiral or outspiral, depending on precisely how the disk is modeled~\cite{Baruteau_2014,Paardekooper:2023}.  Migration is referred to as ``Type I'' in the astrophysics literature if the secondary is small and it does not clear a gap~\cite{Baruteau_2014,Paardekooper:2023}, ``Type II'' if it opens a full gap, and ``Type III'' if it opens a partial gap, leading to complicated disk-secondary interactions~\cite{Baruteau_2014,Paardekooper:2023}. The details of the gap opening depend sensitively on the structure and thermodynamics of the disk and the mass ratio (see e.g.~\cite{Baruteau_2014,Paardekooper:2023}).

An EMRI with a SCO interacting with a thin accretion disk that surrounds the supermassive black hole is quite similar to the proto-planetary disk case, except that the SCO's motion, the spacetime, and the disk dynamics are relativistic.  
For typical mass ratios of relevance to EMRIs, a gap is not expected to open in the inner regions of the disk~\cite{Kocsis_2011}. Hence, we can analyze EMRIs using the perturbation theory techniques developed for Type-I migration~\cite{GT-disc-satellite-interaction,1978ApJ...222..850G,GT-Uranus-Ring,2002ApJ...565.1257T,Derdzinski:2018qzv,Derdzinski:2020wlw}.
Indeed, very early studies used Type-I migratory torques from the proto-planetary disk literature to estimate the possibility of constraining the accretion disk properties close to a supermassive black hole using EMRIs~\cite{Yunes:2011ws,Kocsis_2011}, and a decade later, more recent studies have continued to use these Newtonian torques~\cite{Sberna:2022qbn,Speri_2023,Copparoni_2025}.
Broadly, these studies concluded that EMRIs orbiting supermassive black holes of mass $\mathcal{O}\left(10^6 M_{\odot}\right)$ could perhaps be used to constrain certain disk properties, such as the average density and viscosity of sufficiently massive and thin accretion disks~\cite{Yunes:2011ws,Kocsis_2011,Sberna:2022qbn,Speri_2023,Copparoni_2025}.

Estimates from Newtonian gravity used in the pioneering studies mentioned above will hold at sufficiently large distances from the supermassive black hole. However, close to the supermassive black hole, where EMRIs are expected to be detected~\cite{LISA:2022yao}, relativistic effects may be important.
\begin{table*}[t]
    \centering
    \begin{tabular}{|c|p{0.5\linewidth}|}
        \hline
        \textbf{Equation Number} & \textbf{Description} \\
        \hline
        Eq.~\eqref{eq:modified-Delaunay-variables}. & Introduces the relativistic modified Delaunay variables. \\
        \hline
        Eqs.~\eqref{eq:hamiltonian-equations-3-body} and~\eqref{eq:secular-evolution-equation}.
        &
        Hamiltonian and secular evolution equations for the disk-SCO system. \\
        \hline
        Eq.~\eqref{eq:R-dist-function-expr-lin-e}. & Schematic expression for the relativistic disturbing function in a small eccentricity expansion. \\
        \hline
        Eqs.~\eqref{eq:location-corotation-resonance} and \eqref{eq:Lindblad-location}. & Location of the relativistic Lindblad and corotation resonances. \\
        \hline
        Eqs.~ \eqref{eq:Lindblad-E-expr-actual}, \eqref{eq:lindblad-resonance-total} and \eqref{eq:expr-rel-Lindblad-all-others}. & Differential Lindblad torque on the SCO. \\
        \hline
        Eq.~\eqref{eq:jmax-assumption}. & Ansatz for the torque cutoff parameter. \\
        \hline
    \end{tabular}
    \caption{List of the most important equations in the paper.}
    \label{tab:most-imp-eqns}
\end{table*}
In this paper, we will take the first steps to accurately model the disk-SCO interaction in the vicinity of a supermassive black hole by developing an analytic and relativistically-accurate model. 
A detailed understanding of disk-SCO interactions require a model of both the accretion disk and the SCO's gravitational field.  
The structure of inner, relativistic disks around supermassive black holes is not completely understood (for a review see e.g.~\cite{Blaes:2014, Davis:2020}), due to the  interplay of turbulence in magnetized plasmas, radiation fields, and relativity.  Technical progress toward numerical models that incorporate radiative forces (as a recent example, see \cite{White:2023}), along with parallel developments in neutrino transport are likely to provide a clearer understanding of disk structure, and (specially relevant to this discussion) a statistical characterization of fluctuations in the gravitational field due to turbulence in the disk. In advance of this understanding, however, we will here use analytic models to uncover aspects of the disk-SCO interaction.  

Our approach mirrors early studies in the proto-planetary disk literature~\cite{GT-disc-satellite-interaction,GT-Uranus-Ring}, where the disk is assumed to consist of particles moving in closed orbits under the influence of the supermassive black hole, and interacting with particles in the accretion disk via resonant energy and momentum exchange~\cite{Murray-Dermott-Book,GT-disc-satellite-interaction,GT-Uranus-Ring}. 
These effective models provide an analytical tool to understand the nature of the disk-SCO interaction, and they can guide numerical and semi-analytical studies that model the fluid properties of the disk~\cite{1993Icar..102..150K,1993ApJ...419..155A,Ogilvie-Lubow-wake,Miranda_2019,fairbairn2025pushinglimitseccentricityplanetdisc,Baruteau_2014,Paardekooper:2023}. 
Our analysis relies on the following key ingredients: 
\begin{itemize}
    \item [I.] \textbf{analytic treatment of the interaction between the SCO and a particle in the thin disk:} The energy and angular momentum exchanged in disk-SCO interactions are driven significantly by the interaction of the SCO with the resonances induced by the disk close to the orbit of the SCO~\cite{GT-disc-satellite-interaction}. 
    We leverage this insight to model the interaction between a particle in the disk and the SCO using the singular potential of the disk and the SCO from self-force theory~\cite{Poisson_2011}. 
    Physically, this approach is equivalent to analyzing the disk-SCO interaction in a local Fermi frame that follows the trajectory of the SCO. In such a local frame, the interaction between the disk particle and the SCO is effectively Newtonian~\cite{GT-disc-satellite-interaction}.
    Using this insight helps us avoid explicit global metric reconstruction and significantly simplifies the analysis.
    \item [II.] \textbf{Hamiltonian perturbation theory and secular evolution using action-angle variables:} We build a Hamiltonian that describes the interaction between the SCO and the disk. In the Newtonian limit, the interaction Hamiltonian is built out of the so-called disturbing function~\cite{Murray-Dermott-Book}, which represents the Newtonian potential between two objects moving on the background of a much larger primary. Our Hamiltonian interaction potential generalizes the disturbing function to full general relativity. 
    Once the Hamiltonian is obtained, we use standard techniques from Hamiltonian mechanics~\cite{GT-disc-satellite-interaction,Ogilvie_2007,K-Cole-perturbation-theory} to understand the secular evolution of the orbital elements of the disk and the SCO. 
    \item [III.] \textbf{analytic expressions for generalized Fourier expansions of the relativistic disturbing function:} A key step in our analysis is obtaining analytic, generalized Fourier expansions of the relativistic disturbing function, using action-angle variables that reduce to the modified Delaunay variables in the Newtonian limit~\cite{Murray-Dermott-Book}. This simplification allows us to obtain analytic expressions for the secular evolution of the orbital elements of the SCO and the disk using Hamiltonian perturbation theory.
\end{itemize}

As an application of our formalism, we analyze the interaction between a SCO on an equatorial, circular orbit interacting with a thin accretion disk. 
We provide analytic expressions for the locations of the Lindblad and corotation resonances near the orbit of the SCO, as well as the energy and angular momentum exchanged at these resonances. Our analysis reveals that relativistic effects shift the location of the inner Lindblad resonances closer to the SCO than the outer Lindblad resonances, potentially resulting in a change in the direction of the torque exerted by the disk on the SCO if the density gradients are not too steep. The relativistic effects become extremely strong near the innermost stable circular orbit, and we show that Newtonian expressions for the torque are inaccurate when the orbit of the SCO is close to the supermassive black hole.
We also analyze the impact of different accretion disk profiles on the torque and power exchanged between the disk and SCO.
Finally, we provide an order-of-magnitude estimate of the total accumulated phase due to disk-SCO interactions for binaries detectable in the LISA band. 
Our results are broadly consistent with previous studies~\cite{Kocsis_2011,Speri_2023}, but for EMRIs closer than 5 Schwarzschild radii from the supermassive black hole, the use of relativistic torque expressions is crucial.

Our work differs from previous studies in various ways. As mentioned above, most previous works modeled the SCO-disk interaction through expressions derived from Newtonian gravity. In particular, Ref.~\cite{Hirata_2011,Hirata_2011_II} analyzed the problem of disk-SCO interaction using a semi-analytic approach, which involved the integration of the Teukolsky equation, combined with an energy conservation argument to derive expressions for the resonance strengths of resonances that are sufficiently \textit{far} from the orbit of the SCO (see Fig. 1 and Table 2 of~\cite{Hirata_2011_II}).
The integration of the Teukolsky equation and the subsequent metric reconstruction (energy balance argument) is only required if one is interested in understanding resonances that are sufficiently far away from the orbit of the SCO.
The strongest resonances that contribute to the disk-SCO interaction are the resonances that are close to the location of the SCO~\cite{GT-disc-satellite-interaction}. Moreover, to obtain the total differential torque on the system, one has to sum over all the \textit{close resonances} and understand the differences in the location of the Lindblad resonances in detail.
In this work, our focus will be on obtaining the total differential torque due to the resonances close to the orbit of the SCO analytically, a discussion complementary to~\cite{Hirata_2011_II}.
The locations of the resonances computed using our framework agree with the ones obtained in~\cite{Hirata_2011_II}.
Computing the resonance strength consistently for far-away resonances requires higher-order Taylor series expressions for the singular potential of the SCO, which we do not compute.
Hence, we have not made a direct comparison to the resonance strengths for the far-away resonances obtained from the semi-analytical approach of~\cite{Hirata_2011,Hirata_2011_II}.
Recent work~\cite{Duque:2025yfm} has reproduced some of our results using the methods of~\cite{Hirata_2011,Hirata_2011_II}, strengthening the robustness of our conclusions.

The rest of the paper explains our framework and the results presented above in detail, and is organized as follows:
In Sec.~\ref{sec:Qualitative-explanation-Intro}, we use expressions from Newtonian gravity to provide a \textit{qualitative} explanation of why the relativistic effects predict a torque reversal and show how the torque increases in magnitude as we approach the innermost stable circular orbit.
In Sec.~\ref{sec:resonances-and-rel-disturb-func}, we present a formalism to model the general interaction between the disk and the SCO by using self-force theory and Hamiltonian mechanics. 
This section contains most of our technical calculations and presents the expressions for the relativistic disturbing function.
Next, we restrict our attention to analyzing Lindblad and corotation resonances for an EMRI in a circular and equatorial orbit in Sec.~\ref{sec:corotation-and-Lindblad-resonances}.
The technical details leading to the torque reversal phenomena, described qualitatively in Sec.~\ref{sec:Qualitative-explanation-Intro}, are presented in this section.
We also provide an independent derivation of the leading-order expression for the one-sided Lindblad torques using a local model~\cite{GT-disc-satellite-interaction,Gammie_2004}.
In Sec.~\ref{sec:results}, we quantify the difference between the Newtonian torque and the relativistic torque, explore the impact of different accretion disk profiles on the energy exchanged during the disk-SCO interaction, and compare it with gravitational-wave energy loss.
Our conclusions are discussed in Sec.~\ref{sec:conclusions}.
The appendices provide the details of the derivation of our results.
A review of orbital mechanics in Schwarzschild spacetime using action-angle variables is presented in Appendix~\ref{sec:geodesic-motion-Schw}.
A list of the most important equations in the paper is provided in Table~\ref{tab:most-imp-eqns}. We use geometric units throughout, with $ G=1=c$. 
\section{Qualitative explanation of the torque reversal effect and impact of relativistic corrections}\label{sec:Qualitative-explanation-Intro}
\begin{figure*}[thp!]
    \centering
    \includegraphics[width=1.5\columnwidth]{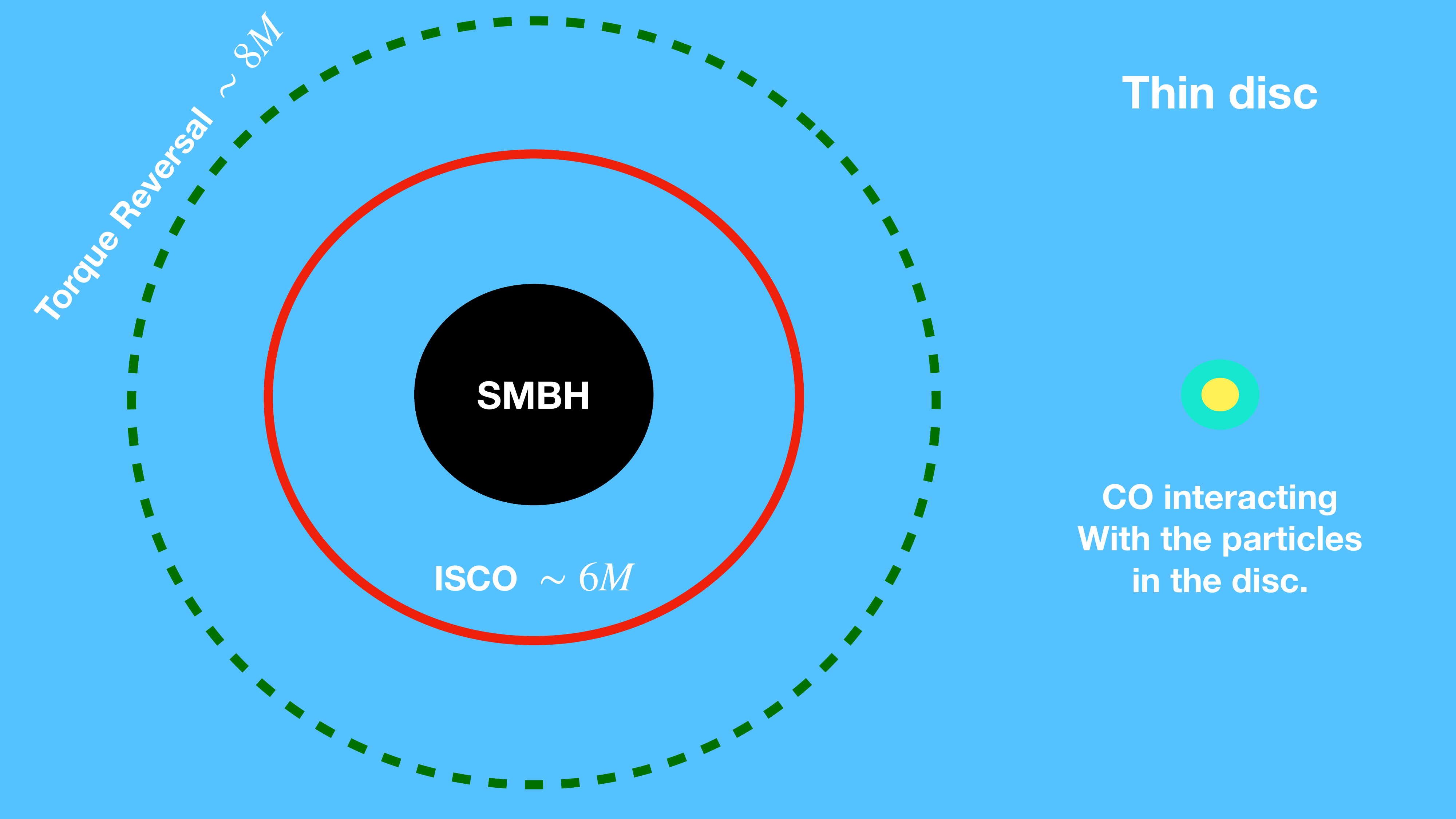}
    \caption{Cartoon (not to scale) depicting the interaction between a small SCO (yellow) on an equatorial circular orbit interacting with a thin accretion disk (blue background) surrounding a supermassive black hole (black circle). 
    The thin cyan disk surrounding the SCO is used to highlight the fact that the dominant source of energy and angular momentum exchange are resonances close to the SCO.
    We also highlight two important surfaces where relativistic effects are important. The green dashed circle shows the approximate location at which relativistic effects change the sign of differential Lindblad torque on the SCO. The relativistic effects become extremely large near the innermost stable circular orbit, shown by a solid red circle.}
    \label{fig:cartoon}
\end{figure*}

In this subsection, we provide some intuition on the relativistic results obtained in the main body for readers familiar with the Newtonian treatment of disk-SCO interactions, by extrapolating the Newtonian formula derived in the proto-planetary literature~\cite{Ward-1986}. 
For simplicity, we focus on describing the important relativistic effects active in disk-SCO interactions when the SCO is on an equatorial circular orbit at radius $r = p' M$. 
We denote the surface density of the accretion disk by $\Sigma(r)$, the mass of the SCO by $m$, and the mass ratio by $q=m/M \ll 1$, where $M$ is the supermassive black hole mass. A cartoon depicting this system and highlighting where relativistic effects are important is shown in Fig.~\ref{fig:cartoon}.

The general expressions derived in Newtonian gravity by Goldreich and Tremaine~\cite{GT-disc-satellite-interaction} already provides valuable intuition on relativistic effects.
The dominant source of torque for the disk-SCO configuration described above arises from the exchange of energy and angular momentum at Lindblad resonances.
The location of Lindblad resonances $r_{\mathrm{Lr}} = p_{\mathrm{Lr}} M$ in Newtonian gravity is given by the roots of~\cite{GT-disc-satellite-interaction,Ward-1986}
\begin{align}\label{eq:GT-formula-location-of-LR-Resonances}
    \Omega(p_{\mathrm{Lr}}) - \Omega(p') &= - \frac{k}{j} \kappa (p_{\mathrm{Lr}})
\end{align}
where $\Omega$ is the angular velocity of the object on a circular orbit around the supermassive black hole, $\kappa$ is the epicyclic frequency of the circular orbit, $k=\pm 1$ denotes whether the resonance is inside $(k=-1)$ or outside $(k=1)$ the orbit of the SCO and $j$ is the mode-number of the resonances.
For an orbit around a non-rotating supermassive black hole, the values of the orbital frequency and the epicyclic frequencies are given by~\cite{Novikov-Thorne,Abramowicz_2013,Gammie_2004}
\begin{subequations}\label{eq:Omega-and-Kappa-Schwarzschild}
\begin{align}
    \Omega(p) &= \frac{1}{M p^{3/2}} \,,\\
    \kappa(p) &= \Omega \sqrt{\frac{p-6}{p}}\,.
\end{align}
\end{subequations}

The above expressions can be used in Eq.~\eqref{eq:GT-formula-location-of-LR-Resonances} to obtain the approximate locations of the closely-spaced resonances (i.e.~those with $j \gg 1$), which are the dominant contributions to the interactions. Solving Eq.~\eqref{eq:GT-formula-location-of-LR-Resonances} in the large $j$ limit one finds
\begin{align}\label{eq:Lindblad-location-intro}
    p_{\mathrm{Lr}} &= p' 
    \Bigg[1
    + \frac{2 k \sqrt{p'-6}}{3\sqrt{p'} j} 
    +
    k^2
    \frac{18-p'}{9p' j^2}
    \Bigg]
    +
    \mathcal{O}(j^{-3})\,.
\end{align}
This tells us that the inner and outer Lindblad resonances are symmetrically located about the location of the SCO to $\mathcal{O}(j^{-1})$.
The asymmetry in the location of resonances occurs at $\mathcal{O}(j^{-2})$ and the outer Lindblad resonances $(k=1)$ are closer to the orbit than the inner Lindblad resonances for $p'>18$. 
Closer to the supermassive black hole, i.e.~for $p'<18$, the inner Lindblad resonances are closer to the SCO. Generally, the closer resonances exert the strongest torques on the SCO, and in the Newtonian limit (i.e.~when $p' \gg 1$), the closer outer Lindblad resonances exert the strongest torque~\cite{Ward-1986,Ward-1997}. However, Eq.~\eqref{eq:Lindblad-location-intro} hints that this behavior no longer holds for COs located closer to the supermassive black hole. As we show below, this change in the inner and outer Lindblad resonances can potentially lead to a reversal in the direction of the torque experienced by the SCO.

Let us consider why Lindblad torques are strong near the innermost stable circular orbit and understand the torque reversal phenomenon by examining the Newtonian expressions. Using the techniques outlined in~\cite{GT-disc-satellite-interaction,Ward-1986}, one can obtain the differential Lindblad torque (i.e.~the sum of the total torque due to outer and inner Lindblad resonances). 
To write these expressions down, we need to first assume a value for the maximum of the mode-number $j$, which regulates the torque. These cutoff parameters are phenomenological here, but they can be obtained by modeling the microphysical nature of the disk~\cite{GT-disc-satellite-interaction, 1993ApJ...419..155A, Ward-1986}. 
Suppose that we can assume that the inner and outer torque cutoff parameters are the same; then, the torque is given by~\cite{Ward-1986}
\begin{widetext}
\begin{align}\label{eq:LR-torque-Newt-Intro}
    \left< \dot{L}'_{z} \right>_{\mathrm{approx}}
    &\propto
    \frac{8 M^2 p'{} q^2 \Sigma(p') }{27 F_0^3}
    \bigg[2 K_0\left(\frac{2 F_0}{3}\right)+F_0 K_1\left(\frac{2 F_0}{3}\right)\bigg]
    \times
    \nonumber\\
    &
    \bigg[
    4 F_0 \left(7 p'{} \frac{d F_0}{dp'}-6 F_0\right) K_1\left(\frac{2 F_0}{3}\right)+\left(8 p'{} \left(F_0^2+3\right) \frac{d F_0}{dp'}-3 F_0 \left(2 F_0^2+5\right)\right) K_0\left(\frac{2 F_0}{3}\right)
    \bigg]
    \nonumber\\
    &
    -
    \frac{d\Sigma}{dp'}
    \frac{16 M^2 p'{}^2 q^2 }{9 F_0^2}
    \bigg[2 K_0\left(\frac{2 F_0}{3}\right)+F_0 K_1\left(\frac{2 F_0}{3}\right)\bigg]^2
    \,,
\end{align}
\end{widetext}
where we have ignored a positive proportionality constant related to the torque cutoff parameters, $\dot{L}_{z}'$ is the rate of change of the angular momentum of the SCO in the $z$-direction, the angular brackets denote an average over the orbital period of the SCO, $F_0 \equiv \kappa(p')/\Omega(p')$, and $K_{0,1}$ are modified Bessel functions. The subscript $\mathrm{approx}$ highlights the fact that the above equation is an approximation that extrapolates the values of the epicycle frequencies from Newtonian gravity to relativity. The fully relativistic expressions for the differential Lindblad torque is derived later in the paper (see Eqs.~\eqref{eq:relativistic-Lindblad-torque}, \eqref{eq:expr-rel-Lindblad-all-others}  and \eqref{eq:Lindblad-E-expr-actual}).
From Eq.~\eqref{eq:Omega-and-Kappa-Schwarzschild}, we see that $\kappa \to 0$ at the innermost stable circular orbit, where $p \to 6$. This means that $F_0 \to 0$ and the modified Bessel functions diverge as we approach the innermost stable circular orbit in Eq.~\eqref{eq:LR-torque-Newt-Intro}, 
signaling that the torque is largest near the innermost stable circular orbit. As we show below, relativistic redshift and time dilation effects only amplify the divergence.

The torque reversal phenomena can be qualitatively understood by examining Eq.~\eqref{eq:LR-torque-Newt-Intro}. Observe that the term proportional to $d\Sigma/dp'$ is always negative. If the surface density increases as we move out, then this term always remains negative. One can show that unless the density gradients are large, this term is subdominant to the term proportional to $\Sigma(p')$ in Eq.~\eqref{eq:LR-torque-Newt-Intro}.
The term proportional to $\Sigma(p')$ can be positive or negative depending on the sign of
\begin{align*}
    \bigg[
    &
    4 F_0 \left(7 p'{} \frac{d F_0}{dp'}-6 F_0\right) K_1\left(\frac{2 F_0}{3}\right)
    \nonumber\\
    &+\left(8 p'{} \left(F_0^2+3\right) \frac{d F_0}{dp'}-3 F_0 \left(2 F_0^2+5\right)\right) K_0\left(\frac{2 F_0}{3}\right)
    \bigg]
    \,.
\end{align*}
Using Eq.~\eqref{eq:Omega-and-Kappa-Schwarzschild}, we can show that this term is negative only if $p'>9.94$.
Therefore, we expect a change in the sign of the torque around $p'=9.94$. 
As we show in Sec.~\ref{sec:results}, relativistic effects push this value to $p'\sim 8$.
The only two caveats to this analysis are the following. First, the torque cutoff parameters for the inner and outer Lindblad resonances are significantly different when including relativistic effects, implying that Eq.~\eqref{eq:LR-torque-Newt-Intro} is no longer valid. Second, if the density gradient is too large and negative, then it can counterbalance the torque reversal. We study both these points in the next sections. 


\section{Relativistic black hole accretion disk interactions}\label{sec:resonances-and-rel-disturb-func}

In this section, we describe a general formulation to treat the interaction between a SCO of mass $m$ moving in the geometry of a supermassive black hole of mass $M$, with $q \equiv m/M \ll 1$ and interacting with a thin accretion disk of density $\Sigma$.
We assume that the SCO perturbs the accretion disk weakly and does not open a gap. This restriction is valid given that $q\ll1$ and that, in the inner regions of the supermassive black hole, the accretion disk is expected to be thin~\cite{Abramowicz_2013}.
Therefore, our analysis focuses on Type-I migration effects due to black hole and accretion disk interactions~\cite{2023ASPC..534..685P,2014prpl.conf..667B}.

Type-I migration can be treated using three different methods.
In the first method, which we describe in this section, 
the SCO and the mass element $d\mu$ of the accretion disk move on geodesics of the central Schwarzschild black hole and are influenced by each other's gravitational field, which we describe using self-force theory~\cite{Poisson_2011}. 
The interaction between the SCO and the mass element in the accretion disk leads to orbital resonances that transfer energy and angular momentum between them~\cite{GT-disc-satellite-interaction}. This transfer of energy and angular momentum leads to secular changes in the orbital elements of the SCO and the accretion disk~\cite{GT-disc-satellite-interaction}.

In the second method, one analyzes the three-body interaction in a local Fermi frame that moves in an orbit close to the SCO and the accretion-disk mass element~\cite{GT-disc-satellite-interaction}.
In this local Fermi frame, the interaction is almost Newtonian, and we can easily deduce the rate of change of angular momentum and eccentricity from the formula derived in Newtonian gravity~\cite{GT-disc-satellite-interaction}.
The three-body interaction is Newtonian only to leading order in the separation between the two objects, and thus, the Fermi frame analysis only allows us to obtain the angular momentum exchange to leading order in the separation.
In principle, it is possible to extend the local Fermi calculation to higher orders, but it quickly becomes cumbersome to do so.
We instead work exclusively in the global coordinate frame in this section, deferring a discussion of the local calculation to Sec.~\ref{sec:corotation-and-Lindblad-resonances}, where we present the relativistic torques due to different types of resonances.

The two methods described above do not account for the presence of pressure gradients, magnetic fields, viscosity and self-gravity of the fluid, and thus, they can be seen as effective local descriptions of the SCO-disk interaction. A more physical method that can accurately capture the interaction between the fluid elements in the disk and the SCO object would involve understanding in detail the density and pressure perturbations induced on the fluid due to the motion of SCO object. We leave such an analysis to future work, although for calculations in Newtonian gravity, we refer the reader to~\cite{1993ApJ...419..155A,1993Icar..102..150K,2002ApJ...565.1257T,fairbairn2025pushinglimitseccentricityplanetdisc} and references therein.

The rest of this section is organized as follows.
We first describe the Hamiltonian for the three-body interaction between the central black hole, the SCO and the accretion disk in Sec.~\ref{sec:perturbed-motion}.
We then present a general formulation to treat the dynamics of resonant interactions in Sec.~\ref{sec:resonance-evolution}.
Finally, we discuss the properties of the interaction potential for orbits in nearly circular and equatorial orbits in Sec.~\ref{sec:disturbing-function-relativistic}.

\subsection{Perturbed geodesic motion of the smaller black hole}\label{sec:perturbed-motion}
We here derive the general form of the Hamiltonian representing the three-body interaction between a SCO, a Schwarzschild black hole, and an accretion-disk mass element. The SCO is assumed to move on a geodesic worldline $z^{\mu'}(t)$ around the black hole. The accretion-disk mass element $d\mu$ is assumed to move on a different geodesic worldline $z^{\mu}(t)$ around the black hole. We assume these two worldlines do not intersect.

First, let us derive an expression for perturbed geodesic motion in a general spacetime, which will be useful later~\cite{Hirata_2011}. 
Geodesic motion in a spacetime with metric $g_{\mu\nu}$ and coordinate time parameterization can be generated by the Hamiltonian (per unit mass) $\mathscr{H} = -p_{t}$.
For timelike geodesics, we have the normalization condition $g^{\mu \nu} p_{\mu} p_{\nu} = -1$.
To understand the perturbation away from geodesic motion (and induced by the metric perturbation), we must perturb the Hamiltonian due to a perturbation in the metric. Since the Hamiltonian of the unperturbed problem is just $p_t$, we must then perturb the normalization condition in $p_{t} \to p_{t} + \delta p_{t}, g^{\mu\nu} \to g^{\mu\nu} + \delta g^{\mu\nu}$:
\begin{align*}
    &2 p_t \delta p_t g^{tt} + p_t p_t \delta g^{tt} + 2 \delta g^{ti} p_{t} p_i + 2 \delta p_t g^{ti} p_i + p_i p_j \delta g^{ij} = 0
\end{align*}
to linear order.
This identity can be simplified to 
\begin{align}
    \delta p_{t}
    &=
    -p_{t}\frac{\delta g^{tt} p_{t} + 2 \delta g^{ti} p_{i}}{2 \left(g^{tt} p_{t} + g^{ti} p_{i}\right) }
    -
    \frac{p_{i} p_{j} \delta g^{ij}}{2 \left(g^{tt} p_{t} + g^{ti} p_{i}\right) }
    \nonumber\\
    &=
    -\frac{p_{\mu} \delta g^{\mu\nu} p_{\nu}}{2 \left(g^{t \mu} p_{\mu} \right)}
    \,.
\end{align}
Using this result, we obtain the following perturbed Hamiltonian for perturbed geodesic motion:
\begin{align}\label{eq:deltaH-v1}
    \mathscr{H} &= -p_{t} - \delta p_{t} = -p_{t} + \frac{p_{\mu} \delta g^{\mu\nu} p_{\nu}}{2 \left(g^{t \mu} p_{\mu} \right)} \,,
\end{align}
where it is now understood that $p_{\mu}$ stands for the background (specific) four-momentum (with $-p_{t}$ given by Eq.~\eqref{eq:Hamiltonian-Schw}) and $g_{\mu \nu}$ for the background Schwarzschild metric. 
Note that the metric perturbation $\delta g_{\mu\nu}[z(t)]$ appearing in the above equation is to be evaluated on the perturbed worldline $z^{\mu}$ (introducing only higher-order error).

We now specialize to the case where the worldline $z^{\mu}$ is that of the accretion disc mass element and $z^{\mu'}(t)$ is the worldline of the SCO. We focus our attention on the worldline of the accretion disc mass element and assume that the source of the metric perturbation is the gravitational field of the disc and the SCO.  
Consider a general spacetime point $x^{\mu}$ near the worldline $z^{\mu}$.
The metric perturbation at this point can be split into the following contributions~\cite{Poisson_2011}
\begin{align}\label{eq:metric-pert-split-v1}
    \delta g_{\mu \nu}(x) = h_{\mu\nu}^{\mathrm{S},d\mu}(x;z) + h_{\mu\nu}^{\mathrm{R},d\mu}(x;z) + h_{\mu \nu}^{\mathrm{int}}(x;z') \,,
\end{align}
where $h_{\mu\nu}^{\mathrm{S},d\mu}(x;z)$ is the singular field of the mass element of the disk, $h_{\mu\nu}^{\mathrm{R},d\mu}(x;z)$ is the regular field of the mass element of the disk, and $h_{\mu \nu}^{\mathrm{int}}(x;z')$ is an interaction term that is generated by the SCO.
The singular self-field of the mass element diverges when evaluated on its worldline, at $x^\mu = z^{\mu}(t)$, but one can show that it does not contribute to the motion of the mass element~\cite{Poisson_2011}.  
We can therefore ignore this contribution when we evaluate the metric perturbation on $z^{\mu}(t)$.  
The regular field arises from the effects of gravitational radiation and leads to a self-force on the mass element. 
We can treat this effect separately from the interaction between the accretion disk and the SCO in linear theory. Hence, we can ignore the contribution from the regular field as well, and concentrate on the interaction perturbation only.

Now approximate the interaction potential assuming that $z^{\mu}$ and $z^{\mu'}$ are sufficiently close together.
Let us decompose $h_{\mu \nu}^{\mathrm{int}}\left(z; z'\right)$ using the singular and regular fields of the SCO 
\begin{align}
    h_{\mu \nu}^{\mathrm{int}}\left(z; z'\right) = h_{\mu\nu}^{\mathrm{S,SCO}}(z;z') + h_{\mu\nu}^{\mathrm{R,SCO}}(z;z')\,,
\end{align}
where $h_{\mu\nu}^{\mathrm{S,SCO}}(z;z')$ and $h_{\mu\nu}^{\mathrm{R,SCO}}(z;z')$ are the singular and the regular field contributions from the SCO.
Since the worldlines are sufficiently close together, $h_{\mu\nu}^{\mathrm{S}}(z;z')$ behaves like a Newtonian potential\footnote{This is because the worldlines are close to each other in spacetime, and thus, their relative velocity is also small. In the freely falling frame of either worldline, gravity is exactly Newtonian.
} 
\begin{align*}
    h_{\mu\nu}^{\mathrm{S}}(z;z') = \mathcal{O} \left(\frac{1}{|z-z'|}\right)
\end{align*}
and is the dominant contribution to the interaction potential. The effect of the regular field $h_{\mu\nu}^{\mathrm{R}}(z;z') = \mathcal{O}(1)$ is important, but we can ignore it when understanding the interaction quasi-adiabatically\footnote{
The singular (Newtonian-like point source potential) of the SCO  is the dominant contribution to the change in the motion of the accretion-disk mass element. The regular field of the SCO is due to gravitational radiation from the SCO, and although this does contribute, it does so at very high PN order relative to the contribution from the singular field of the SCO.}.
We therefore approximate
\begin{align}
    h_{\mu \nu}^{\mathrm{int}}\left(z; z'\right) = h_{\mu\nu}^{\mathrm{S}}(z;z') + \mathcal{O}(1) \,.
\end{align}
Analytic expressions for the singular field near the worldline of an object are available from~\cite{Pound_2014,Heffernan_2012}. 
To use these expressions, we recall certain facts from the theory of bi-tensors without proving them here; for a comprehensive review, see~\cite{Poisson_2011}.

Consider an arbitrary base point $x'{}^{\mu}$ on the worldline $z'{}^{\mu}$. 
The parallel propagator from point $x'{}^{\mu}$ to $z^{\mu}$ is denoted by $g^{\alpha'}_{\beta}(z^{\mu},x'{}^{\mu})$. Let $\sigma(x^{\mu},x'{}^{\mu})$ denote Synge's worldline function and define~\cite{Pound_2014,Heffernan_2012}
\begin{subequations}
\begin{align}
    &P_{\mu' \nu'} \equiv g_{\mu' \nu'} + p_{\mu'}p_{\nu'} \,,\\
    &\sigma_{\mu'} \equiv \partial_{\mu'} \sigma(x,z') \,,\\
    &\textsf{s} \equiv \sqrt{P^{\mu' \nu'}\sigma_{\mu'} \sigma_{\nu'}} \,,\\
    &\textsf{r} \equiv p_{\mu'} \sigma^{\mu'} \,.
\end{align}
\end{subequations}
With these definitions, the singular field $h_{\mu\nu}^{\mathrm{S}}(z;z')$ in Lorenz gauge can be written as
\begin{align}\label{eq:sing-func-expr-1}
    h^{\mathrm{S}}_{\mu\nu}(z;z')
    &=
    \frac{2 \, m}{\textsf{s}} g^{\alpha'}_{\mu} g^{\beta'}_{\nu} \left(g_{\alpha' \beta'} + 2 p_{\alpha'} p_{\beta'} \right)
    +
    \mathcal{O}(|z-z'|) \,.
\end{align}
Observe that the base point $x'{}^{\mu}$ in the above equation is arbitrary but, to ensure fast converge of the approximation used above, it has to sufficiently close to $z^{\mu'}$. 
To simplify Eq.~\eqref{eq:sing-func-expr-1} further, we use near-coincidence expansions and define
\begin{align}\label{eq:delta-z-def}
    \Delta z^{\alpha'} \equiv z^{\alpha} - x^{\alpha'}\,.
\end{align}
In our analysis, we choose $x'{}^{\mu}$ to have the same time coordinate as $z^{\mu}(t)$, and this implies that $\Delta t' = t- t' = 0$. Using standard techniques from the theory of bi-tensors~\cite{Poisson_2011}, we can obtain the Taylor-series expansions for the propagator and $\sigma_{\mu'}$,
\begin{align}
    g^{\alpha'}_{\beta} 
    &= \delta^{\alpha'}_{\beta'} 
    + \Gamma^{\alpha'}_{\beta' \gamma'} \Delta z^{\gamma'}
    +
    \mathcal{O}\left[ \left(\Delta z\right)^2 \right] \,,\\
    \sigma_{\mu'}
    &=
    -g_{\mu' \alpha'} \Delta z^{\alpha'} - \frac{1}{2} \Gamma_{\mu' \alpha' \beta'} \Delta z^{\alpha'} \Delta z^{\beta'} + +
    \mathcal{O}\left[ \left(\Delta z\right)^3 \right] \,.
\end{align}
Substituting these expansions in Eq.~\eqref{eq:sing-func-expr-1}, we can approximate the singular potential as
\begin{align}\label{eq:sing-func-expr-2}
    h^{\mathrm{S}}_{\mu\nu}(z;z')
    &=
    2\frac{\left(m\right) }{\textsf{s}_0} A_{\mu'\nu'} (z;z')
    +\mathcal{O}(\Delta z).
\end{align}
where
\begin{subequations}\label{eq:bitensor-defs}
\begin{align}
    &\textsf{s}_0^2 \equiv P_{\mu' \nu'} \Delta z^{\mu'} \Delta z^{\nu'} - P_{\delta' (\gamma'} \Gamma^{\delta'}_{\alpha'\beta')} \Delta z^{\alpha'} \Delta z^{\beta'} \Delta z^{\gamma'} \,,\\
    &A_{\mu'\nu'} (z;z') = \mathcal{A}_{\mu'\nu'}^{(0)} + \mathcal{A}^{(1)}_{\mu' \nu' \gamma'} \Delta z^{\gamma'} 
    \,,
\end{align}
\end{subequations}
and the operators $\mathcal{A}_{\mu'\nu'}^{(0)}$ and $\mathcal{A}_{\mu'\nu'}^{(1)}$ are defined as
\begin{subequations}\label{eq:A-operators}
\begin{align}
    &\mathcal{A}_{\alpha' \beta'}^{(0)} = g_{\alpha' \beta'} + 2 p_{\alpha'} p_{\beta'} \,,\\
    &\mathcal{A}_{\mu'\nu'}^{(1)} = 2 \mathcal{A}_{\alpha' (\mu'}^{(0)} \Gamma^{\alpha'}_{\nu') \gamma'} \,,
\end{align}
\end{subequations}
with the metric and its associated Christoffel symbol evaluated on the Schwarzschild background at $x'{}^{\mu}$. 
Finally, we substitute Eq.~\eqref{eq:sing-func-expr-2} into Eq.~\eqref{eq:deltaH-v1} to obtain the total Hamiltonian 
\begin{align}
    \int d \mu \, \mathscr{H} &= \int d\mu (H - \mathcal{R} )\,,
\end{align}
where $H= -p_{t}$ and $\mathcal{R}(z,z')$ is the relativistic analog of the disturbing function in classical mechanics~\cite{Murray-Dermott-Book}
\begin{align}\label{eq:disturbing-function-for-disk}
    \mathcal{R} &\equiv \frac{p^{\mu} A_{\mu' \nu'} p^{\nu}}{p^{t} \textsf{s}_0(z,z')} \,.
\end{align}
The Hamiltonian for the SCO is obtained by repeating the same steps as above and is given by 
\begin{align}
     H'_{\mathrm{SCO}} = H'  - \int d\mu \, \mathcal{R}' \,,
\end{align}
where $H'$ is the Hamiltonian of SCO that describes geodesic motion and the disturbing function $\mathcal{R}'$ is obtained by replacing $z \leftrightarrow z'$ in Eq.~\eqref{eq:disturbing-function-for-disk}.

Given a general set of action-angle variables $(\mathbf{P}_{i},\mathbf{Q}^{i})$ and $(\mathbf{P}_{i'},\mathbf{Q}^{i'})$, the equations of motion for the whole system can be obtained through Hamilton's equations, namely
\begin{subequations}\label{eq:hamiltonian-equations-3-body}
\begin{align}
    \dot{\mathbf{P}}_{i} &= \epsilon_{\mathrm{int}}m \frac{\partial \mathcal{R}}{\partial \mathbf{Q}^{i}} \,,\\
    \dot{\mathbf{Q}}^{i} &= \frac{\partial H}{\partial \mathbf{P}_{i}} - \epsilon_{\mathrm{int}}m\frac{\partial \mathcal{R}}{\partial \mathbf{P}^{i}} \,,\\
    \dot{\mathbf{P}}_{i'} &= \epsilon_{\mathrm{int}}\int d \mu \frac{\partial \mathcal{R}'}{\partial \mathbf{Q}^{i'}} \,,\\
    \dot{\mathbf{Q}}^{i'} &= \frac{\partial H'}{\partial \mathbf{P}_{i'}} - \epsilon_{\mathrm{int}}\int d \mu \frac{\partial \mathcal{R'}}{\partial \mathbf{P}^{i'}} \,,
\end{align}
\end{subequations}
where the over dots denote time derivatives, and we have introduced an order counting parameter $\epsilon_{\mathrm{int}}$ to recall that the interaction between the accretion disk and the SCO is treated as a perturbation.
\subsection{Resonance and secular evolution}\label{sec:resonance-evolution}
We now derive general formula for the secular variation of the orbital elements of the system described in Eq.~\eqref{eq:hamiltonian-equations-3-body}.
The formalism presented here is standard in Hamiltonian perturbation theory~\cite{GT-Uranus-Ring,K-Cole-perturbation-theory,Ogilvie_2007}.
We include this section, to provide a detailed derivation of the secular evolution equation for readers not familiar with the astrophysics literature~\cite{GT-Uranus-Ring}.
Readers familiar with these techniques can directly skip to Eqs.~\eqref{eq:evol-E-p-e-prime} and \eqref{eq:evol-ang-z-prime} where the final expressions are presented.

Begin by noting that the action-angle variables allow us to write down a general Fourier expansion for the disturbing function,
\begin{align}\label{eq:disturbing-function-fourier-expansion}
    \mathcal{R} &= \frac{1}{2}\sum_{j,j'} R_{\vec{\mathscr{j}},\vec{\mathscr{j}}'} e^{i \Phi\left(\vec{\mathscr{j}},\vec{\mathscr{j}}'\right)}
\end{align}
where $\vec{\mathscr{j}}$ and $\vec{\mathscr{j}}'$ are three dimensional vectors, and
\begin{align}\label{eq:Phi-j-jp-def}
   \Phi(\vec{\mathscr{j}},\vec{\mathscr{j}}') &\equiv \mathscr{j}_{k} \mathbf{Q}^k + \mathscr{j}_{k'} \mathbf{Q}^{k'} \,,
\end{align}
while
\begin{align}\label{eq:disturbing-function-fourier-coeffs}
   R_{\vec{\mathscr{j}},\vec{\mathscr{j}}'} &\equiv \frac{2}{(2\pi)^6} \int_{\left[-\pi, \pi \right]^6} d \Vec{\mathbf{Q}} d \Vec{\mathbf{Q}}' \mathcal{R} e^{-i \Phi(\vec{\mathscr{j}},\vec{\mathscr{j}}')} \,,
\end{align}
and we assume that
\begin{align}
    R_{\vec{\mathscr{j}},\vec{\mathscr{j}}'} =R_{-\vec{\mathscr{j}},-\vec{\mathscr{j}}'} \,,
    \qquad 
    R'_{\vec{\mathscr{j}},\vec{\mathscr{j}}'} =R'_{-\vec{\mathscr{j}},-\vec{\mathscr{j}}'} \,.
\end{align}
From the above assumptions, we conclude that $R_{\vec{\mathscr{j}},\vec{\mathscr{j}}'} $ is a real function, and a similar analysis reveals that $R'_{\vec{\mathscr{j}},\vec{\mathscr{j}}'}$ is also real.

The dynamics between the disk and the SCO proceeds via resonant interactions, so we present below a general formalism to treat such interactions that generalizes~\cite{ogilvie-gordon-2007,GT-Uranus-Ring}.
Assume that a particular argument $\Phi(\vec{\mathscr{j}},\vec{\mathscr{j}}')$ (with $\vec{\mathscr{j}}$ and $\vec{\mathscr{j}}'$ not simultaneously zero) is close to resonance
\begin{align}\label{eq:resonance-condition-Phi-dot}
    \dot{\Phi}(\vec{\mathscr{j}},\vec{\mathscr{j}}') \approx 0\,.
\end{align}
To understand how this resonant interaction leads to changes in the orbital elements, we perturbatively expand the action-angle variables. 
For example,
\begin{align}
    \mathbf{P}_{i} = \mathbf{P}_{i}^{(0)} + \epsilon_{\mathrm{int}} \mathbf{P}_{i}^{(1)} + \epsilon_{\mathrm{int}}^2 \mathbf{P}_{i}^{(2)} + \ldots\,,
\end{align}
and similar expansions hold for the other action and angle variables.
At zeroth order in perturbation theory, the actions $\mathbf{P}_{i}^{(0)}$ and $\mathbf{P}_{i'}^{(0)}$ are constants and the angles grow linearly with time
\begin{align}
    \mathbf{Q}^{i}_{(0)} = \nu^{i} t + c^{i} \,, \mathbf{Q}^{i'}_{(0)} = \nu^{i'} t + c^{i'} \,,
\end{align}
where $\nu^i = \partial H/\partial \mathbf{P}_{i}$ are the frequencies of the system [Eq.~\eqref{eq:frequencies-iota-equal-zero}] and $c^{i}$ and $c^{i'}$ are constants.
At first order in perturbation theory, we can use Eq.~\eqref{eq:hamiltonian-equations-3-body}, 
to obtain the following equations:
\begin{subequations}\label{eq:hamiltonian-equations-3-body-first-order}
\begin{align}
    \dot{\mathbf{P}}_{l}^{(1)} &= m \mathrm{Re}\left[R\frac{\partial }{\partial \mathbf{Q}^{l}} e^{i \Phi} \right] \,,\\
    \dot{\mathbf{Q}}^{l}_{(1)} &= \frac{\partial^2 H}{\partial \mathbf{P}_{l} \partial \mathbf{P}_{k}} \mathbf{P}_{k}^{(1)} - m \mathrm{Re}\left[e^{i \Phi}\frac{\partial R}{\partial \mathbf{P}^{l}}\right] \,,\\
    \dot{\mathbf{P}}_{l'}^{(1)} &= \int d \mu \mathrm{Re}\left[R'\frac{\partial }{\partial \mathbf{Q}^{l'}} e^{i \Phi}\right] \,,\\
    \dot{\mathbf{Q}}^{l'}_{(1)} &= \frac{\partial^2 H'}{\partial \mathbf{P}_{l'} \partial \mathbf{P}_{k'}}\mathbf{P}_{k'}^{(1)} - \int d \mu \mathrm{Re}\left[e^{i \Phi}\frac{\partial R'}{\partial \mathbf{P}^{l'}}\right] \,.
\end{align}
\end{subequations}
In the above equations we have suppressed the dependence of $(\vec{\mathscr{j}},\vec{\mathscr{j}}')$ in $R_{\vec{\mathscr{j}},\vec{\mathscr{j}}'}$ and $R'_{\vec{\mathscr{j}},\vec{\mathscr{j}}'}$.

The formal solution to the above equations are given by
\begin{subequations}\label{eq:hamiltonian-equations-3-body-first-order-sols}
\begin{align}
    \mathbf{P}_{l}^{(1)} &= m \mathrm{Re}\left[ f_0 R e^{i \Phi} \mathscr{j}_{l}\right] \,,\\
    \mathbf{Q}^{l}_{(1)} &= m \mathrm{Re}\left[ i e^{i \Phi}\frac{\partial(R f_0)}{\partial \mathbf{P}_l}\right] \,,\\
    \mathbf{P}_{l'}^{(1)} &= \int d\mu \mathrm{Re}\left[ f_0 R' e^{i \Phi} \mathscr{j}_{l'}\right] \,,\\
    \mathbf{Q}^{l'}_{(1)} &= \int d\mu \mathrm{Re}\left[ i e^{i \Phi}\frac{\partial(R' f_0)}{\partial \mathbf{P}_{l'}}\right] \,.
\end{align}
\end{subequations}
One could add arbitrary constants to the above solutions, but these can be absorbed into the zeroth-order solutions.
In the above expressions,
\begin{align}
    f_0 &\equiv \frac{1}{\dot{\Phi} - i \gamma} \,
\end{align}
and $\gamma$ is a small real parameter that we have introduced to dissipate the first-order resonance~\cite{GT-Uranus-Ring}. Observe that if the resonance is exact ($\gamma=0$), then the treatment we have provided above does not work, as the first-order solution would diverge. 
The treatment of such a resonance requires a detailed analysis of the resonant structure (see e.g.~Chapter 5 of~\cite{K-Cole-perturbation-theory} and Chapter 8 of~\cite{Murray-Dermott-Book}). 
In accretion disk physics, an exact resonance is prevented from collective effects, such as viscosity in the disk, and the phenomenological treatment presented above has been used extensively and shown to be valid~\cite{GT-Uranus-Ring,MEYERVERNET1987157,ogilvie-gordon-2007}.
Observe that the average of all the first-order solutions [Eq.~\eqref{eq:hamiltonian-equations-3-body-first-order-sols}] evaluates to zero because they depend on $e^{i \Phi}$, where we have defined the average via
\begin{align}
    &\left< F(\boldsymbol{Q}_{1},\ldots, \boldsymbol{Q}_n) \right>
    = 
    \frac{1}{(2\pi)^n} \int_{\left[-\pi, \pi \right]^n} d \Vec{\mathbf{Q}}_{1} d \Vec{\mathbf{Q}}_{2} \ldots d  \Vec{\mathbf{Q}}_{n}  f\,,
\end{align}
for any given function $F(\boldsymbol{Q}_{1},\ldots, \boldsymbol{Q}_n)$ of independent angle variables $(\boldsymbol{Q}_{1},\ldots, \boldsymbol{Q}_n)$.

We now consider the second-order perturbative solutions and obtain the secular evolution equations for the action variables.
The second-order equations are obtained by perturbing Eq.~\eqref{eq:hamiltonian-equations-3-body} and are given by 
\begin{subequations}
\begin{align}
    \label{eq:dotPl-v1}
    \dot{\mathbf{P}}_{l}^{(2)}
    &=m \mathrm{Re}\bigg[ \frac{\partial^2 (R e^{i \Phi})}{\partial \mathbf{Q}^{l} \partial \mathbf{Q}^{k'} } \mathbf{Q}^{k'}_{(1)} 
    + 
    \frac{\partial^2 (R e^{i \Phi})}{\partial \mathbf{Q}^{l} \partial \mathbf{P}_{k'} } \mathbf{P}_{k'}^{(1)}
    +
    \nonumber\\
    &+
    \frac{\partial^2 (R e^{i \Phi})}{\partial \mathbf{Q}^{l} \partial \mathbf{Q}^{k} } \mathbf{Q}^{k}_{(1)} 
    + 
    \frac{\partial^2 (R e^{i \Phi}) }{\partial \mathbf{Q}^{l} \partial \mathbf{P}_{k} } \mathbf{P}_{k}^{(1)}
    \bigg] \,\\
    \dot{\mathbf{P}}_{l'}^{(2)}
    &=
    \int d\mu \mathrm{Re}\bigg[ \frac{\partial^2 (R' e^{i\Phi})}{\partial \mathbf{Q}^{l'} \partial \mathbf{Q}^k } \mathbf{Q}^k_{(1)} + 
    \frac{\partial^2 (R' e^{i \Phi})}{\partial \mathbf{Q}^{l'} \partial \mathbf{P}_k } \mathbf{P}_k^{(1)}
    +
    \nonumber\\
    &+
    \frac{\partial^2 (R' e^{i\Phi})}{\partial \mathbf{Q}^{l'} \partial \mathbf{Q}^{k'} } \mathbf{Q}^{k'}_{(1)} + 
    \frac{\partial^2 (R' e^{i\Phi})}{\partial \mathbf{Q}^{l'} \partial \mathbf{P}_{k'} } \mathbf{P}_{k'}^{(1)}
    \bigg]\,.
\end{align}
\end{subequations}
We can simplify these equations by using Eq.~\eqref{eq:hamiltonian-equations-3-body-first-order-sols} and by taking the average. We present the details of the calculation in Appendix~\ref{appendix:secular-deriv}.
The final expressions are 
\begin{subequations}
\begin{align}
    \label{eq:Pldot-Plprimedot-1}
    \left<\dot{\mathbf{P}}_{l}\right>
    &= \frac{m^2 \mathscr{j}_{l} \mathscr{j}_{k}}{2} \frac{\partial }{\partial \mathbf{P}_{k}} \left[  \frac{(R)^2 \gamma}{\dot{\Phi}^2 + \gamma^2}\right]\,,\\
    \left<\dot{\mathbf{P}}_{l'}\right>
    &=
    \frac{m \mathscr{j}_{l'} \mathscr{j}_{k}}{2} \int d\mu \frac{\partial }{\partial \mathbf{P}_{k}} \left[ \frac{RR' \gamma}{\dot{\Phi}^2 + \gamma^2}\right] 
    \label{eq:Pldot-Plprimedot-2}
    \,.
\end{align}
\end{subequations}
If the parameter $\gamma$ is small, we can take the $\gamma \to 0$ limit of the above equations and use the identity
\begin{align}
    \lim_{\gamma \to 0} \frac{\gamma}{\gamma^2 + \dot{\Phi}^2} &= \pi \delta (\dot{\Phi})
\end{align}
to obtain
\begin{subequations}\label{eq:secular-evolution-equation}
\begin{align}
    \left<\dot{\mathbf{P}}_{l}\right>
    &= \frac{\pi m^2 \mathscr{j}_{l} \mathscr{j}_{k}}{2} \frac{\partial }{\partial \mathbf{P}_{k}} \left[  (R_{\vec{\mathscr{j}},\vec{\mathscr{j}}'})^2 \delta(\dot{\Phi}_{\vec{\mathscr{j}},\vec{\mathscr{j}}'}) \right]\,,\\
    \left<\dot{\mathbf{P}}_{l'}\right>
    &=
    \frac{\pi m \mathscr{j}_{l'} \mathscr{j}_{k}}{2}\int d\mu \frac{\partial }{\partial \mathbf{P}_{k}} \left[ R_{\vec{\mathscr{j}},\vec{\mathscr{j}}'}R'_{\vec{\mathscr{j}},\vec{\mathscr{j}}'} \delta(\dot{\Phi}_{\vec{\mathscr{j}},\vec{\mathscr{j}}'}) \right] \,,
\end{align}
\end{subequations}
where we have restored the dependence on the resonant argument.
This completes our derivation of the secular evolution equations.

With the secular evolution equations at hand, we can derive the evolution equations for functions of conserved quantities. 
For any functions $F(\mathbf{P})$ and $F'(\mathbf{P}')$, we have the following identities
\begin{subequations}\label{eq:general-evolution-eqn-for-elements}
\begin{align}
    \left<\dot{F}(\mathbf{P}) \right> = \frac{\partial F}{\partial \mathbf{P}_{k}} \left<\dot{\mathbf{P}}_k\right> \,,\\
    \left<\dot{F}'(\mathbf{P}') \right> = \frac{\partial F'}{\partial \mathbf{P}_{k'} } \left<\dot{\mathbf{P}}_k'\right>\,.
\end{align}
\end{subequations} 
In our analysis below, we use the relativistic modified Delaunay variables $(\Lambda,\lambda), (P,-\varpi)$ and $(Q,\mathscr{q})$ defined in Eq.~\eqref{eq:modified-Delaunay-variables} as the action-angle variables $(\mathbf{P},\mathbf{Q})$ of the system.
The resonant phase [Eq.~\eqref{eq:Phi-j-jp-def}] in modified Delaunay variables is given by
\begin{align}
    \Phi(\vec{\mathscr{j}}, \vec{\mathscr{j}}') = \mathscr{j}_{1} \lambda - \mathscr{j}_{2} \varpi - \mathscr{j}_{3} \Omega + 
    \mathscr{j}_{1'} \lambda' - \mathscr{j}_{2'} \varpi' - \mathscr{j}_{3'} \Omega'\,.
\end{align}
The angular momentum of the SCO along the $z$-axis is related to the modified Delaunay elements through
\begin{align}
    L'_{z} \equiv m \mathscr{L}'_{z} = m \left(\Lambda' - P' - Q'\right)\,.
\end{align}
We can use this relation with Eq.~\eqref{eq:general-evolution-eqn-for-elements} to obtain the secular evolution of the angular momentum of the SCO along the $z$-axis, namely 
\begin{align}\label{eq:evol-ang-z-prime}
   \left< \dot{L}_{z}' \right> &=  
   \frac{\pi m^2 (\mathscr{j}_{1'} - \mathscr{j}_{2'} - \mathscr{j}_{3'}) \mathscr{j}_{k}}{2} 
   \nonumber\\
   &
   \times\int d\mu \frac{\partial }{\partial \mathbf{P}_{k}} \left[ R_{\vec{\mathscr{j}},\vec{\mathscr{j}}'}R'_{\vec{\mathscr{j}},\vec{\mathscr{j}}'} \delta(\dot{\Phi}_{\vec{\mathscr{j}},\vec{\mathscr{j}}'}) \right]\!.
\end{align}
The secular evolution of the angular momentum along the z-direction of the disk can also be obtained using the same steps
\begin{align}
    \left< \dot{L}_{z} \right> &= \frac{\pi m^2 (\mathscr{j}_{1}- \mathscr{j}_{2} - \mathscr{j}_{3}) \mathscr{j}_{k}}{2}  
    \nonumber\\
    &
    \times \int d\mu  \frac{\partial }{\partial \mathbf{P}_{k}} \left[ (R_{\vec{\mathscr{j}},\vec{\mathscr{j}}'})^2 \delta(\dot{\Phi}_{\vec{\mathscr{j}},\vec{\mathscr{j}}'}) \right].
\end{align}
In Newtonian gravity, $R' = R$ and d'Alembert relation~\cite{2002mcma.book.....M} implies that $\mathscr{j}_{1} - \mathscr{j}_{2} - \mathscr{j}_{3} + \mathscr{j}_{1'} - \mathscr{j}_{2'} - \mathscr{j}_{3'} = 0$. Therefore, 
\begin{align}
     \left< \dot{L}_{z}' \right>_{\mathrm{Newt}} + \left< \dot{L}_{z} \right>_{\mathrm{Newt}} = 0\,.
\end{align}
In general relativity, the gravitational field itself carries angular momentum and the simple law of conservation of angular momentum is not valid. However, as we show below, the conservation statement is valid for certain closely spaced resonant interactions.

In our analysis below, we also need the evolution of the energy $E' = m \mathscr{E}$ [Eq.~\eqref{eq:E-andL-in-p-and-e}], the dimensionless semi-latus rectum $p'$, and the eccentricity $e'$ of the SCO in the limit of small eccentricity. These can be obtained through Eqs.~\eqref{eq:general-evolution-eqn-for-elements}, \eqref{eq:E-andL-in-p-and-e} and \eqref{eq:Lambda-P-Q-small-eccentricity-expansion}, and we find
\begin{widetext}
\begin{subequations}\label{eq:evol-E-p-e-prime}
\begin{align}
    &\left<\dot{E}'\right> = 
    \lim_{e\to 0, e' \to 0}
    \frac{m}{M(p')^{3/2}} 
    \left< \dot{\Lambda}' \right>
    +
    \frac{m\sqrt{p'-6}-\sqrt{p'}}{M (p')^2}
    \left< \dot{P}' \right>
    \,,\\
    &\left<\dot{p}'\right> = 
    \lim_{e\to 0, e' \to 0}
    \frac{2 (p'{}-3)^{3/2}}{M (p'{}-6)}
    \left< \dot{\Lambda}' \right>
    +
    \bigg[
    -\frac{2 \sqrt{(p'{}-3) p'{}} \left(\sqrt{(p'{}-6) p'{}^3}+p'{}-3 \sqrt{(p'{}-6) p'{}}-2\right)}{M (p'{}-6)^{3/2} p'{}}
    \bigg]
    \left< \dot{P}' \right>
    ,\\
    &\left<\dot{e}'\right> = 
    \lim_{e\to 0, e' \to 0}
    \bigg[ 
    -\frac{e'{} \sqrt{p'{}-3} (p'{} ((p'{}-12) p'{}+66)-108)}{2 M (p'{}-6)^2 (p'{}-2) p'{}}
    \bigg]
    \left< \dot{\Lambda}' \right>
    +
    \bigg[
    \frac{\sqrt{p'{}-3} (p'{}-2)}{e'{} M \sqrt{p'{}-6} p'{}^{3/2}}
    \bigg]
    \left< \dot{P}' \right>
    \,.
\end{align}
\end{subequations}
\end{widetext}
Equations~\eqref{eq:evol-E-p-e-prime} and \eqref{eq:evol-ang-z-prime} provide the secular evolutions of the conserved quantities of the SCO due to the interaction with the disk. Note that the mass element of the disk $d\mu$  or the surface density $\Sigma$ are free parameters in these expressions; we quantify its effect using different accretion disk models in Sec.~\ref{sec:results}.

\subsection{Expression for the relativistic disturbing function to first order in eccentricity and for motion in the orbital plane}\label{sec:disturbing-function-relativistic}
We here derive explicit expressions for the relativistic disturbing function $\mathcal{R}$ [Eq.~\eqref{eq:disturbing-function-fourier-expansion}] in a small eccentricity expansion when the SCO and the accretion-disk mass element are in the equatorial plane.
In Newtonian gravity, the derivation of an expression for the disturbing function has a long and rich history; see e.g.~Chapter 6 of~\cite{Murray-Dermott-Book} for  explicit expressions to second order in eccentricity.
One can derive explicit expressions for the disturbing function for arbitrary separations in Newtonian gravity, because we know the exact form of the interaction potential.
In general relativity, however, we only have a series expansion in terms of the separation of the object [Eqs.~\eqref{eq:sing-func-expr-2} and \eqref{eq:disturbing-function-for-disk}].
Therefore, we must assume that the SCO and the accretion disk element are sufficiently close together.

In our calculations, we will need to evaluate integrals of the following form: 
\begin{subequations}\label{eq:relativistic-Laplace-Coefficients}
\begin{align}
    b^{j}_{s}(\alpha,a,b) &\equiv
    \frac{1}{\pi}\int_{-\pi}^{\pi} d\Psi \frac{\cos( j \Psi)}{\left[a (1-\alpha)^2 + b\Psi^2 \right]^{s}} \,,\\
    \mathscr{s}^j_{s,k}(\alpha,a,b)
    &\equiv
    \frac{1}{\pi}\int_{-\pi}^{\pi} d\Psi \frac{ \Psi^{2k+1} \sin( j \Psi)}{\left[a (1-\alpha)^2 + b \Psi^2 \right]^{s}} \,,\\
    \mathscr{c}^j_{s,k}(\alpha,a,b)
    &\equiv 
    \frac{1}{\pi}\int_{-\pi}^{\pi} d\Psi \frac{ \Psi^{2k} \cos( j \Psi)}{\left[a (1-\alpha)^2 + b \Psi^2\right]^{s}} \,.
\end{align}
\end{subequations}
These integrals are closely related to the Laplace coefficient used in Newtonian theory when $a=1=b$~\cite{Murray-Dermott-Book}.
Intuitively, we can use $b^{j}_{s}(\alpha,1,1)$ to approximate the Laplace coefficient of a Newtonian disturbing function in the closely-spaced limit~\cite{GT-disc-satellite-interaction}.
The integrals, $\mathscr{s}^j_{s,k}$ and $\mathscr{c}^j_{s,k}$ can then be used approximately to quantify the deviation of the Laplace coefficient in Newtonian gravity from the approximation obtained using $b^{j}_{s}(\alpha,a=1,b=1)$.

The integrals defined above can all be related to each other by certain recursion relations, which we use extensively to simplify our expressions.
First, we note the following relation obtained by differentiation under the integral of $b^j_s$
\begin{align}\label{eq:recursion-bjs}
    \frac{\partial}{\partial \alpha} b^j_{s}
    &= -2 s (\alpha - 1) a b^j_{s+1} \,.
\end{align}
We also have
\begin{subequations}
\begin{align}
    &\frac{\partial^{2k+1}}{\partial j^{2k+1}} b^{j}_{s}(\alpha,a,b)
    =
    \nonumber\\
    &\frac{(-1)^{k+1}}{\pi}\int_{-\pi}^{\pi} d\Psi \frac{ \Psi^{2k+1} \sin( j \Psi)}{\left[a (1-\alpha)^2 + 2b(1- \cos(\Psi))\right]^{s}}  \,,\\
    &\frac{\partial^{2k}}{\partial j^{2k}} b^{j}_{s}(\alpha,a,b)
    =
    \nonumber\\
    &
    (-1)^k\frac{1}{\pi}\int_{-\pi}^{\pi} d \Psi \frac{ \Psi^{2k} \cos( j \Psi)}{\left[a (1-\alpha)^2 + 2b(1- \cos(\Psi))\right]^{s}} \,,
\end{align}
\end{subequations}
which implies
\begin{subequations}\label{eq:sjsk-cjsk-bsj-relation}
\begin{align}
    \mathscr{s}^j_{s,k}(\alpha,a,b)
    &=
    (-1)^{k+1}\frac{\partial^{2k+1}}{\partial j^{2k+1}} b^{j}_{s}(\alpha) \,,\\
    \mathscr{c}^j_{s,k}(\alpha,a,b)
    &=
    (-1)^k
    \frac{\partial^{2k}}{\partial j^{2k}} b^{j}_{s}(\alpha)
    \,.
\end{align}
\end{subequations}
Equations~\eqref{eq:recursion-bjs} and \eqref{eq:sjsk-cjsk-bsj-relation} can be used to generate $\mathscr{s}^j_{s,k}, \mathscr{c}^j_{s,k}$ and $b^j_{s+1}$ from $b^j_{s}$.
Finally, we note another important analytic relation in the limit $|1-\alpha| \ll 1$:
\begin{align}\label{eq:Bessel-function-relationship}
    b^j_{1/2}(\alpha, a, b) &= \frac{2}{\pi \sqrt{b} } K_0 \left(\sqrt{\frac{a}{b}} j |1-\alpha| \right)\,,
\end{align}
where $K_0 (\cdot)$ is a modified Bessel function of the second kind~\cite{GT-disc-satellite-interaction}. 
We provide a derivation of this result in Appendix~\ref{appendix:bessel-function-derivation}.
We use Eq.~\eqref{eq:Bessel-function-relationship} extensively in Sec.~\ref{sec:corotation-and-Lindblad-resonances}.

Our goal is to obtain a small eccentricity expansion of the disturbing functions $\mathcal{R}$ and $\mathcal{R}'$.
Our final result is of the form:
\begin{subequations}\label{eq:R-dist-function-expr-lin-e}
\begin{align}
    \mathcal{R} &= 
    \frac{1}{2}
    \sum_{j=-\infty}^{\infty} 
    \Bigg[R_{(-j,0,0),(j,0,0)} \cos(j\lambda'- j\lambda)
    \nonumber\\
    &+
    R _{(-j-1,-1,0),(j,0,0)} \cos(j\lambda' - (j+1) \lambda  + \varpi)
    \nonumber\\
    &+
    R_{(-j+1,1,0),(j,0,0)} \cos(j\lambda' -(j-1) \lambda - \varpi)
    \nonumber\\
    &+
    R _{(-j-1,0,0),(j,-1,0)} \cos(j \lambda' -(j+1) \lambda + \varpi')
    \nonumber\\
    &+
    R_{(-j+1,0,0),(j,1,0)} \cos(j \lambda' -(j-1) \lambda - \varpi')
    \Bigg]
    \nonumber\\
    &+
    \mathcal{O}(e^2,(e')^2,ee'), \\
    \mathcal{R}' &= 
    \frac{1}{2}
    \sum_{j=-\infty}^{\infty}  
    \Bigg[R'_{(-j,0,0),(j,0,0)} \cos(j\lambda'- j\lambda)
    \nonumber\\
    &
    +
    R' _{(-j-1,-1,0),(j,0,0)} \cos(j\lambda' - (j+1) \lambda  + \varpi)
    \nonumber\\
    &+
    R'_{(-j+1,1,0),(j,0,0)} \cos(j\lambda' -(j-1) \lambda - \varpi)
    \nonumber\\
    &+
    R' _{(-j-1,0,0),(j,-1,0)} \cos(j \lambda' -(j+1) \lambda + \varpi')
    \nonumber\\
    &+
    R'_{(-j+1,0,0),(j,1,0)} \cos(j \lambda' -(j-1) \lambda - \varpi')
    \Bigg]
    \nonumber\\
    &+
    \mathcal{O}(e^2,(e')^2,ee'),
\end{align}
\end{subequations}
where
\begin{subequations}\label{eq:scaling-relation-dist-func}
\begin{align}
    &R_{(-j,0,0),(j,0,0)}, R'_{(-j,0,0),(j,0,0)} = \mathcal{O}(1)   \,,\\
    &R_{(-j,0,0),(j,0,0)}, R'_{(-j,0,0),(j,0,0)} = \mathcal{O}(1) \,,\\
    &R_{(-j-1,-1,0),(j,0,0)},R' _{(-j-1,-1,0),(j,0,0)} = \mathcal{O}(e) \,,\\
    &R_{(-j+1,1,0),(j,0,0)},R'_{(-j+1,1,0),(j,0,0)} = \mathcal{O}(e) \,,\\
\label{eq:dist-for-corotation-1}
    &R_{(-j-1,0,0),(j,-1,0)}, R'_{(-j-1,0,0),(j,-1,0)} = \mathcal{O}(e')\,,\\
\label{eq:dist-for-corotation-2}
    &R_{(-j+1,0,0),(j,1,0)}, R'_{(-j+1,0,0),(j,1,0)} = \mathcal{O}(e')\,.
\end{align}
\end{subequations}
Explicit forms of the coefficients appearing in the above equation are provided in the supplementary \texttt{MATHEMATICA} notebook.
The derivation of these expressions uses the small eccentricity expansion of the orbital elements provided in Appendix~\ref{appendix:first-order-expansion} and the integrals defined in Eq.~\eqref{eq:relativistic-Laplace-Coefficients}.

Let us now schematically describe how we derive an expression for $R_{(-j,0,0),(j,0,0)}$; the derivation of the other terms in Eq.~\eqref{eq:R-dist-function-expr-lin-e} follow the same steps.
From the scaling relation provided in Eq.~\eqref{eq:scaling-relation-dist-func}, we see that to derive an expression for $R_{(-j,0,0),(j,0,0)}$ we can just set $e=0=e'$. 
Using Eqs.~\eqref{eq:orb-elements-expansion-small-eccentricity} and \eqref{eq:four-momentum-small-eccen} in Eqs.~\eqref{eq:delta-z-def} and \eqref{eq:bitensor-defs} we obtain
\begin{subequations}
\begin{align}
    \Delta z^{\alpha'} &= \left(0,M \Delta p',0,\Delta \lambda '\right) \,,\\
    \label{eq:s0-for-corot}
    \textsf{s}_0^2 &= 
    M^2(p')^2\bigg[ 
    \frac{\Delta \lambda'{}^2 (p'-2)}{p'-3}
    +
    \frac{\Delta \Tilde{p}{}'{}^2 p'}{p'-2}
    \nonumber\\
    &+
    \frac{\Delta \lambda'{}^2 \Delta \Tilde{p}{}' (p'-1)}{p'-3}
    -\frac{\Delta \Tilde{p}{}'{}^3 p'}{(p'-2)^2}
    \bigg]
\end{align}
\end{subequations}
where
\begin{align}
    \Delta \Tilde{p}{}' \equiv \frac{p - p'}{p'}\,.
\end{align}
Using the above equations in Eq.~\eqref{eq:A-operators} along with the metric and the Christoffel symbols of the Schwarzschild spacetime, we can simplify Eq.~\eqref{eq:disturbing-function-for-disk} to
\begin{align}
    \mathcal{R} &= 
    \frac{-4 p^{3/2} (p'{}-2) p'{}^{3/2}+p^3 \left(p'{}^2-3 p'{}+2\right)+(p'{}-1) p'{}^3}{\sqrt{p-3} p^{5/2} (p'{}-3) p'{} \textsf{s}_0}
    \nonumber \\
    &+
    \frac{2 \Delta \Tilde{pp}' (p'{}-1) \left(p^{3/2}-p'{}^{3/2}\right)^2}{\sqrt{p-3} p^{5/2} (p'{}-3) p'{} \textsf{s}_0}
    \,.
\end{align}
Observe that only the effective combination $\Delta \lambda' = \lambda - \lambda'$ of the angle variables appears in the expression above through $\textsf{s}_0$ [Eq.~\eqref{eq:s0-for-corot}]. 
Therefore, we can deduce that the Fourier expansion has to be of the form
\begin{align}
    &\mathcal{R} = \frac{1}{2}\sum_{j} R_{(-j,0,0),(j,0,0)} \cos(j\lambda' - j\lambda)
\end{align}
where
\begin{widetext}
\begin{align}\label{eq:R-corot-v1}
    R_{(-j,0,0),(j,0,0)} &= 
    \bigg[
    \frac{-4 p^{3/2} (p'{}-2) p'{}^{3/2}+p^3 \left(p'{}^2-3 p'{}+2\right)+(p'{}-1) p'{}^3}{\sqrt{p-3} p^{5/2} (p'{}-3) p'{} (p' M) }
    +
    \frac{2 \Delta \Tilde{p}' (p'{}-1) \left(p^{3/2}-p'{}^{3/2}\right)^2}{\sqrt{p-3} p^{5/2} (p'{}-3) p'{} (p'M)}
    \bigg]
    \nonumber\\
    &\times
    \frac{1}{\pi}
    \int_{\Delta \lambda =-\pi}^{\Delta \lambda = \pi} d \Delta \lambda  \frac{ (p' M)}{\textsf{s}_0} \cos( j \Delta \lambda)
    \,.
\end{align}  
\end{widetext}
Note that we can simplify the integral appearing in the expression above by using Eqs.~\eqref{eq:s0-for-corot} and~\eqref{eq:relativistic-Laplace-Coefficients} to
\begin{align}\label{eq:integral-1-1}
    \frac{1}{\pi}
    \int_{\Delta \lambda =-\pi}^{\Delta \lambda = \pi} d \Delta \lambda  \frac{ (p' M)}{\textsf{s}_0} \cos( j \Delta \lambda)
    &=
    b^{j}_{s}(\alpha,a,b)
\end{align}   
where
\begin{subequations}\label{eq:s-alpha-a-b-for-corot}
\begin{align}
    s &= \frac{1}{2}\,,\qquad
    b = 
    \frac{\Delta \Tilde{p}' (p'-1)}{p'-3}+\frac{p'-2}{p'-3}
    \,, \\
    \alpha &= 
    \frac{p}{p'}
    \,,\qquad
    a = \frac{p'}{p'-2}-\frac{\Delta \Tilde{p}' p'}{(p'-2)^2}\,.
\end{align}
\end{subequations}
The integral in Eq.~\eqref{eq:integral-1-1} can be evaluated analytically when $|p-p'|\ll p'$ by using Eq.~\eqref{eq:Bessel-function-relationship}.
When the analytic expression is substituted in Eq.~\eqref{eq:R-corot-v1}, we obtain the following, final expression for $R_{(-j,0,0),(j,0,0)}$:
\begin{widetext}
\begin{align}\label{eq:R-corot-v2}
    R_{(-j,0,0),(j,0,0)} &= 
    \bigg[
    \frac{-4 p^{3/2} (p'{}-2) p'{}^{3/2}+p^3 \left(p'{}^2-3 p'{}+2\right)+(p'{}-1) p'{}^3}{\sqrt{p-3} p^{5/2} (p'{}-3) p'{} (p' M) }
    +
    \frac{2 \Delta \Tilde{p}' (p'{}-1) \left(p^{3/2}-p'{}^{3/2}\right)^2}{\sqrt{p-3} p^{5/2} (p'{}-3) p'{} (p'M)}
    \bigg]
    \nonumber\\
    &\times
    \frac{2}{\pi \sqrt{b} } K_0 \left(\sqrt{\frac{a}{b}} j |1-\alpha| \right)
    \,,
\end{align}  
\end{widetext}
where $s,\alpha,a,b$ are given in Eq.~\eqref{eq:s-alpha-a-b-for-corot}.
\section{Corotation and Lindblad Resonances in a nearly circular and coplanar system}\label{sec:corotation-and-Lindblad-resonances}

We are now in a position to understand the different types of resonances that can be active during the interaction between the SCO and the accretion disk by analyzing the properties of the relativistic disturbing function.
In this section, we restrict our analysis to interactions where the disk and the SCO are on nearly circular [i.e., $e, e' \ll 1$ ] and coplanar orbits in the equatorial plane. 
The formalism outlined in Sec.~\ref{sec:perturbed-motion} can be extended to analyze more general configurations if one wishes, but we leave that to future work.

The treatment of resonances presented here closely follows the Newtonian gravity literature; see e.g.~Chapter~10 of~\cite{Murray-Dermott-Book}.
Recall that resonances occur when any argument $\Phi(\vec{\mathscr{j}},\vec{\mathscr{j}}')$ satisfies $\dot{\Phi}(\vec{\mathscr{j}},\vec{\mathscr{j}}')=0$ [Eq.~\eqref{eq:resonance-condition-Phi-dot}].
From the form of the disturbing function in Eq.~\eqref{eq:R-dist-function-expr-lin-e}, notice that one can have two distinct types of resonances
\begin{subequations}
\begin{align}
    \label{eq:corotation-resonance-condition}
    \dot{\Phi}_{\mathrm{cr}} &= j \dot{\lambda} ' - ( j+k) \dot{\lambda}  + k \dot{\varpi}' = 0\,,\quad k =0, \pm 1 \,,\\
    \label{eq:Lindblad-resonance-condition}
    \dot{\Phi}_{\mathrm{Lr}} &= j \dot{\lambda} ' - (j + k) \dot{\lambda} + k \dot{\varpi} = 0\,,\quad  k = \pm 1.
\end{align}
\end{subequations}
If $\dot{\Phi}_{\mathrm{cr}} = 0$, the system is said to be in a \textit{corotation resonance} and when $\dot{\Phi}_{\mathrm{Lr}} = 0$ the system is in a \textit{Lindblad resonance}.
This terminology originates from galactic dynamics~\cite{Binney-Tremaine-Book}, where one analyzes the pattern speed of a perturbing potential; see e.g.~Sec.10.3 of~\cite{Murray-Dermott-Book} for a discussion of this terminology. 

The organization of the rest of this section is as follows. In Secs.~\ref{sec:corotation-resonace-expressions} and \ref{sec:Lr-resonace-expressions}, we analyze corotation and Lindblad resonances and provide analytic expressions for the differential torque and evolution of eccentricity of the SCO. We present an independent check of our calculations by using a local Fermi frame to derive the leading-order, one-sided Lindblad torque on the SCO in Sec.~\ref{sec:local-model-expressions}.
\subsection{Corotation resonance}\label{sec:corotation-resonace-expressions}
Let us first analyze the corotation resonance. The location of a corotation resonance can be obtained from Eq.~\eqref{eq:corotation-resonance-condition} by using Eq.~\eqref{eq:frequencies-iota-equal-zero}. For large values of $j$, the location of the resonance is given by
\begin{widetext}
\begin{align}\label{eq:location-corotation-resonance}
    p_{\mathrm{cr}} &= 
    \begin{cases}
        p',\,\quad  k=0\,, \\
        p'\bigg[
        1
        +
        \frac{2 k \sqrt{\frac{p'{}-6}{p'{}}}}{3 j}
        +
        \frac{k^2 \left(5 p'{}-6 \sqrt{p'{}-6} \sqrt{p'{}}-30\right)}{9 j^2 p'{}}
        +
        \mathcal{O}\left(j^{-3} \right)
        \bigg]
        \,, \quad k = \pm 1\,.
    \end{cases}
\end{align}
\end{widetext}
From the equation above, we see that when $k=0$, the apparent location of the corotation resonance is exactly at the location of the SCO. In reality, however, the location is shifted from $p'$ because of local fluid properties of the disk, and 
the perturbative approach we use here breaks down during this resonance.
Analyzing the properties of this ``co-orbital'' corotation resonance requires a global analysis or a fully numerical approach, and these torques are prone to saturation unlike Lindblad resonances~\cite{2002ApJ...565.1257T,Baruteau_2014}.
Given these complications, a simple orbital resonance picture is not capable of accurately modeling this resonance, and we leave a detailed analysis to future work.

\begin{figure*}[thp!]
    \centering
    \includegraphics[width=0.45\linewidth]{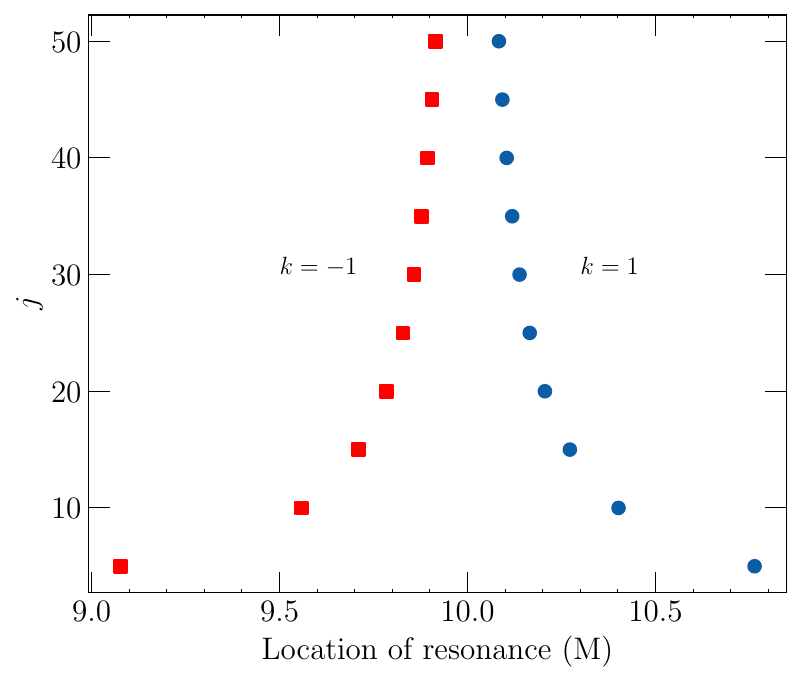}
    \includegraphics[width=0.45\linewidth]{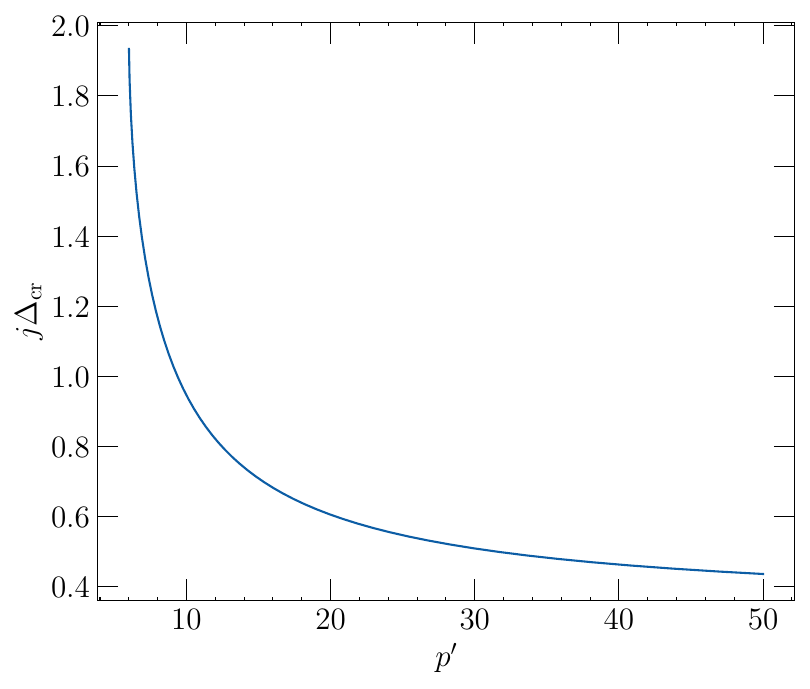}
    \caption{Location of the $k=\pm 1$ corotation resonances as a function of the position of the SCO. In the left panel we show the distribution of the inner (red squares) and outer (blue dots) corotation resonances when the SCO position is fixed at $p'= r/M = 10$ using Eq.~\eqref{eq:location-corotation-resonance}. Observe that the resonances are approximately symmetrically distributed. In the right panel, we quantify the asymmetry in the location of inner and outer resonances by plotting $j \Delta_{\mathrm{cr}}$ [Eq.~\eqref{eq:Delta-corot}] as a function of $p'$. Observe that $j \Delta_{\mathrm{cr}}$ is always positive which indicates that the outer corotation resonances are closer to the SCO than the inner corotation resonances.}
    \label{fig:corot-resonance-locs}
\end{figure*}
The non-co-orbital corotation resonance (i.e.~the one that occurs when $k = \pm 1$, on the other hand, can be studied in more depth.
In the left panel of Fig.~\ref{fig:corot-resonance-locs}, we show the location of the $k=\pm 1$ resonances for $p'=10$ and different values of $j$. 
The blue dots (red squares) located to the right (left) of $p'=10$ represent the location of the $k=1$ ($k=-1$) resonance, and are called outer (inner) corotation resonances.
Observe in Fig.~\ref{fig:corot-resonance-locs} that the corotation resonances are approximately symmetrically located around the location of the SCO ($p'=10$).
This behavior can be understood from Eq.~\eqref{eq:location-corotation-resonance}, from which we see that, to $\mathcal{O}(j)$, the outer ($k=1$) and inner ($k=-1$) corotation resonances are distributed symmetrically around $p'$ for a given value of $j$. 

The $\mathcal{O}(j^{-2})$ correction to the location of the corotation resonances leads to an asymmetry in their position, which we can understand by analyzing the ratio of the separation of the resonance from the location of the SCO, namely
\begin{align}\label{eq:Delta-corot}
    \Delta_{\mathrm{cr}} &\equiv \frac{p' - p_{\mathrm{cr,k=-1}} }{p_{\mathrm{cr,k=1}} - p'} -1 \nonumber\\
    &
    =
    \frac{-5 p'+6 \sqrt{p'-6} \sqrt{p'}+30}{3 j \sqrt{p'-6} \sqrt{p'}}
    +
    \mathcal{O}(j^{-2})
    \,,
\end{align}
as a function of $p'$ and for large values of $j$. If $\Delta_{\mathrm{cr}} >0$ then the outer corotation resonances are closer to the SCO than the inner ones and the opposite is true when $\Delta_{\mathrm{cr}}<0$.
In the Newtonian limit ($p' \gg 1$), we see that $\Delta_{\mathrm{cr}} >0$, and so
\begin{align}
    \Delta_{\mathrm{cr},\mathrm{Newt}} = \frac{1}{3j} + \mathcal{O}(j^{-2}) >0 \,,
\end{align}
which means that the outer corotation resonances are always closer to the SCO than the inner ones.
This remains true when we include relativistic effects, as we see from the right panel of Fig.~\ref{fig:corot-resonance-locs}. 
Note that, as we approach the innermost stable orbit $(p'=6)$,
\begin{align}
    \lim_{p' \to 6} \Delta_{\mathrm{cr}} = \frac{2}{j} \,.
\end{align}
Hence, as relativistic effects become important, the resonances shift closer to the SCO, and we expect a large effect near the innermost stable orbit.

How does a corotation resonance affect the orbital dynamics of the SCO?
For a $k=\pm 1$ corotation resonance, we have that
\begin{align}\label{eq:jvec-corot}
    \vec{\mathscr{j}} = \left(-k-j,0,0 \right)\,, \vec{\mathscr{j}}' = \left(j, -k, 0 \right)\,.
\end{align}
Using this, we see from Eq.~\eqref{eq:secular-evolution-equation} that the modified Delaunay variables $P,Q$ and $Q'$ do not evolve secularly, i.e.~
\begin{align*}
    \left< \dot{P}\right>_{\mathrm{cr}} = 0 = \left< \dot{Q} \right>_{\mathrm{cr}} = 0 = \left<\dot{Q}'\right>_{\mathrm{cr}}\,.
\end{align*} 
Therefore, a $k=\pm 1$ corotation resonance only leads to a secular evolution of $\Lambda, \Lambda'$ and $P'$, at best.
Moreover, the terms in the disturbing functions of the disk and the satellite that are responsible for a $k=\pm 1$ corotation resonance scale as $\mathcal{O}(e')$ [see Eqs.~\eqref{eq:corotation-resonance-condition}, \eqref{eq:dist-for-corotation-1} and \eqref{eq:dist-for-corotation-2}.].
This implies that the secular evolution of $\Lambda$ and $\Lambda'$, which directly quantify the rate of evolution of the semi-major axis of the disk and the SCO, scale with eccentricity and they can be ignored when the eccentricity is zero, see Eq.~\eqref{eq:evol-E-p-e-prime}.

Corotation resonances, however, do lead to a change in the eccentricity of the SCO because of the secular change in $P'$. 
Using Eqs.~\eqref{eq:secular-evolution-equation} and \eqref{eq:jvec-corot}, we see that, in the limit of $(e', e) \to 0$, we have
\allowdisplaybreaks[4]
\begin{align}\label{eq:Pp-evol-corot}
    &\left<\dot{P}'\right>_{\mathrm{cr}}
    =
    \frac{\pi m k (j+k) }{2}\int d\mu \frac{\partial }{\partial \Lambda} \left[ R_{\vec{\mathscr{j}},\vec{\mathscr{j}}'}R'_{\vec{\mathscr{j}},\vec{\mathscr{j}}'} \delta(\dot{\Phi}_{\mathrm{cr}}) \right]
    \nonumber\,,\\
    &=
    \frac{\pi m k (j+k) }{2}\int d\mu \frac{\partial p }{\partial \Lambda} \frac{\partial }{\partial p} \left[ R_{\vec{\mathscr{j}},\vec{\mathscr{j}}'}R'_{\vec{\mathscr{j}},\vec{\mathscr{j}}'} \delta(\dot{\Phi}_{\mathrm{cr}}) \right]
    \nonumber\,,\\
    &=
    \frac{\pi m k (j+k) }{2}
    \!\!
    \int 
    \!\!
    d\mu \frac{\partial p }{\partial \Lambda} \frac{\partial }{\partial p}\left[ R_{\vec{\mathscr{j}},\vec{\mathscr{j}}'}R'_{\vec{\mathscr{j}},\vec{\mathscr{j}}'} \left|\frac{\partial p }{\partial \dot{\Phi}_{\mathrm{cr}}} \right|\delta(p - p_{\mathrm{cr}}) \right].
\end{align}
To integrate the above equation, we use the mass element of a thin disk of surface density $\Sigma$ in the equatorial plane, namely
\begin{align}\label{eq:thin-disk-mass-element}
    d\mu = 2 \pi r \Sigma u^t dr = 2 \pi M^2 p \Sigma \sqrt{\frac{p}{p-3}} dp \,.
\end{align}
Substituting this into Eq.~\eqref{eq:Pp-evol-corot} and integrating, we obtain
\begin{align}\label{eq:P-prime-corot-evolution}
    \left<\dot{P}'\right>_{\mathrm{cr}}
    =
    -
    \frac{\pi m k (j+k) }{2}
    \bigg[
    R_{\vec{\mathscr{j}},\vec{\mathscr{j}}'}R'_{\vec{\mathscr{j}},\vec{\mathscr{j}}'} \left|\frac{\partial p }{\partial \dot{\Phi}_{\mathrm{cr}}} \right|
    \frac{d}{dp}\left[\frac{d\mu}{d\Lambda} \right]
    \bigg]_{p = p_{\mathrm{cr}}}
    \,.
\end{align}
The above equation can be used with Eq.~\eqref{eq:general-evolution-eqn-for-elements} to obtain $\left<\dot{e}'\right>$ [Eq.~\eqref{eq:evol-E-p-e-prime}], which simplifies to
\begin{align}
    \left< \dot{e}' \right>_{\mathrm{cr,k = \pm 1}} &= 
    -
    \frac{\pi m k (j+k) }{2}
    \frac{\sqrt{p'-3} (p'-2)}{e' M \sqrt{p'-6}\, {p'}^{3/2}}\times
    \nonumber\\
    &
    \bigg[
    R_{\vec{\mathscr{j}},\vec{\mathscr{j}}'}R'_{\vec{\mathscr{j}},\vec{\mathscr{j}}'} \left|\frac{\partial p }{\partial \dot{\Phi}_{\mathrm{cr}}} \right|
    \frac{d}{dp}\left[\frac{d\mu}{d\Lambda} \right]
    \bigg]_{p = p_{\mathrm{cr}}} \,.
\end{align}
This equation can be further simplified using the explicit expressions for $R$ and $R'$ in terms of the modified Bessel functions (see the supplementary \texttt{MATHEMATICA} notebook) to obtain the following schematic expression
\begin{align}\label{eq:edot-corot}
    \left< \dot{e}' \right>_{\mathrm{cr,k = \pm 1}}
    &= 
    j^2  \left< \dot{e}' \right>_{\mathrm{cr,k}}^{(0)}
    +
    j \left< \dot{e}' \right>_{\mathrm{cr,k}}^{(1)}
    +
    \mathcal{O}(1)
    \,.
\end{align}
The functions $\left< \dot{e}' \right>_{\mathrm{cr,k}}^{(0)}$ and $\left< \dot{e}' \right>_{\mathrm{cr,k}}^{(1)}$ are independent of $j$ and, in the Newtonian limit, they are given by~\cite{GT-disc-satellite-interaction,Ward-1986} 
\begin{widetext}
\begin{subequations}\label{eq:edot-corot-Newt}
\begin{align}
    \left< \dot{e}' \right>_{\mathrm{cr,k,Newt}}^{(0)}
    &= 
    -2 k e' \sqrt{p'{}} q \left(2 K_0\left(\frac{2}{3}\right)+K_1\left(\frac{2}{3}\right)\right){}^2 \Sigma (p'{})
    -
    \frac{4}{3} k e' \sqrt{p'{}} q \left(2 K_0\left(\frac{2}{3}\right)+K_1\left(\frac{2}{3}\right)\right){}^2 \frac{d \log(\Sigma)}{d \log (p')} \Sigma (p'{})
    \,,\\
    \left< \dot{e}' \right>_{\mathrm{cr,k,Newt}}^{(1)}
    &=
    -2 e' \sqrt{p'{}} q \left(5 K_0\left(\frac{2}{3}\right)+K_1\left(\frac{2}{3}\right)\right) \left(2 K_0\left(\frac{2}{3}\right)+K_1\left(\frac{2}{3}\right)\right) \Sigma (p'{})
    \nonumber\\
    &
    -\frac{4}{9} e' \sqrt{p'{}} q \left(25 K_0\left(\frac{2}{3}\right)+8 K_1\left(\frac{2}{3}\right)\right) \left(2 K_0\left(\frac{2}{3}\right)+K_1\left(\frac{2}{3}\right)\right) \frac{d \log(\Sigma)}{d \log (p')} \Sigma (p'{})
    \nonumber\\
    &-\frac{1}{9} 8 e' \sqrt{p'{}} q \left(2 K_0\left(\frac{2}{3}\right)+K_1\left(\frac{2}{3}\right)\right){}^2 
    \Sigma''(p') p'{}^2
    \,,
\end{align}
\end{subequations}
\end{widetext}
where $q = m/M$ is the mass ratio.
The explicit analytic expressions for $ \left< \dot{e}' \right>_{\mathrm{cr,k}}^{(0,1)}$ including all relativistic corrections are provided in Appendix~\ref{appendix:secular-evol-resonances}.

The total secular rate of change of $e'$ is obtained by summing over all the inner ($k=-1$) and outer ($k=1$) resonances.
One can show that
\begin{align}\label{eq:e-cr-sum-zero}
    \left< \dot{e}' \right>_{\mathrm{cr,+}}^{(0)} + \left< \dot{e}' \right>_{\mathrm{cr,-}}^{(0)}
    =0\,.
\end{align}
The physical reason behind the cancellation of the leading order contributions to $\left< \dot{e}' \right>$ is related to the symmetric location of the inner and outer corotation resonances to $\mathcal{O}(j^{-1})$.
From Eq.~\eqref{eq:edot-corot-Newt}, it is easy to see that this is true in the Newtonian limit. Using the relativistic expressions from Appendix~\ref{appendix:secular-evol-resonances}, one can prove that the same is also true in the relativistic regime.

The next-order corrections to the secular evolution of the eccentricity, however, do not cancel because of the asymmetry in the location of the inner and outer corotation resonances.
Summing over all the resonances, and using Eq.~\eqref{eq:e-cr-sum-zero}, we obtain the following total secular change from Eq.~\eqref{eq:edot-corot}:
\begin{widetext}
\begin{align}\label{eq:edot-due-to-corot}
    \left< \dot{e}' \right>_{\mathrm{cr,tot}} 
    &= 
    \sum_{j\gg 1}
    \left< \dot{e}' \right>_{\mathrm{cr,k =  1}} +  \left< \dot{e}' \right>_{\mathrm{cr,k =  -1}}
    =
    \sum_{j \gg 1}
    j^2
    \bigg[\left< \dot{e}' \right>_{\mathrm{cr,+}}^{(0)} + \left< \dot{e}' \right>_{\mathrm{cr,-}}^{(0)} \bigg]
    +
    j 
    \bigg[\left< \dot{e}' \right>_{\mathrm{cr,+}}^{(1)} + \left< \dot{e}' \right>_{\mathrm{cr,-}}^{(1)} \bigg]
    +
    \mathcal{O}(1)
    \,, \nonumber \\
    &=
    \sum_{j = 1}^{j_{\mathrm{max,cr,+}}}
    \bigg[ j^2 \left< \dot{e}' \right>_{\mathrm{cr,+}}^{(0)}
    +
    j \left< \dot{e}' \right>_{\mathrm{cr,+}}^{(1)}
    \bigg]
    +
    \sum_{j=1}^{j_{\mathrm{max,cr,-}}}
    \bigg[ j^2 \left< \dot{e}' \right>_{\mathrm{cr,-}}^{(0)}
    +
    j \left< \dot{e}' \right>_{\mathrm{cr,-}}^{(1)}
    \bigg]
    +
    \mathcal{O}(1)
    \,,
    \nonumber\\
    &\approx 
    \left(
    \frac{j_{\mathrm{max,cr,+}}^3}{3}-\frac{j_{\mathrm{max,cr,-}}^3}{3}
    \right)\left< \dot{e}' \right>_{\mathrm{cr,+}}^{(0)}
    +
    \frac{j_{\mathrm{max,cr,+}}^2}{2}\left< \dot{e}' \right>_{\mathrm{cr,+}}^{(1)} + \frac{j_{\mathrm{max,cr,-}}^2}{2}\left< \dot{e}' \right>_{\mathrm{cr,-}}^{(1)}  \,.
\end{align}
\end{widetext}
In the last line, we used the asymptotic approximation $\sum_{j} j^n \approx j_{\mathrm{max,cr,\pm}}^{n+1}/(n+1)$, where $j_{\mathrm{max,cr,\pm}}$ are cutoff parameters that regulate the location of the inner and outer corotation resonances and Eq.~\eqref{eq:e-cr-sum-zero}. The cutoff parameters are related to the scale height of the disk, and we discuss them in more detail in Sec.~\ref{sec:results}.

\subsection{Lindblad resonance}\label{sec:Lr-resonace-expressions}
We now analyze Lindblad resonances [Eq.~\eqref{eq:Lindblad-resonance-condition}], whose location can be obtained from Eq.~\eqref{eq:Lindblad-resonance-condition} by using Eq.~\eqref{eq:frequencies-iota-equal-zero}. For large values of $j$, the locations of the resonances are given by
\begin{align}\label{eq:Lindblad-location}
    p_{\mathrm{Lr}} &= p' 
    \Bigg[1
    + \frac{2 k }{3 j} \left[1- \frac{6}{p'} \right]^{1/2}
    -
    \frac{k^2}{9 j^2}
    \left[1- \frac{18}{p'} \right]
    \Bigg]
    \nonumber\\
    &+
    \mathcal{O}(j^{-3})
    \,.
\end{align}
Similar to the $k=\pm 1$ corotation resonances, the Lindblad resonances are located symmetrically inside and outside the SCO to $\mathcal{O}(j^{-1})$, and the $\mathcal{O}(j^{-2})$ contribution leads to location asymmetry.
In analogy with corotation resonances, we refer to the $k=1$ ($k=-1$) Lindblad resonance as an ``outer Lindblad resonance'' (``inner Lindblad resonance'').

The location of the Lindblad resonances possesses a striking feature  that is absent from corotation resonances. Let us use 
\begin{align}\label{eq:Delta-LR}
    \Delta_{\mathrm{Lr}} &\equiv \frac{p' - p_{\mathrm{Lr,k=-1}} }{p_{\mathrm{Lr,k=1}} - p'} -1 
    =
    \frac{p'-18}{3 j \sqrt{p'-6} \sqrt{p'}}
    +
    \mathcal{O}(j^{-2})
    \,,
\end{align}
to quantify the ratio of asymmetry, as in the corotation case [Eq.~\eqref{eq:Delta-corot}]. 
In the Newtonian limit, we have
\begin{align}
    \Delta_{\mathrm{Lr,Newt}} = \frac{1}{3j} + \mathcal{O}(j^{-2}) > 0\,,
\end{align}
which implies that the outer Lindblad resonances are closer to the SCO than the inner ones.
However, when we include relativistic corrections, Eq.~\eqref{eq:Delta-LR} shows that $\Delta_{\mathrm{Lr}}>0$ only if $p'>18$.
Therefore, the inner Lindblad resonances shift closer to the SCO as the latter's orbit crosses $p'=18$.
We will show in Sec.~\ref{sec:results} that this effect leads to the reversal in the direction of the torque on the SCO due to the accretion disk.

Let us now understand how the Lindblad resonance leads to the secular evolution of the orbital elements of the SCO. From Eq.~\eqref{eq:Lindblad-resonance-condition}, we have that
\begin{align}\label{eq:jvec-LR}
    \vec{\mathscr{j}} = \left(-k-j,-k,0 \right)\,, \quad \vec{\mathscr{j}}' = \left(j, 0, 0 \right)\,.
\end{align}
Using the above equations, we see from Eq.~\eqref{eq:secular-evolution-equation} that the modified Delaunay variables $P',Q$ and $Q'$ do not evolve secularly, while $\Lambda, P$ and $\Lambda'$ do. 
The evolution of $\Lambda, P$ and $\Lambda'$ can be obtained through the same manipulations that helped us obtain Eq.~\eqref{eq:P-prime-corot-evolution}. The final expressions in the limit $(e, e') \to 0$ are given by
\begin{subequations}
\begin{align}
    &\left<\dot{\Lambda}\right>_{\mathrm{Lr}} = 
    \frac{\pi m^2 (k+j) k}{2} \frac{\partial e }{\partial P} \frac{\partial}{\partial e} \left[  (R_{\vec{\mathscr{j}},\vec{\mathscr{j}}'})^2 \delta(\dot{\Phi}_{\mathrm{Lr}}) \right]
    \,,\\
    &\left< \dot{P}\right>_{\mathrm{Lr}} = 
    \frac{\pi m^2 k^2}{2} \frac{\partial e }{\partial P} \frac{\partial}{\partial e} \left[  (R_{\vec{\mathscr{j}},\vec{\mathscr{j}}'})^2 \delta(\dot{\Phi}_{\mathrm{Lr}}) \right]
    \,,\\
    &\left< \dot{\Lambda}'\right>_{\mathrm{Lr}} =
    \nonumber\\
    & 
    -\frac{\pi m j k}{2} 
    \bigg[
    \frac{\partial e }{\partial P} \frac{\partial}{\partial e} \left[ R_{\vec{\mathscr{j}},\vec{\mathscr{j}}'} R'_{\vec{\mathscr{j}},\vec{\mathscr{j}}'} \right] \left|\frac{\partial p}{\partial \dot{\Phi}_{\mathrm{Lr}}} \right| \frac{d\mu}{dp}
    \bigg]_{p = \mathrm{p}_{\mathrm{Lr}}}
    \,,
\end{align}
\end{subequations}
Using these equations, we obtain the following secular evolution of the orbital elements $E',{L}_{z}', p'$ and $e'$ of the SCO (see, Eqs.~\eqref{eq:evol-ang-z-prime} and \eqref{eq:evol-E-p-e-prime}):
\begin{widetext}
\begin{subequations}
\begin{align}
    &\left<\dot{E}'\right>_{\mathrm{Lr}} = 
    -\frac{\pi m^2 j k}{2 M (p')^{3/2}} 
    \bigg[
    \frac{\partial e }{\partial P} \frac{\partial}{\partial e} \left[ R_{\vec{\mathscr{j}},\vec{\mathscr{j}}'} R'_{\vec{\mathscr{j}},\vec{\mathscr{j}}'} \right] \left|\frac{\partial p}{\partial \dot{\Phi}_{\mathrm{Lr}}} \right| \frac{d\mu}{dp}
    \bigg]_{p = \mathrm{p}_{\mathrm{Lr}}}
    \,,\\
    &\left< \dot{L}'_{z}\right>_{\mathrm{Lr}} = 
    -\frac{\pi m^2 j k}{2} 
    \bigg[
    \frac{\partial e }{\partial P} \frac{\partial}{\partial e} \left[ R_{\vec{\mathscr{j}},\vec{\mathscr{j}}'} R'_{\vec{\mathscr{j}},\vec{\mathscr{j}}'} \right] \left|\frac{\partial p}{\partial \dot{\Phi}_{\mathrm{Lr}}} \right| \frac{d\mu}{dp}
    \bigg]_{p = \mathrm{p}_{\mathrm{Lr}}}
    \,,\\
    &\left< \dot{p}'\right>_{\mathrm{Lr}} = 
    -\frac{\pi m j k (p'-3)^{3/2}}{M(p'-6)} 
    \bigg[
    \frac{\partial e }{\partial P} \frac{\partial}{\partial e} \left[ R_{\vec{\mathscr{j}},\vec{\mathscr{j}}'} R'_{\vec{\mathscr{j}},\vec{\mathscr{j}}'} \right] \left|\frac{\partial p}{\partial \dot{\Phi}_{\mathrm{Lr}}} \right| \frac{d\mu}{dp}
    \bigg]_{p = \mathrm{p}_{\mathrm{Lr}}}
    \,,\\
    &\left< \dot{e}'\right>_{\mathrm{Lr}} = 
    \bigg[ 
    \frac{e'{} \sqrt{p'{}-3} (p'{} ((p'{}-12) p'{}+66)-108)}{2 M (p'{}-6)^2 (p'{}-2) p'{}}
    \bigg]
    \frac{\pi m j k}{2} 
    \bigg[
    \frac{\partial e }{\partial P} \frac{\partial}{\partial e} \left[ R_{\vec{\mathscr{j}},\vec{\mathscr{j}}'} R'_{\vec{\mathscr{j}},\vec{\mathscr{j}}'} \right] \left|\frac{\partial p}{\partial \dot{\Phi}_{\mathrm{Lr}}} \right| \frac{d\mu}{dp}
    \bigg]_{p = \mathrm{p}_{\mathrm{Lr}}}
    \,.
\end{align}
\end{subequations}
\end{widetext}
The above equations can be simplified using the thin-disk mass element [Eq.~\eqref{eq:thin-disk-mass-element}] and the analytic expressions for the disturbing functions $R$ and $R'$ provided in the supplementary \texttt{MATHEMATICA} file. 
The final expressions after these simplifications lead to the following simpler forms for $k=\pm 1$:
\begin{subequations}
\begin{align}
    &\left<\dot{E}'\right>_{\mathrm{Lr},k} = 
    j^2  \left< \dot{E}' \right>_{\mathrm{Lr,k}}^{(0)}
    +
    j \left< \dot{E}' \right>_{\mathrm{Lr,k}}^{(1)}
    +
    \mathcal{O}(1)
    \,,\\
    \label{eq:Lindblad-torque-k-1}
    &\left< \dot{L}'_{z}\right>_{\mathrm{Lr}} = 
    j^2  \left< \dot{L}'_z \right>_{\mathrm{Lr,k}}^{(0)}
    +
    j \left< \dot{L}'_z \right>_{\mathrm{Lr,k}}^{(1)}
    +
    \mathcal{O}(1)
    \,,\\
    \label{eq:dotp-Lr}
    &\left< \dot{p}'\right>_{\mathrm{Lr}} = 
    j^2  \left< \dot{p}' \right>_{\mathrm{Lr,k}}^{(0)}
    +
    j \left< \dot{p}' \right>_{\mathrm{Lr,k}}^{(1)}
    +
    \mathcal{O}(1)
    \,,\\
    &\left< \dot{e}'\right>_{\mathrm{Lr}} = 
    j^2  \left< \dot{e}' \right>_{\mathrm{Lr,k}}^{(0)}
    +
    j \left< \dot{e}' \right>_{\mathrm{Lr,k}}^{(1)}
    +
    \mathcal{O}(1)
    \,.
\end{align}
\end{subequations}
In the Newtonian limit these expressions reduce to~\cite{Ward-1986}
\begin{widetext}
\begin{subequations}
\begin{align}
    \left< \dot{E}' \right>_{\mathrm{Lr,k,Newt}}^{(0)}
    &=
    -k
    \frac{4 M q^2 \left(2 K_0\left(\frac{2}{3}\right)+K_1\left(\frac{2}{3}\right)\right){}^2 \Sigma}{3 \sqrt{p'{}}}
    \,,\\
\label{eq:Edot-Newt-expr-LR}
    \left< \dot{E}' \right>_{\mathrm{Lr,k,Newt}}^{(1)}
    &=
    -\frac{8 M q^2 \left(2 K_0\left(\frac{2}{3}\right)+K_1\left(\frac{2}{3}\right)\right){}^2 }{9 \sqrt{p'{}}}
    \frac{d \log(\Sigma)}{d \log (p')} \Sigma (p'{})
    \nonumber\\
    &
    -\frac{4 M q^2}{9 \sqrt{p'{}}}\Sigma (p'{})
    \left(2 K_0\left(\frac{2}{3}\right)+K_1\left(\frac{2}{3}\right)\right) \left(7 K_0\left(\frac{2}{3}\right)+8 K_1\left(\frac{2}{3}\right)\right) 
    \,,
\end{align}
\end{subequations}
\end{widetext}
where $q$ is the mass ratio and
\begin{subequations}
\begin{align}
\label{eq:Ldot-Newt-expr-LR}
    \left< \dot{L}'_z \right>_{\mathrm{Lr,k,Newt}}^{(0,1)}
    &=
    M(p')^{3/2}
    \left< \dot{E}' \right>_{\mathrm{Lr,k,Newt}}^{(0,1)}
    \,,\\
    \left< \dot{p}' \right>_{\mathrm{Lr,k,Newt}}^{(0,1)}
    &=
    \frac{2 (p')^2}{q M }
    \left< \dot{E}' \right>_{\mathrm{Lr,k,Newt}}^{(0,1)}
    \,,\\
    \left< \dot{e}' \right>_{\mathrm{Lr,k,Newt}}^{(0,1)}
    &=
    -
    \frac{e' p'}{2 q M}
    \left< \dot{E}' \right>_{\mathrm{Lr,k,Newt}}^{(0,1)}
    \,.
\end{align}
\end{subequations}

The expressions for $\left< \dot{E}' \right>_{\mathrm{Lr,k}}^{(0)}$ and $\left< \dot{E}' \right>_{\mathrm{Lr,k}}^{(1)}$ including all relativistic corrections are given by
\begin{widetext}
\begin{subequations}\label{eq:Lindblad-E-expr-actual}
\begin{align}
    &\left< \dot{E}' \right>_{\mathrm{Lr,k}}^{(0)}
    =
    -\frac{16 k M (p'{}-3)^4 \sqrt{p'{}} q^2 \Sigma (p'{}) }{3 (p'{}-2)^2 ((p'{}-6) p'{})^{3/2}}
    \left[K_0\left(\frac{\kappa }{2 | A| }\right)+\frac{\sqrt{p'{}-6} }{2 \sqrt{p'{}-3}}K_1\left(\frac{\kappa }{2 | A| }\right)\right]^2
    \,,\\
\label{eq:Edot-Relativistic-expr-LR}
    &\left< \dot{E}' \right>_{\mathrm{Lr,k}}^{(1)}
    =
    -\frac{32 M (p'{}-3)^4 q^2}{9 (p'{}-6) (p'{}-2)^2 p'{}^{3/2}}
    \left[K_0\left(\frac{\kappa }{2 | A| }\right)+\frac{\sqrt{p'{}-6} }{2 \sqrt{p'{}-3}}K_1\left(\frac{\kappa }{2 | A| }\right)\right]^2
    \frac{d \log(\Sigma)}{d \log (p')} \Sigma (p'{})
    \nonumber\\
    &-
    \frac{8 M (p'{}-3)^2 \left(p'{}^2-5 p'{}+6\right) q^2}{27 (p'{}-6)^2 (p'{}-2)^5 p'{}^{3/2}}
    \left(21 p'{}^4-331 p'{}^3+1626 p'{}^2-3348 p'{}+2376 \right)\Sigma(p')\times
    \nonumber\\
    &
    \left[K_0\left(\frac{\kappa }{2 | A| }\right)+\frac{\sqrt{p'{}-6} }{2 \sqrt{p'{}-3}}K_1\left(\frac{\kappa }{2 | A| }\right)\right]
    \bigg[ 
    K_0\left(\frac{\kappa }{2 | A| }\right)
    +
    \frac{\left(4 \left(12 p'{}^4-190 p'{}^3+927 p'{}^2-1800 p'{}+1188\right)^2\right)}{\left(21 p'{}^4-331 p'{}^3+1626 p'{}^2-3348 p'{}+2376\right)^2}
    \times
    \nonumber\\
    &
    \left(\frac{3 (p'{} (p'{} (2 p'{}+5)-264)+396)}{8 p'{} (p'{} (2 p'{} (6 p'{}-95)+927)-1800)+9504}+\frac{7}{8}\right)
    \frac{\sqrt{p'{}-6}}{\sqrt{p'{}-3}}
    K_1\left(\frac{\kappa }{2 | A| }\right)
    \bigg]
    \,.
\end{align}
\end{subequations}
\end{widetext}
where
\begin{align}
    \frac{\kappa}{2|A|} = \frac{2 \sqrt{(p'{}-6) (p'{}-3)}}{3 (p'{}-2)}\,.
\end{align}
The expressions for $\left< \dot{L}'_z \right>_{\mathrm{Lr,k}}^{(0,1)}, \left< \dot{p}' \right>_{\mathrm{Lr,k}}^{(0,1)}$ and $\left< \dot{e}' \right>_{\mathrm{Lr,k}}^{(0,1)}$ can be derived from Eq.~\eqref{eq:Lindblad-E-expr-actual} using
\begin{subequations}\label{eq:expr-rel-Lindblad-all-others}
\begin{align}
\label{eq:Ldot-Relativistic-expr-LR}
    &\left< \dot{L}'_z \right>_{\mathrm{Lr,k}}^{(0,1)}
    =
    M (p')^{3/2} \left< \dot{E}' \right>_{\mathrm{Lr,k}}^{(0,1)}
    \,,\\
\label{eq:pdot-Relativistic-expr-LR}
    &\left< \dot{p}' \right>_{\mathrm{Lr,k}}^{(0,1)}
    =
    \frac{2 (p'{}-3)^{3/2} p'{}^{3/2}}{M (p'{}-6) q} \left< \dot{E}' \right>_{\mathrm{Lr,k}}^{(0,1)}
    \,,\\
    &\left< \dot{e}' \right>_{\mathrm{Lr,k}}^{(0,1)}
    =
    \frac{e'{} \sqrt{p'{}-3} \sqrt{p'{}} (108-p'{} ((p'{}-12) p'{}+66))}{2 M (p'{}-6)^2 (p'{}-2) q} 
    \nonumber\\
    &\hspace{2cm} \times \left< \dot{E}' \right>_{\mathrm{Lr,k}}^{(0,1)}
    \,.
\end{align}
\end{subequations}

The total secular rate of change of orbital elements is obtained by summing over all the inner ($k=-1$) and outer ($k=1$) resonances. Similar to Eq.~\eqref{eq:e-cr-sum-zero}, the $\mathcal{O}(j^2)$ contributions sum to zero and we obtain
\begin{widetext}
\begin{subequations}\label{eq:lindblad-resonance-total}
\begin{align}
    \left<\dot{E}'\right>_{\mathrm{Lr,tot}} 
    &= 
    \sum_{j\gg 1}
    \left<\dot{E}'\right>_{\mathrm{Lr,+}} + \left<\dot{E}'\right>_{\mathrm{Lr,-}}
    = 
    \bigg(\frac{j_{\mathrm{max,Lr,+}}^3}{3} - \frac{j_{\mathrm{max,Lr,-}}^3}{3}\bigg)\left< \dot{E}' \right>_{\mathrm{Lr,+}}^{(0)}
    +
    \frac{j_{\mathrm{max,Lr,+}}^2}{2}\left< \dot{E}' \right>_{\mathrm{Lr,+}}^{(1)} 
    \nonumber\\
    &+ \frac{j_{\mathrm{max,Lr,-}}^2}{2}\left< \dot{E}' \right>_{\mathrm{Lr,-}}^{(1)}
    \,,\\
    \label{eq:relativistic-Lindblad-torque}
    \left<\dot{L}'_z\right>_{\mathrm{Lr,tot}} 
    &= 
    \sum_{j\gg 1}
    \left<\dot{L}'_z\right>_{\mathrm{Lr,+}} + \left<\dot{L}'_z\right>_{\mathrm{Lr,-}}
    = 
    \bigg(\frac{j_{\mathrm{max,Lr,+}}^3}{3} - \frac{j_{\mathrm{max,Lr,-}}^3}{3}\bigg)\left< \dot{L}'_z \right>_{\mathrm{Lr,+}}^{(0)}
    +
    \frac{j_{\mathrm{max,Lr,+}}^2}{2}\left< \dot{L}'_z \right>_{\mathrm{Lr,+}}^{(1)} 
    \nonumber\\
    &+ \frac{j_{\mathrm{max,Lr,-}}^2}{2}\left< \dot{L}'_z \right>_{\mathrm{Lr,-}}^{(1)}
    \,,\\
    \left<\dot{p}'\right>_{\mathrm{Lr,tot}} 
    &= 
    \sum_{j\gg 1}
    \left<\dot{p}'\right>_{\mathrm{Lr,+}} + \left<\dot{p}'\right>_{\mathrm{Lr,-}}
    = 
    \bigg(\frac{j_{\mathrm{max,Lr,+}}^3}{3} - \frac{j_{\mathrm{max,Lr,-}}^3}{3}\bigg)\left< \dot{p}' \right>_{\mathrm{Lr,+}}^{(0)}
    +
    \frac{j_{\mathrm{max,Lr,+}}^2}{2}\left< \dot{p}' \right>_{\mathrm{Lr,+}}^{(1)} + \frac{j_{\mathrm{max,Lr,-}}^2}{2}\left< \dot{p}' \right>_{\mathrm{Lr,-}}^{(1)}
    \,,\\
    \left<\dot{e}'\right>_{\mathrm{Lr,tot}} 
    &= 
    \sum_{j\gg 1}
    \left<\dot{e}'\right>_{\mathrm{Lr,+}} + \left<\dot{e}'\right>_{\mathrm{Lr,-}}
    = 
    \bigg(\frac{j_{\mathrm{max,Lr,+}}^3}{3} - \frac{j_{\mathrm{max,Lr,-}}^3}{3}\bigg)\left< \dot{e}' \right>_{\mathrm{Lr,+}}^{(0)}
    +
    \frac{j_{\mathrm{max,Lr,+}}^2}{2}\left< \dot{e}' \right>_{\mathrm{Lr,+}}^{(1)} + \frac{j_{\mathrm{max,Lr,-}}^2}{2}\left< \dot{e}' \right>_{\mathrm{Lr,-}}^{(1)}
    \,.
\end{align}
\end{subequations}
\end{widetext}
In the above equations, $j_{\mathrm{max,Lr,}\pm}$ are cutoff parameters that regulate the inner and outer Lindblad resonances.
Equation~\eqref{eq:relativistic-Lindblad-torque}, when combined by the analytic expressions for $\left< \dot{L}'_z \right>_{\mathrm{Lr,\pm}}^{(0,1)}$ from Eq.~\eqref{eq:Lindblad-E-expr-actual}, 
describes the differential Lindblad torque exerted on the SCO due to its interaction with the disk.
To the best of our knowledge, this is the first time an analytic expression for the relativistic torque has been presented in literature.
In the Newtonian limit, our expression reduces to the ones derived in~\cite{GT-disc-satellite-interaction,Ward-1986}.
We caution the reader that our expression for the evolution of eccentricity due to Lindblad resonances is not accurate because the dominant correction to the torque arises from $\mathcal{O}(e^3)$ terms in the disturbing function which cannot be calculated here due to the truncation of the expansion of the disturbing function to $\mathcal{O}(e)$~\cite{Ward-1988}.
\subsection{Leading order Lindblad torque from a local model}\label{sec:local-model-expressions}
Consider a local Fermi frame close to the orbit of the SCO. 
In this frame, the relative velocity between the SCO and the disk material close to the orbit of the SCO is non-relativistic and the gravitational interaction is Newtonian.
Therefore, we can use the expressions for the torque already derived in Newtonian gravity~\cite{GT-disc-satellite-interaction} to obtain the torque on the SCO in this frame by using a coordinate transformation from the local frame to the global coordinate frame.
The main drawback of this method is that, while we can easily obtain the torque to leading order (the $\mathcal{O}(j^2)$ contribution in Eq.~\eqref{eq:Lindblad-torque-k-1}) in the interaction by recycling the expressions from Newtonian gravity, deriving the next-to-leading order terms (the $\mathcal{O}(j)$ contribution in Eq.~\eqref{eq:Lindblad-torque-k-1}) that contributes to the total differential Lindblad torque [Eq.~\eqref{eq:relativistic-Lindblad-torque}] is cumbersome.
Nevertheless, the method provides a direct consistency check of our calculation.

Let us begin by stating some results regarding Fermi coordinates for a geodesic worldline $x^{\mu} = \tilde{z}^{\mu}(t)$.
We denote the Fermi normal coordinates near this worldline by $(\hat{t}, \hat{x}^i)$.
Fermi normal coordinates are constructed by parallel transporting an orthonormal tetrad
$(\tilde{u}^{\mu}, e^{\mu}_{a})$; see e.g.~Sec.~9.2 of~\cite{Poisson_2011} for more details.
The components of the metric in Fermi normal coordinates are given by [Eq. 9.2 of~\cite{Poisson_2011}]
\begin{align}\label{eq:metric-FNC}
    &g_{\hat{t}\hat{t}} = -1 - R_{\hat{t}c\hat{t}d} \hat{x}^c \hat{x}^d  + \mathcal{O}(s^3) \,,\\
    &g_{\hat{t} a} = -\frac{2}{3} R_{\hat{t}cad}\hat{x}^c \hat{x}^d 
    + \mathcal{O}(s^3) \,,\\
    &g_{ab} = \delta_{ab} - \frac{1}{3} R_{acbd}\hat{x}^c \hat{x}^d
    + \mathcal{O}(s^3) \,.
\end{align}
where $R_{\hat{t}c\hat{t}d}$, $R_{\hat{t}cad}$ and $R_{acbd}$ are the frame components of the Riemann tensor evaluated on the worldline
\begin{subequations}
\begin{align}
    &R_{\hat{t}c\hat{t}d} = \left. R_{{\alpha} {\gamma} {\beta} {\delta}} \tilde{u}^{\alpha} e^{\gamma}_{c} \tilde{u}^{\beta} e^{\delta}_{d}
    \right|_{x=\tilde{z}}
    \,,\\
    &R_{\hat{t}cad} = \left. R_{\alpha \gamma \beta \delta} \tilde{u}^{\alpha} e^{\gamma}_{c} e^{\beta}_{c} e^{\delta}_{d} \right|_{x=\tilde{z}} \,,\\
    &R_{acbd} = \left.R_{\alpha \gamma \beta \delta} e^{\alpha}_{a} e^{\gamma}_{c} e^{\beta}_{c} e^{\delta}_{d}\right|_{x=\tilde{z}} \,,
\end{align}
\end{subequations}
and $s = \delta_{ab} \hat{x}^{a} \hat{x}^{b}$.
In the Fermi normal coordinates, the metric is Minkowski to leading order.
The relation between the global coordinate system and the Fermi normal coordinate system can be obtained from Eq.~(9.5) of~\cite{Poisson_2011}.

Instead of going over the details of the construction of Fermi normal coordinates for general geodesics, we restrict attention to an equatorial circular orbit located at $r = \Tilde{p} M$ in Schwarzschild coordinates.
The worldline is
\begin{align}
    \tilde{z}^{\mu}(\tilde{\tau}) = 
    \bigg( 
    \sqrt{\frac{\tilde{p}}{\tilde{p}-3}} (\tilde{\tau}-\tilde{\tau}_0),
     M \tilde{p},
    \frac{\pi}{2},
    \frac{(\tilde{\tau}-\tilde{\tau}_0)}{M \tilde{p}\sqrt{\tilde{p}-3}}
    + \tilde{\phi}_0
    \bigg)
    \,,
\end{align}
where $\tilde{\tau}$ is the proper time and the geodesic is at $\phi = \tilde{\phi}_0$ and $t=0$ when $\tilde{\tau}=\tilde{\tau}_0$.
The four-velocity vector and covector of the circular geodesic orbit are given by
\begin{subequations}
\begin{align}
    \tilde{u}^{\mu} &= \left(\frac{\sqrt{\Tilde{p}}}{\sqrt{\tilde{p}-3}},0,0, \frac{1}{M\tilde{p}\sqrt{\tilde{p}-3}} \right)\,,\\
    \tilde{u}_{\mu} &= \left( -\frac{\tilde{p}-2}{\sqrt{\tilde{p}(\tilde{p}-3)}},0, \frac{M \tilde{p}}{\sqrt{\tilde{p}-3}}\right)
    \,.
\end{align} 
\end{subequations}
A tetrad that is parallel propagated along this geodesic is~\cite{Parker-Luis}
\begin{subequations}
\begin{align}
    &e^{\tilde{\mu}}_0 = \tilde{u}^{\mu} \,,\\
    &e^{\tilde{\mu}}_{1} = 
    \bigg(-\frac{\sqrt{\tilde{p}}\sin\left[\tilde{\Omega} (\tilde{\tau} - \tilde{\tau}_0)\right]}{\sqrt{(\tilde{p}-2)(\tilde{p}-3)}} 
    ,
    \sqrt{\frac{\tilde{p}-2}{\tilde{p}}} \cos\left[ \tilde{\Omega} (\tilde{\tau} - \tilde{\tau}_0)\right],
    \nonumber\\
    &
    \quad
    0
    ,
    -
    \frac{\sqrt{\tilde{p}-2}}{M \sqrt{\tilde{p}-3}\,\tilde{p}}
    \sin\left[ \tilde{\Omega} (\tilde{\tau} - \tilde{\tau}_0)\right]
    \bigg)
    \,,\\
    &e^{\tilde{\mu}}_{2} = 
    \bigg(\frac{\sqrt{\tilde{p}}\cos\left[\tilde{\Omega} (\tilde{\tau} - \tilde{\tau}_0)\right]}{\sqrt{(\tilde{p}-2)(\tilde{p}-3)}} 
    ,
    \sqrt{\frac{\tilde{p}-2}{\tilde{p}}} \sin\left[ \tilde{\Omega} (\tilde{\tau} - \tilde{\tau}_0)\right]
    ,
    \nonumber\\
    &
    \quad
    0
    ,
    \frac{\sqrt{\tilde{p}-2}}{M \sqrt{\tilde{p}-3}\,\tilde{p}}
    \cos\left[ \tilde{\Omega} (\tilde{\tau} - \tilde{\tau}_0)\right]
    \bigg)
    \,,\\
    &e^{\tilde{\mu}}_{3} = \left(0,0,-\frac{1}{\tilde{p} M}, 0 \right)\,
\end{align}
\end{subequations}
where
\begin{align}
    \tilde{\Omega} = \frac{1}{M \tilde{p}^{3/2}} \,.
\end{align}
The dual tetrad is obtained by 
\begin{align}
    e^{0}_{\mu} = - u_{\mu} \,, e^{a}_{\mu} = \delta^{ab} g_{\mu \nu} e^{\nu}_{b}\,.
\end{align}
Consider a spacetime point in the neighborhood of the worldline with Schwarzschild coordinates $(t,r,\theta,\phi)$.
Using Eq. (9.5) of~\cite{Poisson_2011}, we can obtain the relation between global Schwarzschild coordinates and local Fermi normal coordinates system at this point. The metric can be obtained by using Eq.~\eqref{eq:metric-FNC}.
In our case, we use a comoving Fermi normal coordinate system $(t_c, x_{\mathrm{c}}, y_{\mathrm{c}}, z_{\mathrm{c}})$ defined by 
\begin{subequations}
\begin{align}
    t_c
    &=
    \Tilde{\tau}
    =
    \frac{\Omega^{-1} (\tilde{\phi}_0-\phi )+(\tilde{p}-2) (t-t_0)}{\sqrt{(\tilde{p}-3) \tilde{p}}}
    \,,\\
\label{eq:xc-global-rel}
    x_{\mathrm{c}} 
    &= 
    \hat{x} \cos\left[\tilde{\Omega} (\tilde{\tau} - \tilde{\tau}_0)\right]
    +
    \hat{y} \sin\left[\tilde{\Omega} (\tilde{\tau} - \tilde{\tau}_0)\right]
    \nonumber\\
    &=
    \sqrt{\frac{\tilde{p}}{\tilde{p}-2}}(r- \tilde{p} M)
    \,,\\
    y_{\mathrm{c}} &=  
    -\hat{x} \sin\left[\tilde{\Omega} (\tilde{\tau} - \tilde{\tau}_0)\right]
    +
    \hat{y} \cos\left[\tilde{\Omega} (\tilde{\tau} - \tilde{\tau}_0)\right]
    \nonumber\\
    &=
    -
    \frac{\sqrt{\tilde{p}-2}\bigg[(t - \tilde{t}_0) -  \tilde{\Omega}^{-1} (\phi- \tilde{\phi}_0)\bigg]}{\sqrt{\tilde{p}(\tilde{p}-3)}}
    \,,\\
    z_{\mathrm{c}} &=  \hat{z} = M \tilde{p} \bigg(\frac{\pi}{2} - \theta \bigg)
    \,.
\end{align}
\end{subequations}
The relation between the Fermi normal coordinates and the global Schwarzschild coordinates can be obtained by inverting the above equation, which yields
\begin{subequations}
\begin{align}
    &\hat{x} = x_c \cos\left[\tilde{\Omega} (\tilde{\tau} - \tilde{\tau}_0)\right] - y_c \sin\left[\tilde{\Omega} (\tilde{\tau} - \tilde{\tau}_0)\right]\,,\\
    &\hat{y} = y_c \cos\left[\tilde{\Omega} (\tilde{\tau} - \tilde{\tau}_0)\right] + y_c \sin\left[\tilde{\Omega} (\tilde{\tau} - \tilde{\tau}_0)\right] \,,\\
    &\hat{z} = z_c \,.
\end{align}
\end{subequations}
The components of the metric in comoving Fermi normal coordinate system are provided in Appendix~\ref{appendix:comoving-coordinates}.
For a general discussion of rotating local coordinate systems, see~\cite{Ni-Zimmermann} and Chapter 6 of~\cite{MTW}.

We now assume that the worldine $\tilde{z}^{\mu}$ is close to the worldlines of the SCO ($z^{\mu'}$) and a particle ($z^{\mu}$) in the accretion disk.
The velocity of the SCO and the particle in comoving Fermi normal coordinates are non-relativistic and the geodesic equations for the SCO reduce to 
\begin{subequations}\label{eq:geodesic-SCO-CFNC}
\begin{align}
    &\frac{d t_c^'}{d\tau'} = 1 \,,\\
    &\frac{d^2 x_c^'}{d (t_c^')^2} = 2 \tilde{\Omega} \frac{d y_c^'}{d t_c^'}
    +
    \frac{3 (\tilde{p}-2) x_c}{M^2 (\tilde{p}-3) \tilde{p}^3}
    \,\\
    &\frac{d^2 y_c^'}{d (t_c^')^2} = 
    - 2 \tilde{\Omega} \frac{d y_c^'}{d t_c^'}
    \,\\
    &\frac{d^2 z_c^'}{d (t_c^')^2} = 
    -\tilde{\nu}^2 z_c
    \,.
\end{align}
\end{subequations}
One can derive the above equations using the geodesic equations and truncating the equations to leading order in velocity and curvature (or directly by using Eq. (22) of~\cite{Ni-Zimmermann} and ignoring all relativistic terms of Table I of~\cite{Ni-Zimmermann}).
The geodesic equations of the particle are obtained by replacing $(t'_c, x'_{\mathrm{c}},y'_{\mathrm{c}},z'_{\mathrm{c}}) \to (t_c, x_{\mathrm{c}},y_{\mathrm{c}},z_{\mathrm{c}}) $ in the above equation.

Equation~\eqref{eq:geodesic-SCO-CFNC} is exactly the same as the Newtonian equation derived from a local model in Eqs. (32) and (33) of~\cite{GT-disc-satellite-interaction} if we identify $A$ in their expression with
\begin{align}\label{eq:A-GT-expr}
    A \equiv  -\frac{3 (\tilde{p}-2) }{4 M (\tilde{p}-3) \tilde{p}^{3/2}}\,,
\end{align}
Given that the interaction is Newtonian, we can use the expression from~\cite{GT-disc-satellite-interaction} directly.
We first note that the solution to Eq.~\eqref{eq:geodesic-SCO-CFNC} when $z_c=0$ is given by [Eq. (34) of~\cite{GT-disc-satellite-interaction}]
\begin{align}\label{eq:GT-sol-geodesic}
    x_c' = \alpha'\,, 
    y_c' =  2 A \alpha' t_c '\,,
\end{align}
and a similar expression holds for the particle in the disk.
In Sec. III of~\cite{GT-disc-satellite-interaction}, the change in $\alpha'$ due to the interaction of the particle and the disk is calculated during one encounter.
The end result is Eq.~(65) of their paper, which in our notation is
\begin{align}\label{eq:Delta-alpha-2}
    \Delta_2 \alpha' &= \frac{m d\mu \kappa^2}{32 |A|^5 B (\alpha' - \alpha)^5} \times
    \nonumber\\
    &\bigg[
    \frac{2\tilde{\Omega}}{\kappa} K_0 \left( 
    \frac{\kappa}{2 |A|} \right)
    +
    K_1 
    \left( 
    \frac{\kappa}{2 |A|} \right)
    \bigg]^2
\end{align}
where the operator $\Delta_2 \alpha$ denotes the second order secular change of $\alpha$, $d\mu$ is the accretion disk mass element [Eq.~\eqref{eq:thin-disk-mass-element}], $A$ was defined in Eq.~\eqref{eq:A-GT-expr}, and $\kappa$ is the epicyclic frequency
\begin{align}
    \kappa &\equiv
    \frac{\sqrt{\tilde{p}-6}}{M \sqrt{\tilde{p}-3} \tilde{p}^{3/2}}
    \,.
\end{align}
To calculate $dp'/dt$ using the method of~\cite{GT-disc-satellite-interaction} we note that, from Eqs.~\eqref{eq:GT-sol-geodesic} and \eqref{eq:xc-global-rel},
\begin{align}
    \Delta_2 \alpha' = \sqrt{\frac{\tilde{p}}{\tilde{p}-2}} \Delta r'
    =
    \sqrt{\frac{\tilde{p}}{\tilde{p}-2}} M\Delta_2 p'
    \,.
\end{align}
Next, we calculate
\begin{align}\label{eq:dpdt-GT}
    \frac{dp'}{dt}
    =
    \lim_{\tilde{p} \to p'}
    \frac{\Delta_2 p'}{\Delta t_{\mathrm{enc}}}
    \,,
\end{align}
where $\Delta t_{\mathrm{enc}}$ is the time elapsed in one encounter.
To calculate $\Delta t_{\mathrm{enc}}$, we note that, in global coordinates, the particle and the disk are at the same location relative to each other within a time
\begin{align}
    \Delta t_{\mathrm{enc}} 
    = 
    \frac{2\pi}{|\Omega(p) - \Omega(p')|} 
    \approx
    \frac{2\pi}{| \Delta r' \frac{d \Omega}{dr'}|}
    \approx 
    \frac{4 \pi (p')^{5/2} M^{2}}{3|\Delta r'|}
    \,.
\end{align}
Substituting this in Eq.~\eqref{eq:dpdt-GT} and approximating $\Sigma[p] = \Sigma[p']$ one obtains
\begin{align}
    \frac{dp'}{dt} = 
    -\frac{64 k m (p'{}-3)^{11/2} p'{}^2 \Sigma(p')}{81 M (p'{}-6) (p'{}-2)^2 (p-p'{})^4} dp
    \,,
\end{align}
after simplifying. We now integrate this expression over the resonances to find
\begin{align}
    &\left<\dot{p'} \right>_{k,\mathrm{local}}
    =
    k
    \int_{p = p' 
    \Bigg[1
    + \frac{2 k\sqrt{p'-6}}{3\sqrt{p'} j_{\mathrm{max,Lr,k}}}\Bigg]} ^{k\times\infty}
    \frac{dp'}{dt} 
    \nonumber\\
    &=
    -\frac{8 k j_{\mathrm{max,Lr,k}}^3 m (p'-3)^{11/2} \sqrt{p'} \Sigma (p')}{9 M (p'-6)^{5/2} (p'-2)^2}
    \times \nonumber\\
    &
    \bigg[2 K_0\left(\frac{\kappa}{2|A|}\right)
    +\frac{\sqrt{p'-6}}{\sqrt{p'-3}} K_1\left(\frac{\kappa}{2|A|}\right)\bigg]^2
    \,.
\end{align}
One can show that the above equation is exactly equal to the sum over resonances of the leading-order term in Eq.~\eqref{eq:dotp-Lr}
\begin{align*}
    \left<\dot{p'} \right>_{k,\mathrm{local}}
    =\sum_{j\gg 1}j^2  \left< \dot{p}' \right>_{\mathrm{Lr,k}}^{(0)}
    \,,
\end{align*}
using Eq.~\eqref{eq:pdot-Relativistic-expr-LR}.
We also note that the method described here can be immediately generalized to Kerr spacetime using Eq.~\eqref{eq:Delta-alpha-2} using the values of the epicyclic frequencies for Kerr geodesics~\cite{Gammie_2004}.
\section{Impact of relativistic corrections to accretion-disk environmental effects on an extreme mass-ratio inspiral}\label{sec:results}
In this section, we estimate the impact of relativistic corrections on the torque and power exchanged from orbit of the SCO due to its interaction with the disk. Given the large uncertainties in the properties of the accretion disk near the innermost stable circular orbit of a supermassive black hole, we choose to model the surface density of the accretion disk using a simple, parameterized model. Therefore, our results should be interpreted as order-of-magnitude estimates that highlight the impact of relativistic corrections in disk-SCO interactions close to a supermassive black hole.

The organization of this section is as follows.
We first introduce the parameterized disk model in Sec.~\ref{sec:accretion-disk-models}, and then compare the Newtonian torque formula to full relativistic expressions in Sec.~\ref{sec:Newt-rel-comparison}.
Finally, we discuss how the secular evolution of the energy of the SCO due to its interaction with the disk compares with the decay of energy due to gravitational wave emission in Sec.~\ref{sec:compare-disk-and-GW}.
\begin{figure*}[thp!]
    \centering
    \includegraphics[width=0.49\linewidth]{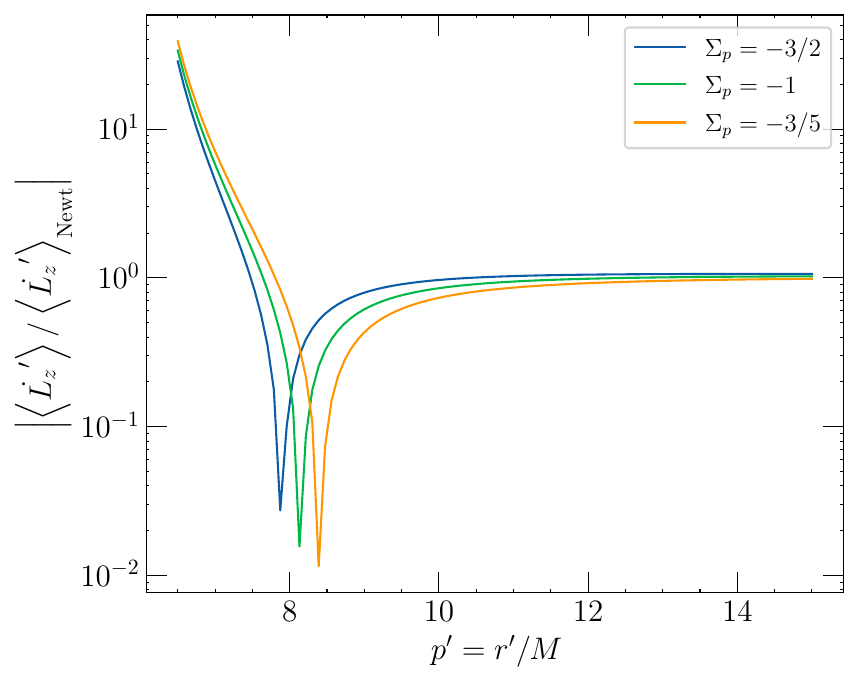}
    \includegraphics[width=0.49\linewidth]{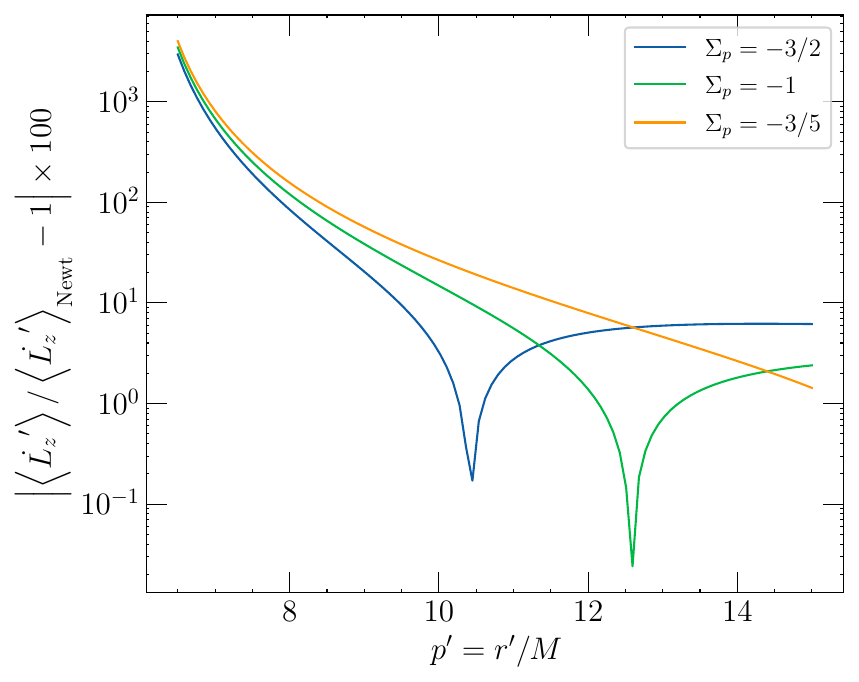}
    \caption{Comparison between Newtonian and relativistic torque expressions. The left panel shows the absolute value of the ratio of the differential relativistic Lindblad torque [Eq.~\eqref{eq:Ldot-Relativistic-expr-LR}] to the Newtonian torque [Eq.~\eqref{eq:Ldot-Newt-expr-LR}] as a function of the position of the SCO for different values of $\Sigma_{p}$. Observe that the ratio steeply increases as we approach the innermost stable circular orbit. The ratio also changes sign around $p' \sim 8$ because of the transition in the distribution of the inner and outer Lindblad resonances.
    In the right panel, we show the percentage difference between the Newtonian and relativistic torques.}
    \label{fig:torque-Newt-Rel}
\end{figure*}
\begin{figure*}[thp!]
    \centering
    \includegraphics[width=0.49\linewidth]{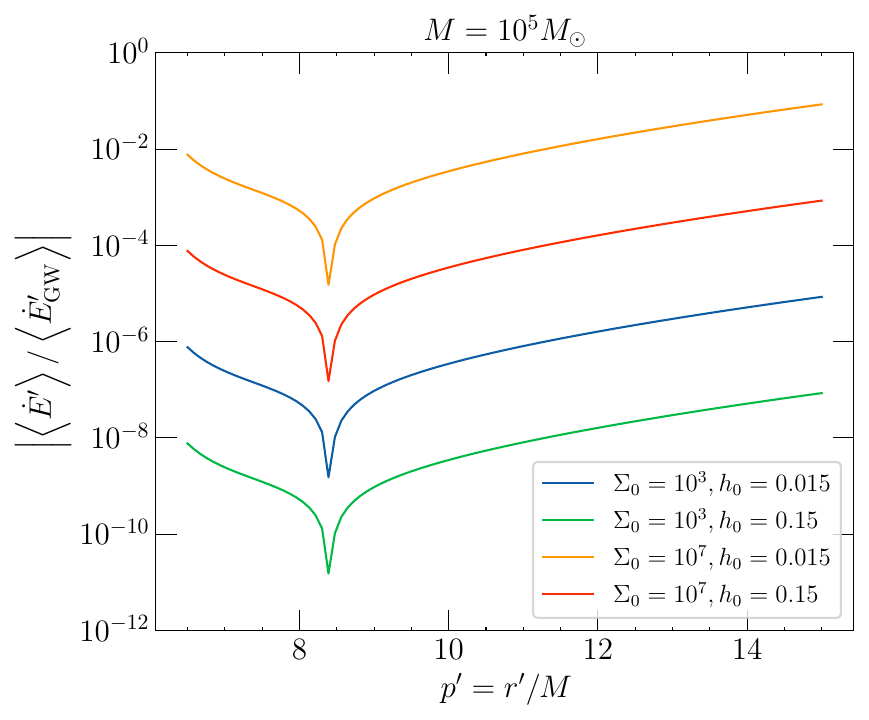}
    \includegraphics[width=0.49\linewidth]{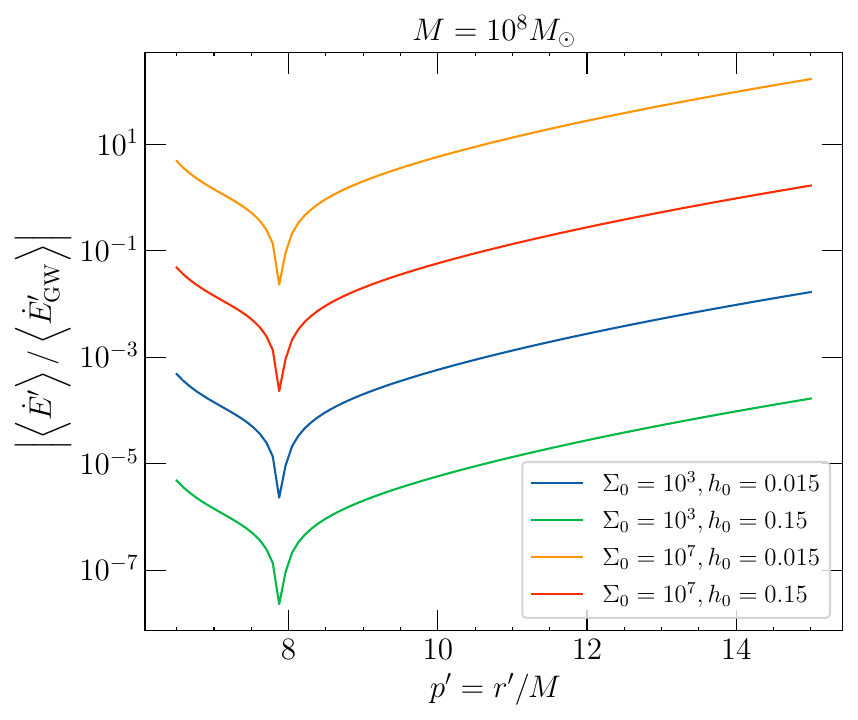}
    \caption{Comparison between energy exchange due to disk-SCO interaction to gravitational-wave emission for a supermassive black hole of mass $10^5 M_{\odot}$ (left) and $10^8 M_{\odot}$ (right) for different values of $\Sigma_{0}$ and $h_0$. From the left plot, we see that for disk surface densities characteristic of $\alpha$-disks $\Sigma_{0} \sim 10^3$, the energy exchange due to disk-SCO interaction is 6-8 orders of magnitude smaller than the loss due to gravitational-wave emission. For $\beta$-disk like values $\Sigma_{0} \sim 10^7$, the ratio increases to $10^{-2}-10^{-5}$ . 
    In right plot, we see that the ratio increases significantly because of the increase in supermassive black hole mass [Eq.~\eqref{eq:scaling-relationship-Edot-Ldot-edot}] and we see that for $\beta$-disk like surface densities the energy exchange due to the disk can compete with gravitational-wave loss.}
    \label{fig:dEDT_disk_GW}
\end{figure*}
\subsection{Parameterized accretion disk model}\label{sec:accretion-disk-models}

We are interested in analyzing the effect of disk-SCO interactions close to the supermassive black hole.  For EMRIs, this typically means orbital separations  $r\lesssim15 M$. In this regime, for high accretion rates where the disk is relatively massive, we expect the disk to be in a radiation-pressure-domianted regime, with opacity dominated by electron scattering~\cite{Abramowicz_2013}. 

Modeling relativistic, radiation-dominated disks is an area of active research.  Radiation pressure dominated disk simulations likely have a multi-layer structure, with a dense, lower radiation-to-gas-pressure layer at the midplane \cite{turner_2004, zhang2025radiationgrmhdmodelsaccretion}. They are also likely to be turbulent, exhibiting strong spiral features with surface density contrast of order $1$ \cite{jiang_2019_super}.  Absent a clear physical model we restrict our attention to phenomenological  accretion-disk models ~\cite{duque2025constrainingaccretionphysicsgravitational} that can provide order-of-magnitude estimates of the size of corrections induced from disk-SCO interactions.

Consider an accretion disk with surface density $\Sigma$, scale height $H(r)$ and aspect ratio $h(r) \equiv H(r)/r$.
We parameterize the surface density and the aspect ratio in terms of four constants $(\Sigma_{0},\Sigma_{p}, h_{0},\Sigma_{h})$\footnote{Our model has one more free parameter than that used in~\cite{duque2025constrainingaccretionphysicsgravitational}.}~\cite{duque2025constrainingaccretionphysicsgravitational}
\begin{align}\label{eq:disk-parameterization}
    &\Sigma = \Sigma_{0} \left(\frac{r}{10 M} \right)^{-\Sigma_{p}} \mathrm{g \; cm^{-2}}\,,\qquad
    h = h_0 \left(\frac{r}{10 M} \right)^{-\Sigma_{h}}
    \,.
\end{align}
The free parameters in the model $(\Sigma_{0}, \Sigma_{p}, h_0, \Sigma_{h})$ can be related to analytic models of $\alpha$- and $\beta$-disks [see Eqs. (33) and (34) of~\cite{Kocsis_2011}] via
\begin{widetext}
\begin{subequations}
\begin{align}
    &\Sigma_{0} = 
    \begin{cases}
        538.5 \left(\frac{\alpha}{0.1}\right)^{-1} \left(\frac{\dot{m}}{0.1} \right)^{-1} \mathrm{g \; cm^{-2}},\qquad \text{for $\alpha$ disks}\,,\\
        1.262 \times 10^{6} \left(\frac{\alpha}{0.1}\right)^{-4/5} \left(\frac{\dot{m}}{0.1} \right)^{3/5} \left(\frac{M}{10^5 M_{\odot}}\right)^{1/5} \mathrm{g \; cm^{-2}},\qquad \text{for $\beta$ disks}
        \,,
    \end{cases}
    \\
    &\Sigma_{p} = 
    \begin{cases}
        -3/2,\qquad \text{for $\alpha$ disks}\,,\\
        -3/5,\qquad \text{for $\beta$ disks}
        \,,
    \end{cases}
    \\
    &h_{0} = 0.15 \left(\frac{\dot{m}}{0.1} \right) \quad {\rm{and}} \quad \Sigma_{h} = 1, \qquad \text{for $\alpha$ and $\beta$ disks}\,,
\end{align}
\end{subequations}
\end{widetext}
where $M$ is the mass of the supermassive black hole, $\alpha$ is the turbulent viscosity parameter, and $\dot{m} = \dot{M}/\dot{M}_{\mathrm{edd}}$ is the accretion rate of the central black hole relative to the Eddington rate.  

Simulations of gas-dominated, magnetized disks find a wide range of possible $\alpha$ values, $\sim 0.001-0.2$, depending on the strength of vertical field and other factors (e.g. \cite{balbus_2003}), and we  consider models with $\alpha$ in this range.  The distribution of $\dot{m}$ is very broad, with only a small fraction of supermassive black holes accreting near $\dot{m} \sim 1$ (e.g. \cite{aird_2013}).  The thin disk model likely only applies for $10^{-3} \lesssim \dot{m} \lesssim 1$; lower and higher accretion rate systems probably transition to a geometrically thick, optically thin, radiatively inefficient state \cite{yuan_2014}. Here we consider models with $\dot{m} \sim 10^{-3} - 1$.  

The model defined above is capable of interpolating across
$\alpha$- and $\beta$-disks, which have different properties.
If $\dot{m} = 0.1$, then the values of $\Sigma_{0}$ are approximately in the range $\left[10^2, 10^4\right]$ for $\alpha$ disks and $\left[10^6, 10^8\right]$ for $\beta$ disks, when $M \in [10^5 M_{\odot}, 10^8 M_{\odot}]$.
Therefore, in our analysis, we assume that $\Sigma_{0} \in \left\{10^3, 10^7 \right\}$.
The scale height parameter $h_0 = 0.15$, if we set $\dot{m}=0.1$, but to discuss its impact on the SCO dynamics, we assume that $h_{0} \in \left\{ 0.015,0.15\right\}$.
Unless stated otherwise, we set $\Sigma_{h} = 1$.
We caution the reader that the large-$j$ approximation used to derive Eq.~\eqref{eq:lindblad-resonance-total} breaks-down in the regime when $h_0 \sim 0.15$, leading to a relative error of $\mathcal{O}(h_0) \sim \mathcal{O}(15 \%)$ (see, Eq.~\eqref{eq:jmax-assumption}).
However, the qualitative effects that we observe should persist when $h_0 \sim 0.15$, and we include this value of $h_0$ to highlight the differences in power and torque across the parameter space.

As described in Sec.~\ref{sec:corotation-and-Lindblad-resonances}, the secular evolution equations depend on $(\Sigma(p'), d\Sigma(p')/dp', j_{\mathrm{max,Lr,\pm}},j_{\mathrm{max,Cr,\pm}})$.
We can obtain $(\Sigma(p'), d\Sigma(p')/dp')$ from Eq.~\eqref{eq:disk-parameterization}.
Obtaining the cutoff parameters, on the other hand, is more subtle as it depends sensitively on the disk model~\cite{GT-disc-satellite-interaction,Ward-1986,1993ApJ...419..155A}. 
In this work, we adopt the common assumption that the cutoff parameters for all the Lindblad and corotation resonances are equal to each other, and they are approximately set by the ratio of the orbital velocity of the SCO to the local speed of sound~\cite{Ward-1986}.
More precisely, we set
\begin{align}\label{eq:jmax-assumption}
    j_{\mathrm{max}}
    &=
    j_{\mathrm{max,Lr,\pm}}
    =
    j_{\mathrm{max,Cr,\pm}}
    =
    \left.\frac{r \Omega}{c_{s}}\right|_{r=p'M}
    \approx
    \left.\frac{r \Omega}{H \Omega}\right|_{r=p'M}
    \nonumber\\
    &=
    \left.\frac{1}{h}\right|_{r=p'M}
    =
    \frac{1}{h_0} \left(\frac{p'}{10} \right)^{\Sigma_{h}}
    \,,
\end{align}
where we have used that $c_{s} = H \Omega$ and Eq.~\eqref{eq:disk-parameterization} in the first and second lines, respectively.

With these assumptions, we can calculate the ratio of the Newtonian to the relativistic torque for the differential Lindblad resonance, using the analytic expressions of Appendix~\ref{appendix:secular-evol-resonances}. 
While one can obtain exact expressions from Eqs.~\eqref{eq:edot-due-to-corot}, \eqref{eq:lindblad-resonance-total} and Appendix~\ref{appendix:secular-evol-resonances}, it is beneficial to understand the characteristic scaling relation for the power and the torque on the SCO due to the disk-SCO interactions to interpret the results presented in the next two subsections. These approximate, scaling relations are given by
\begin{subequations}\label{eq:scaling-relationship-Edot-Ldot-edot}
\begin{align}
    &\left<\dot{E}' \right> = \left<\dot{E}'\right>_{\mathrm{Lr,tot}} \propto 
    M q^2 \Sigma j_{\mathrm{max}}^2
    \propto \frac{M q^2 \Sigma_{0}}{h_0^2} (p')^{\Sigma_{h} - \Sigma_{p}}
    \,,\\
    &\left<\dot{L}'_z \right> = \left<\dot{L}'_z\right>_{\mathrm{Lr,tot}}
    \propto
    \frac{M^2 q^2 \Sigma_{0}}{h_0^2} (p')^{\Sigma_{h} - \Sigma_{p}}
    \,.
\end{align}
\end{subequations}
In the above equations, the right-hand sides depend only on the position of the SCO ($p'$) and $\Sigma_{p}$, because of the dependence on density gradients $d \Sigma /dp'$; see e.g.~Eq.~\eqref{eq:Edot-Relativistic-expr-LR}.

In our analysis below, we will also need the expression for the secular evolution of the energy of the SCO in the absence of an accretion disk and due to gravitational-wave emission alone.
The characteristic scaling of this quantity for an EMRI on a circular orbit is given by
\begin{align}\label{eq:GW-loss}
    &\left<\dot{E}' \right>_{\mathrm{GW}} = -\frac{32}{5} q^2 (p')^{-5} \left(1
    + \mathcal{O}\left(\frac{1}{p'}\right)\right)\,,
\end{align}
Higher PN order corrections to the above expressions are available in~\cite{Fujita_2012,blanchet2024postnewtoniantheorygravitationalwaves}.
\subsection{Comparison between Newtonian and relativistic effects}\label{sec:Newt-rel-comparison}
We here compare the expressions we derived for the secular change of orbital parameters with expressions from Newtonian gravity.
The main results of this subsection are presented in Fig.~\ref{fig:torque-Newt-Rel}. Since both the Newtonian [Eqs.~\eqref{eq:Edot-Newt-expr-LR} and \eqref{eq:Ldot-Newt-expr-LR}] and relativistic torques [Eqs.~\eqref{eq:Edot-Relativistic-expr-LR} and \eqref{eq:Ldot-Relativistic-expr-LR}] obey the scaling relation given in Eq.~\eqref{eq:scaling-relationship-Edot-Ldot-edot}, all the dependence on $\Sigma_{0}$, $\Sigma_{p}$, and $\Sigma_{h}$ cancels and the ratio only depends on
\begin{align*}
    \frac{d \Sigma}{d \log p'} = -\Sigma_{p}\,.
\end{align*}
In the left panel of Fig.~\ref{fig:torque-Newt-Rel} we present the absolute value of the ratio the torque exchanged with the disk using relativistic expression to the Newtonian expression as a function of the location $p'$ for various values of $\Sigma_{p}$.
From Eqs.~\eqref{eq:Edot-Newt-expr-LR} and \eqref{eq:Ldot-Newt-expr-LR}, the Newtonian torque is always negative. The relativistic torque, however, changes sign due to the shift in the concentration of the inner and outer Lindblad resonances, see Sec.~\ref{sec:Lr-resonace-expressions}. 
Generally, the closer the resonance to the SCO orbit, the stronger the torque it exerts.
Therefore, in the Newtonian limit, where the outer Lindblad resonances is closer there is a net negative torque on the orbit of the SCO.

The shift in resonance position due to relativistic effects causes a competition between the strength of torque exerted by the inner and outer Lindblad resonances. At around $p' \sim 8$ the inner Lindblad resonances overwhelm the outer Lindblad resonances, leading to a net positive torque if the position of the SCO is less that $8M$. 
As we see from the left panel of Fig.~\ref{fig:torque-Newt-Rel}, the behavior observed here originates from the relative shift in the location of the inner and outer Lindblad resonances is only mildly sensitive to the value of $\Sigma_{p}$.
However, we should caution the reader that this change in the sign of the torque might not persist if the assumption we made in Eq.~\eqref{eq:jmax-assumption} does not hold, i.e.,
\begin{align}
    j_{\mathrm{max,Lr,-}} \ll j_{\mathrm{max,Lr,+}} \,.
\end{align}
In this case, the dominant contribution to the torque still comes from the outer Lindblad resonances, leading to a net negative torque, as in the Newtonian limit.
Understanding this situation in detail requires one to perform global calculations of disk-SCO interactions by modeling the fluid structure of the disk, so we leave this to future work.
Another situation where the torque change might not occur is if $\Sigma_{p} < -14$. In this case, the contribution from the density gradients can lead to a large and net negative effect [see Eq.~\eqref{eq:Edot-Relativistic-expr-LR}], and thus, lead to an overall, negative sign for the torque for $p'>6.5$.

In the right panel of Fig.~\ref{fig:torque-Newt-Rel}, we present the percentage change between the relativistic and the Newtonian torque expressions.
Observe that when the SCO is at separations $p'>15$, the difference is of the order of a few percent, but it steeply rises as the SCO approaches the innermost stable circular orbit, because small changes in angular momentum have a large effect here.
This figure highlights the importance of using a consistent relativistic approach to model the disk torque when the location of the SCO is smaller than $\mathcal{O}(20 M)$.

\subsection{Secular evolution of the energy of the SCO due to interactions with the disk and gravitational-wave emission}\label{sec:compare-disk-and-GW}
Next, we compare the ratio of the energy extracted from the orbit due to interactions with the disk, versus due to gravitational-wave emission. 
Our main results are presented in Fig.~\ref{fig:dEDT_disk_GW}. 
For simplicity, we here set $\Sigma_{p} = -3/2$, as the ratio is only sensitive to $\Sigma_{0}$ and $h_{0}$.
In these plots, the gravitational-wave flux is evaluated using the 22 PN accurate analytic expression derived in~\cite{Fujita_2012}, which is available in~\cite{BHPToolkit}.

Figure~\ref{fig:dEDT_disk_GW} shows the ratio of the energy extracted from the orbit due to interactions with the disk, versus that lost to gravitational-wave emission, for a supermassive black hole mass of $M=10^5 M_{\odot}$ (left panel) and one of mass $M = 10^8 M_{\odot}$ (right panel).
For a $10^5 M_{\odot}$ black hole and all values of $\Sigma_{0}$, the energy loss due to the disk-SCO interaction is around 3--10 orders of magnitude smaller than that due to gravitational-wave emission.
The disk-SCO energy exchange scales linearly with the mass of the supermassive black hole [Eq.~\eqref{eq:scaling-relationship-Edot-Ldot-edot}], and therefore, the ratio is 3 orders of magnitude larger for a supermassive black hole of mass $10^8 M_{\odot}$, as we see in the right panel.
This implies that, for $\beta$-disks $\Sigma_{0}$, the energy exchange due to the disk-SCO interaction can compete (and potentially exceed) the gravitational-wave energy loss for EMRIs around supermassive black holes of mass $10^8 M_{\odot}$ or larger. 
When $p'<8$, the orbit actually absorbs energy from the disk (i.e.~the sign of the torque flips), and therefore, there is a potential for floating orbits, i.e.~orbits in which the disk-SCO interaction balances the gravitational-wave radiation-reaction, forcing the SCO to momentarily stay at a fixed orbital separation (emitting approximately monochromatic gravitational waves). This, however, only occurs if the SCO is sufficiently close to the supermassive black hole ($p'<8)$, the disk density is sufficiently large $\Sigma_{0} \sim 10^7$, and the disk is sufficiently thin $h_{0} \sim 0.015$ in supermassive black holes of mass $10^8 M_{\odot}$.

\begin{figure*}
    \centering
    \includegraphics[width=0.49\linewidth]{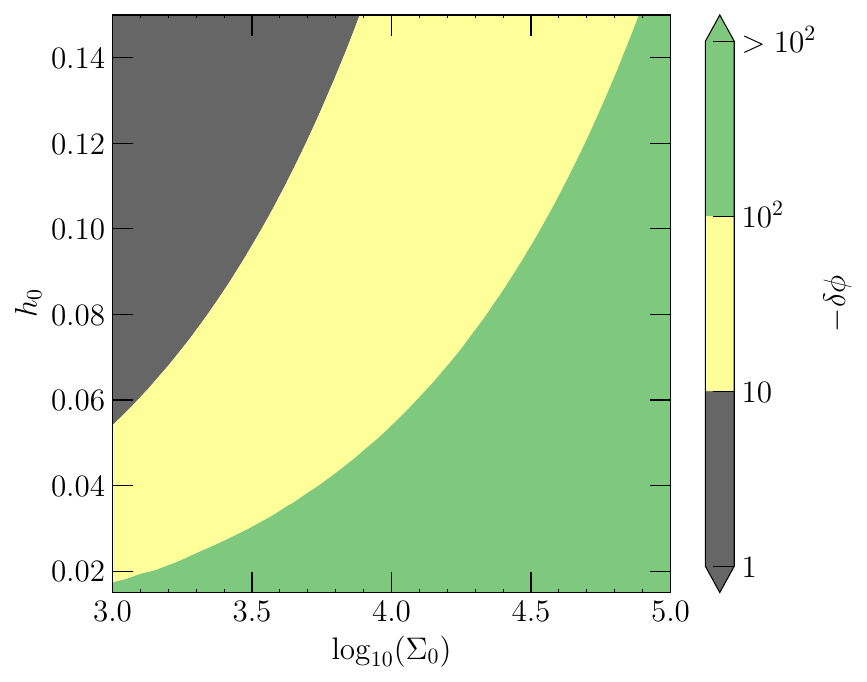}
    \includegraphics[width=0.49\linewidth]{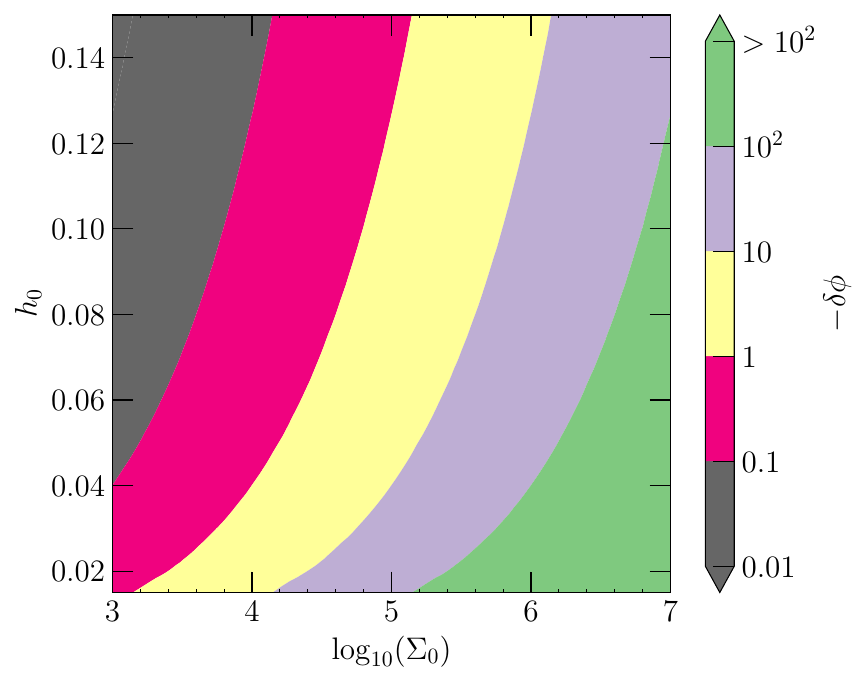}
    \caption{Total accumulated phase due to disk-SCO interactions for fiducial EMRI systems with parameters listed in Table~\ref{tab:EMRI-params}. Observe that for system I (left panel) the total accumulated phase is large in a significant portion of the $\Sigma_{0},h_{0}$ parameter space due to the longer duration it spends in the LISA band. The total phase accumulated for system II (right panel) is greater than $\mathcal{O}(10)$ radians only when $\Sigma_{0} > \mathcal{O}(10^5)$. However, system II is highly relativistic due to its closeness to the supermassive black hole and we require the relativistic expression for the energy loss due disk-SCO to accurately model this system, see Eq.~\eqref{eq:sys-II-diff}.}
    \label{fig:dephasing-estimate}
\end{figure*}
\subsection{Qualitative estimate of the change in the number of cycles due to relativistic disk-SCO interactions in the LISA band for an EMRI}
Let us carry out a qualitative analysis of the number of additional cycles caused by the accretion disk-SCO on the EMRI waveform when modeling the former relativistically. To perform this calculation, we model the total loss of orbital energy of the SCO using an adiabatic approximation, namely
\begin{subequations}\label{eq:GW-phase-Energy-Balance}
\begin{align}
    \frac{d E'}{dt} &= -\mathcal{F}_{\mathrm{GW}}^{\infty} -\mathcal{F}_{\mathrm{GW}}^{H} + \left< \dot{E}'\right>_{\mathrm{Lr,tot}}
    \equiv \left< \dot{E}' \right>_{\mathrm{tot.}}
    \,, \\
    \frac{dM}{dt} &= \mathcal{F}_{\mathrm{GW}}^{H} \,,
\end{align}
\end{subequations}
where
$\mathcal{F}_{\mathrm{GW}}^{\infty} = -\left<\dot{E}' \right>_{\mathrm{GW}}$, $\mathcal{F}_{\mathrm{GW}}^{H}$ and $\left< \dot{E}'\right>_{\mathrm{Lr,tot}}$ are the fluxes due to the loss of energy to gravitational waves that escape to (spatial) infinity, gravitational-waves that fall into the supermassive black hole event horizon, and due to the disk-SCO interaction, respectively. 
An accurate estimate of the dephasing or of the number of useful cycles caused due to the EMRI system requires fluxes calibrated to numerical simulations, since the EMRI is highly relativistic~\cite{Hughes_2021,Pound_2021}. 
In this work, we are only interested in a \textit{qualitative} estimate, so we use the approximate, high PN order fluxes of~\cite{Fujita_2012,Shah_2014,Fujita_2009} to evaluate $\mathcal{F}_{\mathrm{GW}}^{\infty}$ and $\mathcal{F}_{\mathrm{GW}}^{H}$; these fluxes are available in the \texttt{PostNewtonianSelfForce} package of~\cite{BHPToolkit}.

To proceed with the calculation, let us define certain standard PN parameters~\cite{blanchet2024postnewtoniantheorygravitationalwaves}, such as
\begin{align}
    x \equiv \left(M \Omega' \right)^{2/3}
\end{align}
where $\Omega'$ is the SCO's orbital frequency. 
The orbital energy of the SCO in terms of $x$ is given by
\begin{align}
    E' &= \frac{m(1-2 x)}{\sqrt{1-3 x}} \,,
\end{align}
to leading-order in the mass-ratio.
Substituting this into Eq.~\eqref{eq:GW-phase-Energy-Balance} and simplifying, we obtain an adiabatic evolution for $\dot{x}$
\begin{align}\label{eq:xdot-eqn-1}
    &\frac{dE'}{dx} \dot{x} + \frac{dE'}{dM} \dot{M}
    =
    \left< \dot{E}' \right>_{\mathrm{tot.}}
    \iff
    \nonumber\\
    &
    \dot{x} 
    =
    \frac{dx}{dE'}
    \bigg(
    \left< \dot{E}' \right>_{\mathrm{tot.}}
    -\frac{dE'}{dM} \dot{M}
    \bigg)
    \,.
\end{align}
We can ignore the second term above in the extreme mass-ratio limit, as it is of $\mathcal{O}((m/M)^3)$,
\begin{align}\label{eq:xdot-eqn}
    \dot{x} 
    =
    \frac{dx}{dE'}
    \left< \dot{E}' \right>_{\mathrm{tot.}}
    = \dot{x}_{\mathrm{GW}} + \delta \dot{x}\,,
\end{align}
where, in the second equality, we split the evolution into gravitational-wave-driven and disk-SCO-driven terms, which are given by
\begin{subequations}
\begin{align}
    \dot{x}_{\mathrm{GW}}
    &=
    \frac{dx}{dE'}
    \bigg(
    -\mathcal{F}_{\mathrm{GW}}^{\infty} -\mathcal{F}_{\mathrm{GW}}^{H}
    \bigg) \,,\\
    \delta \dot{x}
    &=
    \frac{dx}{dE'}
    \left< \dot{E}'\right>_{\mathrm{Lr,tot}}
    \,.
\end{align}
\end{subequations}
For later use, we also define 
\begin{align}
    \delta_{\mathrm{Newt}} \dot{x}
    &=
    \frac{dx}{dE'}
    \left< \dot{E}'\right>_{\mathrm{Lr,tot,Newt}}
    \,.
\end{align}
From Fig.~\ref{fig:dEDT_disk_GW}, we see that, for $\alpha$-disk-like values of $\Sigma_{0}$, $(\delta \dot{x}, \delta_{\mathrm{Newt}} \dot{x}) \ll \dot{x}_{\mathrm{GW}}$. We assume the same is true for $\beta$-disk-like values of $\Sigma_0$, but this approximation might not hold when $h_{0} \sim 0.015$.

\begin{table}[htp]
    \centering
    \begin{tabular}{|c|c|c|}
        \hline
        System & I & II \\ 
        \hline
        $M$ & $10^6 M_{\odot}$ & $10^6 M_{\odot}$ \\
        \hline
        $m$ & $50 M_{\odot}$  & $15 M_{\odot}$ \\
        \hline
        $p'_{\mathrm{i}}$ & $16.84$ & $10.12$ \\
        \hline
        $p'_{\mathrm{f}}$ & $6.5$ & $6.5$ \\
        \hline
        $f_{\mathrm{GW,i}}$ & $9.4 \times 10^{-4} \mathrm{Hz}$ & $2.01 \times 10^{-3} \mathrm{Hz}$ \\
        \hline
        $f_{\mathrm{GW,f}}$ & $3.9 \times 10^{-3} \mathrm{Hz}$ & $3.9 \times 10^{-3} \mathrm{Hz}$ \\
        \hline
        $T_{\mathrm{obs}}$ & $4\, \mathrm{yrs}$ & $1\, \mathrm{yr}$ \\
        \hline
        $\phi_{\mathrm{GW}}$ & $1.05 \times 10^6$ rads & $4.94 \times 10^5$ rads \\
        \hline
    \end{tabular}
    \caption{Parameters of the fiducial EMRI systems used in the phase accumulation calculation.}
    \label{tab:EMRI-params}
\end{table}

We can now use Eq.~\eqref{eq:xdot-eqn} to calculate the total accumulated gravitational-wave phase $\Psi$ of the
dominant, $\ell=2$, harmonic in the stationary phase approximation.
Let $f_{\mathrm{GW}}$ denote the gravitational-wave frequency and $F'$ the (linear) orbital frequency of the SCO.
In the stationary phase approximation, $f_{\mathrm{GW}} = 2 F'$, and thus, the total accumulated phase is
\begin{align}
    \phi_{\mathrm{tot}} = \int_{F'_{\mathrm{i}}}^{F'_{\mathrm{f}}}
    4 \pi F' \frac{dt}{dF'} dF' 
    =
    2
    \int_{x_{\mathrm{i}}}^{x_{\mathrm{f}}}
    \frac{x^{3/2}}{M \dot{x}} dx
    \,,
\end{align}
where $F'_{\mathrm{i}}$ and $F'_{\mathrm{f}}$ are the initial and final frequencies of the SCO in the LISA band.
We can split the above equation into
\begin{align}
    \phi_{\mathrm{tot}} &= \phi_{\mathrm{GW}} + \delta \phi
\end{align}
where $\phi_{\mathrm{GW}}$ and $\delta \phi$ are contributions from only gravitational-wave emission and disk-SCO interactions, respectively.
These quantities are given by
\begin{subequations}
\begin{align}
    &\phi_{\mathrm{GW}}
    =
    2
    \int_{x_{\mathrm{min}}}^{x_{\mathrm{max}}}
    \frac{x^{3/2}}{M \dot{x}_{\mathrm{GW}}} dx
    \,,\\
    &\delta \phi
    =
    2
    \int_{x_{\mathrm{i}}}^{x_{\mathrm{f}}}
    \frac{x^{3/2}}{M \dot{x}_{\mathrm{GW}} + M \delta \dot{x}} dx
    \nonumber\\
    &-
    2
    \int_{x_{\mathrm{min}}}^{x_{\mathrm{max}}}
    \frac{x^{3/2}}{M \dot{x}_{\mathrm{GW}}} dx
    \,.
\end{align}
\end{subequations}
To obtain the difference between a Newtonian estimate and a relativistic estimate, we also define
\begin{align}
    \delta_{\mathrm{Newt}} \phi
    &=
    2
    \int_{x_{\mathrm{i}}}^{x_{\mathrm{f}}}
    \frac{x^{3/2}}{M \dot{x}_{\mathrm{GW}} + M \delta_{\mathrm{Newt}} \dot{x}} dx
    \nonumber\\
    &-
    2
    \int_{x_{\mathrm{min}}}^{x_{\mathrm{max}}}
    \frac{x^{3/2}}{M \dot{x}_{\mathrm{GW}}} dx
    \,.
\end{align}

To assess the differences between a Newtonian and relativistic estimate, we consider two fiducial EMRI systems with parameters listed in Table~\ref{tab:EMRI-params}.
These systems start at $p'_{\mathrm{i}}$ and end at $p'_{\mathrm{f}}$ during an observation time $T_{\mathrm{obs}}$, as estimated using gravitational wave emission only. The table also lists initial and final gravitation-wave frequencies $f_{\mathrm{GW},i/f}$, and the total accumulated phase purely from gravitational-wave emission.
We note that we have selected initial conditions such that the final location of the SCO is at $p' = 6.5$ at the end of the observation time. 
This is done prevent unphysical contributions from the artificial blow-up in the relativistic torque as $p'\to 6$. This blow-up is not expected to physically occur as the SCO and the disc transition to plunge around the ISCO, where our model of resonant interaction does not apply.

The parameters of system I are such that it accumulates a large phase in the region where the relativistic effects are subdominant, when $p'>8$. Since both $\delta_{\mathrm{Newt}} \dot{x}$ and $\delta\dot{x}$ scale the same way with accretion-disk-model parameters, we expect a weak dependence for the relative fractional difference between a Newtonian and a relativistic estimate.
The mean difference over the parameter space is approximately
\begin{align}
    \left.\left( \frac{\delta_{\mathrm{Newt}} \dot{x} 
    - \delta \dot{x}}{\delta_{\mathrm{Newt}} \dot{x}} \right)
    \right|_{\mathrm{sys\,I}}\times 100
    \approx
    6 \%
    \,.
\end{align}
On the other hand, system II accumulates a large phase in the region where the relativistic effects are important, when $p'<8$. In this case, the mean error is approximately
\begin{align}\label{eq:sys-II-diff}
    \left.\left( \frac{\delta_{\mathrm{Newt}} \dot{x} 
    - \delta \dot{x}}{\delta_{\mathrm{Newt}} \dot{x}} \right)
    \right|_{\mathrm{sys\,II}}\times 100
    \approx
    25 \%
    \,.
\end{align}
Therefore, it is clearly important to include relativistic effects in the modeling of accretion-disk environmental effects in EMRIs that are in close orbits around supermassive black holes.

A crude estimate for detectability for systems with signal-to-noise ratio of $\mathcal{O}(10)$ is that $\delta \phi \gtrsim 1/{\rm{SNR}} \sim 1/10$~\cite{Kocsis_2011}. However, this crude estimate usually overestimate the prospect of detectability (as it does not account, for example, for covariances between parameters or the specific shape of the noise spectral density), so a reliable estimate requires a Bayesian (or at the very least a Fisher) analysis~\cite{Speri_2023,duque2025constrainingaccretionphysicsgravitational}. 
Nevertheless, to qualitatively understand what regimes of parameter space can lead to significant dephasing, we plot the total phase accumulated due to disk-SCO interactions in systems I and II, as a function of $\log_{10}(\Sigma_{0})$ and $h_{0}$, in the left and right panels of Fig.~\ref{fig:dephasing-estimate}. 
Observe from the color bar that $-\delta \phi>0$ in these plots because the net effect of disk-SCO interactions is to \textit{speed up} the inspiral of the EMRI system, even though the relativistic effects close to the supermassive black hole change the sign of the torque. However, if one uses a Newtonian expression for the disk-SCO torque, then the magnitude of $\delta \phi$ will be off by up to $40 \%$ for system II, see, Eq.~\eqref{eq:sys-II-diff}.
From the left and right panels of the figure, we see that, for disk densities $\Sigma_{0} \sim 10^3$, the disk-SCO interaction causes very small dephasing $h_{0} < 0.05$ for system I. For system II, the dephasing caused by $\Sigma_{0} < 10^4$ is too small to leave a measurable imprint for signal-to-noise ratios of $\mathcal{O}(10)$. We note that $\Sigma_{0} > \mathcal{O}(10^5)$ causes significant dephasing in both systems and could be potentially observable. These results are broadly consistent with previous  calculations, but they improve them significantly by now allowing for a relativistic treatment of disk-SCO interactions~\cite{Yunes:2011ws,Kocsis_2011,Speri_2023}.
\section{Conclusions}\label{sec:conclusions}
In this work, we have developed a formalism to study the interactions between a thin disk and a stellar mass SCO interacting in the background of a non-spinning supermassive black hole.
Our main results are the following: 
\begin{itemize}
    \item We demonstrated that techniques from Hamiltonian mechanics, combined with self-force theory, can be used to model the interaction between a closely separated SCO and a thin disk using a relativistic disturbing function, similar to the approach in Newtonian theory~\cite{Murray-Dermott-Book}. 
    We also provided analytic expressions for the relativistic disturbing function in a small eccentricity expansion in Sec.~\ref{sec:disturbing-function-relativistic}.
    \item We extended the resonance dissipation formalism from Newtonian theory~\cite{GT-disc-satellite-interaction,Ogilvie_2007} to understand the interaction between the disk and SCO in nearly circular and equatorial orbits. One of our main results is the explicit, analytic, and relativistically-accurate expressions for the secular evolution of the SCO, provided in Eqs.~ \eqref{eq:Lindblad-E-expr-actual}, \eqref{eq:lindblad-resonance-total} and \eqref{eq:expr-rel-Lindblad-all-others}.
    \item We showed that the relativistic effects shift the location of the inner Lindblad resonances toward the SCO more than they shift the location of the outer Lindblad resonances, leading to a change in the direction of the torque on the SCO if (i) the inner and the outer torque cutoff parameters are equal, and (ii) the orbital separation between the supermassive black hole and the SCO is $\lesssim 8 M$. Moreover, relativistic effects enhance the value of the disk-induced torque on the SCO by a few 10 -- 1000 \%, if the SCO's orbital separation is $\lesssim 10 M$.
    \item Using a simple, parameterized, accretion disk model, we compared the torque and energy exchange due to disk-SCO interaction to that due to gravitational-wave emission. We show that for supermassive black holes of mass $10^5 M_{\odot}$, the energy exchanged due to the SCO-disk interaction is always significantly smaller than gravitational-wave energy loss for $\alpha$ and $\beta$ disk-like profiles, and thus, the former can be treated as a small perturbation.
    However, for a supermassive black hole of mass $10^8 M_{\odot}$, the energy exchanged due to the disk-SCO interaction can be comparable to the energy loss due to gravitational-wave emission in the regions close to the innermost stable circular orbit.
\end{itemize}

There are several future directions that can extend this work. 
A first direction is to understand the effects of the spin of the supermassive black hole on the disk-SCO interaction. 
The spin of the supermassive black hole can greatly influence the location of the innermost stable circular orbit, and therefore, it can potentially increase or decrease the impact of relativistic corrections on the torque formula.
This calculation would involve obtaining the disturbing function in a Kerr spacetime and applying the resonance dissipation framework developed in Sec.~\ref{sec:resonance-evolution} to find the secular evolution of the orbital elements due to the resonant interaction.
Given the existence of action-angle variables in a Kerr spacetime, we expect this calculation to be straightforward for EMRIs with small eccentricity and inclination angle.

A second direction for future work is to understand the impact of eccentricity and orbital inclination on the disk-SCO interactions using our formalism.
This would allow us to model the impact of relativistic effects on highly-eccentric and highly-inclined SCO orbits, which may occur in some EMRIs. 
For small inclination EMRIs, one could extend the calculation of the relativistic disturbing function in a small eccentricity expansion to higher orders to achieve this goal. However, for general inclinations, this calculation may not be feasible, and one might need to integrate the SCO-disk Hamiltonian directly to obtain reliable results.

Another interesting avenue where our formalism could be applied is the study of the impact of scalar fields formed through superradiant instability on the secular evolution of the SCO. Once the gravitational potentials of such clouds are obtained, it would be straightforward to use our resonance dissipation picture to understand these interactions. This would be complement numerical and semi-analytic approaches that model the global problem of scalar field-SCO coupling~\cite{dyson2025environmentaleffectsextrememass,li2025extrememassratioinspiralultralight}.

Finally, it is essential to assess the validity of the resonant interaction approximation employed in this work and others by examining a fluid model of the disk. 
Such a comparison would enable a better understanding of the torque reversal found in this work and the impact of co-orbital corotation resonances and non-linear fluid interactions on the secular evolution of the SCO.
Moreover, these calculations will help us understand the impact that realistic disk profiles and magnetic fields have on the disk-SCO interaction. 
One could also use the global calculation to study the impact of disk-SCO interactions on intermediate mass black holes and understand their evolution in the background of a supermassive black hole~\cite{Derdzinski:2018qzv,Derdzinski:2020wlw}.

\acknowledgements
A.~H.~K.~R.~and N.~Y.~acknowledge support from the Simons Foundation through Award No. 896696, the NSF through Grant No. PHY-2207650, NASA through Grant No. 80NSSC22K0806, and the Simons Foundation International through Award No. SFI-MPS-BH-00012593-01.  C.~F.~G. acknowledges support from the NSF through AST 20-34306.  Parts of this work were completed at the Aspen Center for Physics, which is supported by National Science Foundation grant PHY-2210452.

\appendix
\section{Geodesic motion in a Schwarzschild spacetime}\label{sec:geodesic-motion-Schw}
In this appendix, we describe some basic properties of geodesic motion in the Schwarzschild spacetime. 
The material presented in this section is standard material reviewed e.g.~in~\cite{Pound_2021} and we only present it for the sake of completeness, and to set up our notation. 

The outline of this section is as follows:
We first describe the action-angle formulation of geodesic motion in a Schwarzschild spacetime in Sec.~\ref{sec:Schwarzschild-geodesic-action-angle}, and then we describe the relation between the motion in the fundamental and the orbital planes in Sec.~\ref{sec:orb-fund}.
Finally, we derive expressions for the orbital elements in a small eccentricity expansion in Sec.~\ref{sec:small-eccentricity-Schwarzschild-geodesic}.
\subsection{Geodesics and action-angle formulation}\label{sec:Schwarzschild-geodesic-action-angle}
Consider geodesic equation in a Schwarzschild spacetime\footnote{We use Greek letters to denote spacetime-indices and Latin letters to denote spatial indices.} of mass $M$.
The geodesic equation for a worldline $z^{\mu}(\lambda)$ parameterized by $\lambda$ with coordinate four-velocity $u^{\mu}(\lambda) \equiv d z^{\mu} / d\lambda$  can be written as
\begin{align}\label{eq:geodesic-general}
    \frac{d p_{\alpha}}{d \lambda} = -\frac{1}{2} \partial_{\alpha} g^{\rho \sigma} \sqrt{-g_{\alpha \beta} u^{\alpha} u^{\beta}} p_{\rho} p_{\sigma}\,,
\end{align}
where the specific four-momentum of the object is
\begin{align}\label{eq:specfic-four-momentum-def}
    p_{\mu} &\equiv \frac{u_{\mu}}{\sqrt{-g_{\alpha \beta} u^{\alpha} u^{\beta}}} = g_{\nu \mu} \frac{d z^{\nu}}{d\tau} \,,
\end{align}
and $\tau$ is proper time.

If we work with the coordinate time parameterization $\lambda = t$ and Schwarzschild coordinates $x^{\mu} = (t,r,\theta,\phi)$, the worldline traces a curve
\begin{align}
    z^{\mu} = \left(t, r(t), \theta(t), \phi(t) \right),
\end{align}
in a Schwarzschild spacetime. Geodesic motion [Eq.~\eqref{eq:geodesic-general}] in a coordinate time parameterization is generated by the Hamiltonian $\mathscr{H}$
\begin{align}
    \mathscr{H} = -p_{t}\,,
\end{align}
which, in Schwarzschild coordinates, simplifies to
\begin{align}\label{eq:Hamiltonian-Schw}
    \mathscr{H} = \sqrt{f \left(1 + f p_{r}^2 + \frac{1}{r^2} p_{\theta}^2 + \frac{1}{r^2 \sin^2(\theta)} p_{\phi}^2\right)}\,,
\end{align}
where $f = 1- 2M/r$. 

We now write the geodesic equation in action-angle form by using the solutions to the Hamilton-Jacobi equation
\begin{align}\label{eq:Hamilton-Jacobi-Equation}
    \mathscr{H}\left[x^{\nu} , \frac{\partial \mathscr{S}}{\partial x^{\nu}}\right]
    +
    \frac{\partial \mathscr{S}}{\partial t}
    &=0\,,
\end{align}
where $\mathscr{S}$ is Hamilton's principal function.
To solve Eq.~\eqref{eq:Hamilton-Jacobi-Equation}, we use the ansatz
\begin{align}
    \mathscr{S} = - \mathscr{E} t + \mathscr{W}(x)  \,,
\end{align}
where $\mathscr{W}(x)$ is Hamilton's characteristic function and $\mathscr{E}$ is the specific energy of the system. 
Standard techniques in Hamiltonian mechanics~\cite{Goldstein-Safko-book} guide us to the following ansatz for $\mathscr{W}$:
\begin{align}\label{eq:Hamiltons-characteristic-ansatz}
    \mathscr{W} &= \mathscr{L}_{z} \phi + W_R(r) + W_{\theta}(\theta) \,,
\end{align}
where $\mathscr{L}_{z}$ is the angular momentum per unit mass in the $z$ direction.
Substituting this ansatz in Eq.~\eqref{eq:Hamilton-Jacobi-Equation}, we obtain 
\begin{align}
    \frac{r^2}{f} \left(\mathscr{E}^2 - f\right) - r^2 f \left( \frac{d W_R}{dr}\right)^2
    &=
    \left( \frac{d W_{\theta}}{d\theta} \right)^2 + \frac{\mathscr{L}_{z}^2}{\sin^2(\theta)} \nonumber\\
    &= \mathscr{L}^2 \,,
\end{align}
where $\mathscr{L}$ is a separation constant that represents total angular momentum per unit mass.
Solving this differential equation, we obtain
\begin{subequations}
\begin{align}\label{eq:WR-sol}
    W_R
    &=
    \pm 
    \int \frac{dr}{f} \sqrt{ \mathscr{E}^2 - f \left(1+\frac{\mathscr{L}^2 }{r^2}\right)}\,,\\
    \label{eq:Wtheta-sol}
    W_{\theta}
    &=
    \pm \int d \theta \sqrt{\mathscr{L}^2 - \frac{\mathscr{L}_z^2}{\sin(\theta)^2}}\,.
\end{align}
\end{subequations}

Given a solution to the Hamilton-Jacobi equation, we can now obtain the canonical momentum of the system using
\begin{align}
    p_{a} = \frac{\partial \mathscr{S}}{\partial z^{a}}\,.
\end{align}
Substituting Eqs.~\eqref{eq:Hamiltons-characteristic-ansatz},  \eqref{eq:WR-sol} and \eqref{eq:Wtheta-sol} we see that
\begin{subequations}
\begin{align}
    p_{r} &= \frac{\partial W_R}{\partial r} \,,\\
    p_{\theta} &= \frac{\partial W_{\theta}}{\partial \theta}\,,\\
    p_{\phi} &= \mathscr{L}_{z} \,.
\end{align}
\end{subequations}
We are now in a position to compute the action variables $(J_{r},J_{\theta}, J_{\phi})$
\begin{subequations}
\begin{align}
    \label{eq:Jr-v1}
    J_r &= \frac{1}{2 \pi} \oint p_{r} dr\,,\nonumber\\
    &= 
    \frac{1}{\pi} \int_{r_{\mathrm{min}}}^{r_{\mathrm{max}}} \frac{dr}{f} \sqrt{ \mathscr{E}^2 - f \left(1+\frac{\mathscr{L}^2 }{r^2}\right)}
    \,,\\
    \label{eq:Jtheta-v1}
    J_{\theta}
    &= \frac{1}{2 \pi} \oint p_{\theta} d\theta
    = 
    \frac{1}{\pi}\int_{\theta_{0}}^{\pi - \theta_{0}} d \theta \sqrt{\mathscr{L}^2 - \frac{\mathscr{L}_{z}^2}{\sin(\theta)^2}} \nonumber\\
    &=
    \frac{1}{\pi}\int_{\pi - \theta_{0}}^{\theta_{0}} d \theta \sqrt{\mathscr{L}^2 - \frac{\mathscr{L}_{z}^2}{\sin(\theta)^2}} 
    \nonumber\\
    &=\frac{2}{\pi}\int_{\pi/2}^{\theta_{0}} d \theta \sqrt{\mathscr{L}^2 - \frac{\mathscr{L}_{z}^2}{\sin(\theta)^2}}
    = \mathscr{L} - \mathscr{L}_{z} \,, \\
    \label{eq:Jphi-v1}
    J_{\phi} &= \mathscr{L}_{z} \,,
\end{align}
\end{subequations}
where $\theta_{0} = \arcsin{(\mathscr{L}_{z}/\mathscr{L})}$ is the turning point in the polar coordinate $\theta$ space.
The angle variables $q_{a}$ and the frequencies $\nu_{a}$ of the system are obtained from $\mathscr{W}$ and $\mathscr{H}\left(J_{a}\right)$ using
\begin{subequations}
\begin{align}
    \label{eq:angle-q-vars}
    q_{a} &= \frac{\partial \mathscr{W}}{\partial J_{a}}\,, \\
    \label{eq:nu-definition}
    \nu_{a} &= \frac{\partial \mathscr{H}(J)}{\partial J_{a}} = \frac{\partial \mathscr{E}\left[J_{r}, J_{\theta}, J_{\phi}\right]}{\partial J_{a}}
    \,.
\end{align}
\end{subequations}

To evaluate these expressions, we need to analyze the turning points in the $(p_{r}, r)$ plane. Here we are interested in bound orbits, and the invariant cycle defined by these orbits are contours defined by
\begin{align}
    \mathscr{E}^2 = f \left(1 + f p_{r}^2 + \frac{\mathscr{L}^2}{r^2} \right)\,,
\end{align}
where $\mathscr{E}^2<1$ for bounded orbits.
The turning points are obtained by setting $p_{r}=0$ in the above equation, yielding
\begin{align}\label{eq:orbital-plane-E}
    \mathscr{E}^2 = \left(1- \frac{2M}{r}\right) \left(1 + \frac{\mathscr{L}^2}{r^2}\right) \,.
\end{align}

Analyzing the solutions to this equation is standard material covered in many textbooks (e.g. Chapter 3 of~\cite{Chandrasekar-BH-Book}).
For bound orbits, we can adopt a ``Keplerian'' parameterization
\begin{align}\label{eq:Keplerian-parameterization}
    r = \frac{p M }{1 + e \cos(\chi)}
\end{align}
where $p$, $e$ and $\chi$ are relativistic analogs of the semi-latus rectum,  eccentricity and true anomaly.
Substituting the above into Eq.~\eqref{eq:orbital-plane-E}, we see that for bound orbits, the turning points are the pericenter/periapsis $r_{p}$ and the apocenter/apoapsis $r_{a}$, defined by
\begin{subequations}
\begin{align}\label{eq:peri-apo-eqn}
    r_{p} = \frac{pM}{1 + e}\,,\\
    r_{a} = \frac{pM}{1-e}\,.
\end{align}
\end{subequations}
where, $p$ and $e$ are related to $\mathscr{E}$ and $\mathscr{L}$ via
\begin{subequations}\label{eq:E-andL-in-p-and-e}
\begin{align}
    \mathscr{E}^2 &= \frac{(p-2-2e)(p-2+2e)}{p(p-3-e^2)}\,,\\
    \mathscr{L}^2 &= \frac{p^2 M^2}{p-3-e^2} 
    \,.
\end{align}
\end{subequations}
We note that there is also a third root of Eq.~\eqref{eq:orbital-plane-E} given by
\begin{align}
    r_{3} = \frac{2 M p}{p-4}\,.
\end{align}
The roots $r_{a}, r_{p}$ and $r_{3}$ can be used to simplify the bound orbit equation into
\begin{align}
    \mathscr{E}^2 - f \left(1+\frac{\mathscr{L}^2 }{r^2}\right)
    =
    \frac{\left(1 - \mathscr{E}^2\right)}{r^4}\mathscr{P}(r; r_{3}, r_{a}, r_{p})\,.
\end{align}
where
\begin{align}
    \mathscr{P}(r; r_{3}, r_{a}, r_{p}) \equiv r(r-r_{3})(r-r_{p}) (r_{a} - r)\,.
\end{align}
Substituting this into Eq.~\eqref{eq:WR-sol}, the quadrature solution for the radial part of Hamilton's characteristic function becomes
\begin{align}
    &\frac{\partial W_R}{\partial r}
    = 
    \pm \frac{1}{r^2 f} \sqrt{1 - \mathscr{E}^2}
    \sqrt{\mathcal{P}(r; r_{3}, r_{a}, r_{p})}
    \,.
\end{align}
We now simplify the expression for the radial action variable $J_r$ [Eq.~\eqref{eq:Jr-v1}] using the above equation, and one finds
\begin{align}
    J_{r}
    &=
    \frac{\sqrt{1 - \mathscr{E}^2}}{\pi}\int_{r_a}^{r_p}
    \frac{dr}{r^2 f} 
    \sqrt{\mathcal{P}(r; r_{3}, r_{p}, r_{a})}\,,
    \label{eq:Jr-evaled}
\end{align}
where the right-hand side can be evaluated in terms of elliptic functions~\cite{Fujita_2009}.

To evaluate the frequencies of the system, we need to obtain $\mathscr{E}\left[J_{r}, J_{\theta}, J_{\phi} \right]$ by inverting Eq.~\eqref{eq:Jr-evaled}. 
Observe from Eq.~\eqref{eq:Jr-v1} that $\mathscr{E}$ can only depend on $J_{r}$ and the effective combination $\mathscr{L} = J_{\theta} + J_{\phi}$. Let us then first differentiate  $\mathscr{E}$ with respect to $J_r$ and ${\mathscr{L}}$, and then use the above equation to simplify the resulting expressions, so that we obtain
\begin{align}
    \left(\frac{\partial \mathscr{E}}{\partial J_r} \right)^{-1}
    &=
    \frac{\mathscr{E}}{\pi \sqrt{1-\mathscr{E}^2}}
    \int_{r_{p}}^{r_{a}}
    \frac{dr r^2}{f \sqrt{\mathcal{P}(r; r_{3}, r_{a}, r_{p})}}\,,\\
    \frac{\partial \mathscr{E}}{\partial \mathscr{L}}
    &=
    \frac{\mathscr{L}}{\mathscr{E}}
    \frac{\int_{r_p}^{r_a} \frac{dr}{\sqrt{\mathcal{P}}}}{\int_{r_p}^{r_a} \frac{dr r^2}{f \sqrt{\mathcal{P}}}}
    =
    \frac{\partial \mathscr{E}}{\partial J_r}
    \frac{\mathscr{L}}{\pi \sqrt{1-\mathscr{E}^2}} \int_{r_p}^{r_a} \frac{dr}{\sqrt{\mathcal{P}}}
    \,.
\end{align}
Using the above equations in Eq.~\eqref{eq:nu-definition}, one finds
\begin{subequations}\label{eq:frequencies-v1}
\begin{align}
    \nu_{r} &= \frac{\partial \mathscr{E}}{\partial J_r}
    =
    \left[\frac{\mathscr{E}}{\pi \sqrt{1-\mathscr{E}^2}}
    \int_{r_{p}}^{r_{a}}
    \frac{dr r^2}{f \sqrt{\mathcal{P}(r; r_{3}, r_{a}, r_{p})}} \right]^{-1}
    \,,\\
    \nu_{\theta}
    &=
    \nu_{\phi}
    = 
    \frac{\partial \mathscr{E}}{\partial \mathscr{L}}
    =
    \frac{\mathscr{L} \nu_{r}}{\pi \sqrt{1-\mathscr{E}^2}} \int_{r_p}^{r_a} \frac{dr}{\sqrt{\mathcal{P}}}\,.
\end{align}
\end{subequations}
Obtaining the expressions for the angle variables is also possible in terms of elliptic functions~\cite{Fujita_2009}.
We notice that unlike the Newtonian case, the frequency $\nu_{r} \neq \nu_{\theta}$ or $\nu_{\phi}$ due to pericenter precession. 
In fact, one can easily check that
\begin{align}
    \nu_{\theta} -\nu_{r} &= \nu_{r}\left(1-\frac{\mathscr{L}}{\pi \sqrt{1-\mathscr{E}^2}} \int_{r_{p}}^{r_{a}} \frac{dr}{\sqrt{P}} \right) =
    \mathcal{O}
    \left[
    \frac{\nu_{r}}{c^2}\right] \,,
\end{align}
where we have restored factors of $c$ to indicate that this is a 1 post-Newtonian (PN) effect\footnote{A PN expansion is one in weak fields and small velocities, with terms of relative order $(v/c)^{2N}$ representing corrections of $N$PN order\cite{blanchet2024postnewtoniantheorygravitationalwaves}.}.

Equations~\eqref{eq:angle-q-vars} and \eqref{eq:frequencies-v1} complete the action-angle description of the geodesics of Schwarzschild spacetime, but, in our calculations below, we will use a different set of action -angle variables that are relativistic generalizations of the classical Delaunay and modified Delaunay variables of Newtonian celestial mechanics~\cite{Murray-Dermott-Book,2002mcma.book.....M}. 
The first set of action-angle variables $\left(\left\{L,l\right\},\left\{G,g\right\},\left\{H,h\right\}\right)$ are the relativistic Delaunay variables~\cite{witzany2022actionanglecoordinatesblackholegeodesics}, which we define, together with their associated frequencies, as
\begin{subequations}\label{eq:Delaunay-variables}
\begin{align}
    &L \equiv J_{r} + J_{\theta} +J_{\phi} \,, &l& \equiv q_{r}\,, &\nu_{l}&  \equiv \nu_{r} \,,\\
    &G \equiv J_{\theta} + J_{\phi} \,, &g& \equiv q_{\theta} - q_{r}\,, &\nu_{g}& \equiv \nu_{\theta} - \nu_{r} \,,\\
    &H \equiv J_{\phi} \,, &h& \equiv q_{\phi} - q_{\theta} \,, &\nu_{h}& \equiv 0 \,.
\end{align}
\end{subequations}
The relativistic modified Delaunay $\left(\left\{\Lambda,\lambda\right\},\left\{P,\mathscr{p} \right\},\left\{Q, \mathscr{q}\right\}\right)$ variables and their associated frequencies are defined by
\begin{subequations}\label{eq:modified-Delaunay-variables}
\begin{align}
    \Lambda &\equiv L \,, &\lambda&\equiv l+ g + h \,, &\nu_{\lambda}&\equiv \nu_{\theta} \,,\\
    P &\equiv L- G \,, &\mathscr{p}&\equiv -\varpi= - g -h \,, &\nu_{\mathscr{p}}&\equiv \nu_r - \nu_{\theta} \,,\\
    Q &\equiv G - H \,, &\mathscr{q}&\equiv-\Omega\equiv-h\,, &\nu_{\mathscr{q}}&= 0 \,.
\end{align}
\end{subequations}
In the Newtonian limit, the Delaunay and modified Delaunay variables can be obtained explicitly in terms of the orbital elements, see e.g.~Eqs.~(1.68) and (1.69) of ~\cite{2002mcma.book.....M}.
\subsection{Transformation from the orbital plane to the fundamental plane}\label{sec:orb-fund}
It is beneficial to treat the motion of orbital motion in the fundamental plane (i.e.~a general coordinate system attached to the Schwarzschild black hole)
\begin{align}
\label{eq:fund-plane}
    z^{\mu} = (t, r, \theta(t), \phi(t))\,,
\end{align}
and the orbital plane
\begin{align}
\label{eq:orb-plane}
    z^{a}_{\mathrm{orb \, plane}}
    &=
    \left( r(t) ,\varphi(t), \frac{\pi}{2} \right)
    \,.
\end{align}
A series of Euler rotations generate the coordinate transformation from the orbital to the fundamental plane~\cite{Poisson-Will,Murray-Dermott-Book,Warburton_2017}.
Let 
\begin{align}
    X= r \sin(\theta) \cos(\phi) \,,
    Y= r \sin(\theta) \sin(\phi) \,,
    Z= r \cos(\phi) \,,
\end{align}
and 
\begin{align}
    x = r(t) \cos(\varphi(t)) , y=r(t) \sin(\varphi(t)), z = 0\,.
\end{align}
To relate the fundamental plane to the orbital plane, we rotate by Euler angles $\Omega$ (the longitude of the ascending node) and $\iota$ (the inclination angle) to obtain
\begin{align}\label{eq:euler-rotation-v2}
    \begin{pmatrix}
        X \\
        Y \\
        Z
    \end{pmatrix}
    &=
    r
    \begin{pmatrix}
        \cos \Omega \cos( \varphi) - \cos \iota \sin \Omega \sin( \varphi) \\
        \sin \Omega \cos( \varphi) + \cos \iota \cos \Omega \sin(\varphi) \\
        \sin \iota \sin ( \varphi)
    \end{pmatrix}
    \,,
\end{align}
e.g.~see Eq.~(2.122) of~\cite{Murray-Dermott-Book} or Eq.~(3.40) of~\cite{Poisson-Will}. 
To simplify the integrals appearing in Eqs.~\eqref{eq:angle-q-vars} and~\eqref{eq:frequencies-v1}, we make use of the following identities:
\begin{subequations}\label{eq:u-identities}
\begin{align}
    \sin u &= \cot \theta \cot \iota \,,\\
    \tan(u) &= \cos \iota \tan(\varphi) \,.
\end{align}
\end{subequations}
where 
\begin{align}
    u \equiv \phi - \Omega\,.
\end{align}
In this way, the above transformations map between the fundamental and the orbital planes. 
\subsection{Small eccentricity expansion of orbital motion in a Schwarzschild spacetime}\label{sec:small-eccentricity-Schwarzschild-geodesic}
We are now interested in expanding the action-angle variables in the Schwarzschild spacetime in a small eccentricity expansion.
To obtain analytic expansions, we note that, in the orbital plane, the motion is described by Eq.~\eqref{eq:orb-plane} with $r$ given by Eq.~\eqref{eq:Keplerian-parameterization} and $\chi = \chi(t)$. The evolution equations for the orbital elements in the orbital plane are given by~\cite{Pound_2008}
\begin{align}
    \frac{dt}{d\chi}
    &= \frac{r^2}{M(p-2-2 e \cos(\chi))}\sqrt{\frac{(p-2)^2-4 e^2}{p-6-2e \cos \chi}}\\
    \label{eq:varphi-evol}
    \frac{d \varphi}{d \chi}
    &= \sqrt{\frac{p}{p-6-2 e \cos \chi}}\,.
\end{align}
The orbit in the fundamental plane is parameterized by Eq.~\eqref{eq:fund-plane} and we obtain the orbital elements $(\theta(t),\phi(t))$ by using the Euler rotations of Eq.~\eqref{eq:euler-rotation-v2}.
The angle variables $q_{a}$ are obtained from Eq.~\eqref{eq:angle-q-vars}. Substituting Eq.~\eqref{eq:Hamiltons-characteristic-ansatz} into Eq.~\eqref{eq:angle-q-vars}, and using that $\mathscr{L}_{z} = J_{\phi}$ and $\mathscr{L} = J_{\theta} + J_{\phi}$, we obtain
\begin{subequations}
\begin{align}
    q_{r} &= \frac{\partial W_R}{\partial \mathscr{E}} \frac{\partial \mathscr{E}}{\partial J_r}\,,\\
    q_{\theta} &= \frac{\partial W_R}{\partial \mathscr{E}} \frac{\partial \mathscr{E}}{\partial \mathscr{L}}
    +
    \frac{\partial W_R}{\partial \mathscr{L}} 
    +
    \frac{\partial W_{\theta}}{\partial \mathscr{L}}
    \,,\\
    q_{\phi} &= 
    \phi + 
    \frac{\partial W_R}{\partial \mathscr{E}} \frac{\partial \mathscr{E}}{\partial \mathscr{L}}
    +
    \frac{\partial W_R}{\partial \mathscr{L}} 
    +
    \frac{\partial W_{\theta}}{\partial \mathscr{L}}
    +
    \frac{\partial W_{\theta}}{\partial \mathscr{L}_z}
    \,.
\end{align}
\end{subequations}
Using the above equations in Eq.~\eqref{eq:Delaunay-variables}, we obtain the Delaunay elements
\begin{subequations}\label{eq:Deluany-angles-v1}
\begin{align}
    l &= q_r = \frac{\partial W_R}{\partial \mathscr{E}} \frac{\partial \mathscr{E}}{\partial J_r} \,,\\
    g &= q_{\theta} - q_{r}
    \nonumber\\
    &=
    \frac{\partial W_{\theta}}{\partial \mathscr{L}}
    +
    \frac{\partial W_R}{\partial \mathscr{L}}
    +
    \frac{\partial W_R}{\partial \mathscr{E}} \frac{\partial \mathscr{E}}{\partial \mathscr{L}}
    -
    \frac{\partial W_R}{\partial \mathscr{E}} \frac{\partial \mathscr{E}}{\partial J_r} \,,\\
    h &= q_{\phi} - q_{\theta} 
    =
    \phi + \frac{\partial W_{\theta}}{\partial \mathscr{L}_z}
    \,.
\end{align}
\end{subequations}
Observe that\footnote{We picked the negative sign in Eq.~\eqref{eq:Wtheta-sol}; see footnote on page 475 of~\cite{Goldstein-Safko-book}.}
\begin{align*}
    \frac{\partial W_{\theta}}{\partial \mathscr{L}_{z}}
    &=
    \int \frac{ \mathscr{L}_z d \theta}{  \sin^2 \theta\sqrt{\mathscr{L}^2 - \frac{\mathscr{L}_z^2}{\sin(\theta)^2}}} \,,\\
    \frac{\partial W_{\theta}}{\partial \mathscr{L}}
    &=
    -\int \frac{ \mathscr{L} d \theta}{\sqrt{\mathscr{L}^2 - \frac{\mathscr{L}_z^2}{\sin(\theta)^2}}} \,.
\end{align*}
These integrals can be solved by the substitution $\sin(u) = \cot(\theta) \cot \iota$ (see e.g.~page 476 of~\cite{Goldstein-Safko-book}) to obtain
\begin{align*}
    \frac{\partial W_{\theta}}{\partial \mathscr{L}_{z}}
    &=
    -u 
    =\Omega - \phi
    \,,\\
    \frac{\partial W_{\theta}}{\partial \mathscr{L}}
    &=
    \tan^{-1} \left( 
    \frac{\tan(u)}{\cos \iota }\right)
    =
    \varphi \,,
\end{align*}
where we have used the identities in Eq.~\eqref{eq:u-identities} to simplify the expressions above.
Using the above equations in Eq.~\eqref{eq:Deluany-angles-v1}, we can simplify $g$ and $h$ to obtain
\begin{subequations}
\begin{align}
    g &= \varphi +
    \frac{\partial W_R}{\partial \mathscr{L}}
    +
    \frac{\partial W_R}{\partial \mathscr{E}} 
    \left[
    \frac{\partial \mathscr{E}}{\partial \mathscr{L}}
    -
    \frac{\partial \mathscr{E}}{\partial J_r} 
    \right]
    \,,\\
    h &= \Omega \,.
\end{align}
\end{subequations}
One can show that
\begin{align}
    \varphi + \frac{\partial W_R}{\partial \mathscr{L}} = \varphi(\chi=0) \equiv \omega\,.
\end{align}
Therefore,
\begin{subequations}
\begin{align}
    g &= 
    \omega +
    \frac{\partial W_R}{\partial \mathscr{E}} 
    \left[
    \frac{\partial \mathscr{E}}{\partial \mathscr{L}}
    -
    \frac{\partial \mathscr{E}}{\partial J_r} 
    \right]
    \,,\\
    h &= \Omega \,.
\end{align}
\end{subequations}
The radial integrals appearing in $W_R$, $\partial \mathscr{E}/\partial \mathscr{L}$ and $\partial \mathscr{E}/\partial J_r$ can be evaluated by changing variables from $r\to \chi$, using the quasi-Keplerian parameterization. 

In Newtonian gravity, one can show that
$g = \omega$.
In the relativistic problem, we express $(\chi, \omega)$ in terms of $(l,g)$.
Schematically
\begin{subequations}\label{eq:chi-omega-expansions}
\begin{align}
    \chi
    &=
    l
    +
    \sum_{j=1}^{\infty}
    a_{j,\chi}(p,e) \sin (j l) \,,\\
    \omega
    &=
    g
    +
    a_{0,\omega}(p,e) l
    +
    \sum_{j=1}^{\infty}
    a_{j, \omega} (p,e) \sin(j l) \,.
\end{align}
\end{subequations}
The coefficients $a_{j,\chi}$ and $a_{j,\omega}$ have the following scaling with eccentricity:
\begin{align}
    a_{j,\chi} = \mathcal{O}(e^j) \,, \quad 
    a_{j, \omega} = \mathcal{O}(e^j) \,, \quad  j \geq 1\,.
\end{align}
Expressions similar to Eq.~\eqref{eq:chi-omega-expansions} can be obtained for the other orbital elements.
For example, to obtain the sum representation of $\theta$ and $\phi$, we can adopt a small inclination expansion; more specifically, we expand in $s = \sin (\iota /2)$.
Using Eq.~\eqref{eq:euler-rotation-v2}, we obtain expansions for $\cos(\phi), \sin(\phi)$ and $\cos(\theta)$.
These expressions are long and complicated; for Newtonian results, see e.g.~Eqs.~(6.79)-(6.80) of~\cite{Murray-Dermott-Book}.
In our analysis, we do not need these expansions and therefore, we do not present them here. 
Our ultimate goal is to obtain the expansion of the orbital parameters in terms of the modified Delaunay variables [Eq.~\eqref{eq:modified-Delaunay-variables}]. 
Doing this is straightforward using the transformation listed in Eq.~\eqref{eq:modified-Delaunay-variables}, once we have expressions in terms of the Delaunay variables.
We provide explicit expressions for low-order terms in $e$ and for $\iota=0$ in Appendix~\ref{appendix:first-order-expansion}. 
\section{Small eccentricity expansion of orbital elements}\label{appendix:first-order-expansion}
The expansions of the orbital elements in terms of modified Delaunay variables when the inclination angle $\iota = 0$ are 
\begin{subequations}\label{eq:orb-elements-expansion-small-eccentricity}
\begin{align}
    &r = M p-e M p \cos (\lambda -\varpi )
    +
    e^2 \bigg[\frac{M p \left(3 p^2-27 p+50\right)}{2 \left(p^2-8 p+12\right)}
    \nonumber\\
    &-\frac{M p \left(p^2-11 p+26\right) \cos (2 \lambda -2 \varpi )}{2 \left(p^2-8 p+12\right)}\bigg]
   + \mathcal{O}(e^3) 
    \,,\\
    &\phi = \lambda +\frac{2 e (p-3) \sqrt{p} \sin (\lambda -\varpi )}{\sqrt{p-6} (p-2)}
    \nonumber\\
    &+\frac{e^2 \left(5 p^3-64 p^2+250 p-300\right) \sqrt{p} \sin (2 \lambda -2 \varpi )}{4 (p-6)^{3/2} (p-2)^2}
    \nonumber\\
    &+ \mathcal{O}(e^3)\,.
\end{align}
\end{subequations}
The relations between $(\chi,\omega)$ and $(\lambda,\varpi)$ are 
\begin{widetext}
\begin{subequations}
\begin{align}
    \chi &= 
    \lambda -\varpi
    +
    \frac{e \left(2 p^2-19 p+38\right) \sin (\lambda -\varpi )}{p^2-8 p+12}
    +
    \frac{e^2 \left(10 p^4-196 p^3+1357 p^2-3892 p+3892\right)}{8 \left(p^2-8 p+12\right)^2}
    \sin (2 (\lambda -\varpi ))
    +
    \mathcal{O}(e^3)
    \,\\
    \omega &= 
    \frac{\left(-12 p^{3/2}+p^{5/2}+36 \sqrt{p}\right) \varpi }{(p-6)^{5/2}}+\frac{\lambda  \left(\sqrt{p-6}-\sqrt{p}\right)}{\sqrt{p-6}}-\Omega
    +
    e^2 \left(\frac{3 \sqrt{p} \varpi }{4 (p-6)^{5/2}}-\frac{3 \lambda  \sqrt{p}}{4 (p-6)^{5/2}}\right)
    +
    \mathcal{O}(e^3)
    \,.
\end{align}
\end{subequations}
\end{widetext}
The expansion of the modified Delaunay action variables in terms of $(e,p)$ are given by
\begin{subequations}\label{eq:Lambda-P-Q-small-eccentricity-expansion}
\begin{align}
    &\Lambda =\frac{M p}{\sqrt{p-3}} 
    +
    \frac{1}{2} e^2 M \left(\frac{\sqrt{p-6} \sqrt{p}}{\sqrt{p-3} (p-2)}+\frac{1}{(p-3)^{3/2}}\right) p \nonumber\\
    &\hspace{2cm}+\mathcal{O}(e^4)\,,\\
    &P = \frac{e^2 M \sqrt{p-6} p^{3/2}}{2 \sqrt{p-3} (p-2)}  + \mathcal{O}(e^4)\,,\\
    &Q = \mathcal{O}(\sin(\iota/2)^2) \,.
\end{align}
\end{subequations}
The expansion of the frequencies of the system are
\begin{subequations}\label{eq:frequencies-iota-equal-zero}
\begin{align}
    &\nu_{\lambda} = 
    \frac{1}{M p^{3/2}}-\frac{3 e^2 \left(p^2-10 p+22\right)}{2 M (p-6) (p-2) p^{3/2}} +\mathcal{O}(e^4)
    \,,\\
    &\nu_{\varpi} = \frac{\sqrt{p}-\sqrt{p-6}}{M p^2} + 
    \nonumber\\
    &+
    \frac{e^2 \left(\frac{6 p^3-96 p^2+495 p-798}{(p-6)^{3/2}}-\frac{6 \sqrt{p} \left(p^2-10 p+22\right)}{p-6}\right)}{4 M (p-2) p^2}
    +
    \mathcal{O}(e^4)
    \,.
\end{align}
\end{subequations}
The expansion of the specific four momentum [Eq.~\eqref{eq:specfic-four-momentum-def}] is
\begin{subequations}\label{eq:four-momentum-small-eccen}
\begin{align}
    p_{t} &= -\mathscr{E}\,,\\
    p_{r} &= \frac{e \sqrt{p-6} \sqrt{p} \sin (\lambda -\varpi )}{\sqrt{p-3} (p-2)}
    \nonumber\\
    &+
    \frac{e^2 (p-7) \sin (2 \lambda -2 \varpi )}{(p-2) \sqrt{p+\frac{18}{p}-9}}
    +\mathcal{O}(e^3)
    \,,\\
    p_{\theta} &= 0\,,\\
    p_{\phi} &= \mathscr{L}\,,
\end{align}
\end{subequations}
where expressions for $\mathscr{E}$ and $\mathscr{L}$ are provided in Eq.~\eqref{eq:E-andL-in-p-and-e}.
\section{Derivation of secular evolution equations}\label{appendix:secular-deriv}
Let us first concentrate on $\left<\dot{\mathbf{P}}_{l}^{(2)}\right>$.
Using Eq.~\eqref{eq:hamiltonian-equations-3-body-first-order-sols}, we see that the first two terms in Eq.~\eqref{eq:dotPl-v1} simplify to 
\begin{align*}
    &m \mathrm{Re}\bigg[ \frac{\partial^2 (R e^{i \Phi})}{\partial \mathbf{Q}^{l} \partial \mathbf{Q}^{k'} } \mathbf{Q}^{k'}_{(1)} \bigg]
    =
    \nonumber\\
    &
    -m j_{l} j_{k'} \mathrm{Re}\bigg[ R e^{i \Phi}
    \int d\mu \mathrm{Re}\left[ i e^{i \Phi}\frac{\partial(R' f_0)}{\partial \mathbf{P}_{k'}}\right]
    \bigg]
    \,,\\
    &m \mathrm{Re}\bigg[\frac{\partial^2 (R e^{i \Phi})}{\partial \mathbf{Q}^{l} \partial \mathbf{P}_{k'} } \mathbf{P}_{k'}^{(1)}
    \bigg]
    =
    \nonumber\\
    &
    j_{l}
    \mathrm{Re}\bigg[i e^{i \Phi}\frac{\partial R }{\partial \mathbf{P}_{k'} } \int d\mu \mathrm{Re}\left[ f_0 R' e^{i \Phi} j_{l'}\right]
    \bigg]
\end{align*}
The average of these two quantities is zero because the phases $e^{i\Phi}$ inside and outside the integrals are independent of each other, i.e.~
\begin{subequations}\label{eq:sim-Pldot-1}
\begin{align}
    &\left<m \mathrm{Re}\bigg[ \frac{\partial^2 (R e^{i \Phi})}{\partial \mathbf{Q}^{l} \partial \mathbf{Q}^{k'} } \mathbf{Q}^{k'}_{(1)} \bigg]\right>=0 \,,\\
    &\left< m \mathrm{Re}\bigg[\frac{\partial^2 (R e^{i \Phi})}{\partial \mathbf{Q}^{l} \partial \mathbf{P}_{k'} } \mathbf{P}_{k'}^{(1)}
    \bigg]\right>
    = 0\,.
\end{align}
\end{subequations}
The last two terms in Eq.~\eqref{eq:dotPl-v1}, can be simplified to
\begin{widetext} 
\begin{align*}
    &m \mathrm{Re}
    \bigg[\frac{\partial^2 (R e^{i \Phi})}{\partial \mathbf{Q}^{l} \partial \mathbf{Q}^{k} } \mathbf{Q}^{k}_{(1)}
    \bigg]
    =
    -m^2 j_{l} j_{k} \mathrm{Re}
    \bigg[R e^{i \Phi} \mathrm{Re}\left[ i e^{i \Phi}\frac{\partial(R f_0)}{\partial \mathbf{P}_{k}}\right]
    \bigg]
    =
    -\frac{m^2 j_{l} j_{k}}{2}
    \mathrm{Re}
    \bigg[ 
        i R\, e^{2 i \Phi}  \frac{\partial(R f_0)}{\partial \mathbf{P}_{k}}
        -
        i R
        \frac{\partial(R f_0^{*})}{\partial \mathbf{P}_{k}}
    \bigg]
    \nonumber\\
    &
    =
    -\frac{m^2 j_{l} j_{k}}{2}
    \mathrm{Re}
    \bigg[i R\, e^{2 i \Phi}  \frac{\partial(R f_0)}{\partial \mathbf{P}_{k}} \bigg]
    +\frac{m^2 j_{l} j_{k}}{4}
    \bigg[ 
        R
        \frac{\partial[R (i f_0^{*} - i f_0)]}{\partial \mathbf{P}_{k}}    
    \bigg]
    \,,\\
    &m \mathrm{Re}
    \bigg[
     \frac{\partial^2 (R e^{i \Phi}) }{\partial \mathbf{Q}^{l} \partial \mathbf{P}_{k} } \mathbf{P}_{k}^{(1)}    
    \bigg]
    =
    m^2 j_{l} j_{k}
    \mathrm{Re}
    \bigg[i
    e^{i \Phi}
    \frac{ \partial R }{\partial \mathbf{P}_{k} } 
     \mathrm{Re}\left[ f_0 R e^{i \Phi} \right]    
    \bigg]
    =
    \frac{m^2 j_{l} j_{k}}{2}
    \mathrm{Re}
    \bigg[i
    e^{2 i \Phi}
    f_0 R
    \frac{ \partial R }{\partial \mathbf{P}_{k} } 
    +
    i f_0^* R
    \frac{ \partial R }{\partial \mathbf{P}_{k} } 
    \bigg] 
    \nonumber\\
    &=
    \frac{m^2 j_{l} j_{k}}{2}
    \mathrm{Re}
    \bigg[i
    e^{2 i \Phi}
    f_0 R
    \frac{ \partial R }{\partial \mathbf{P}_{k} }
    \bigg]
    +
    \frac{m^2 j_{l} j_{k}}{4}
    \bigg[ 
    R (i f_0^{*} - i f_0) \frac{ \partial R }{\partial \mathbf{P}_{k} } 
    \bigg]
\end{align*}
Averaging the above quantities and taking the sum we obtain
\begin{subequations}\label{eq:sim-Pldot-2}
\begin{align}
    &\left<m \mathrm{Re}
    \bigg[\frac{\partial^2 (R e^{i \Phi})}{\partial \mathbf{Q}^{l} \partial \mathbf{Q}^{k} } \mathbf{Q}^{k}_{(1)}
    \bigg] \right>
    +
    \left<
     m \mathrm{Re}
    \bigg[
     \frac{\partial^2 (R e^{i \Phi}) }{\partial \mathbf{Q}^{l} \partial \mathbf{P}_{k} } \mathbf{P}_{k}^{(1)}    
    \bigg]   
    \right>
    =
    \frac{m^2 j_{l} j_{k}}{4}
    \bigg[ 
        R
        \frac{\partial[R (i f_0^{*} - i f_0)]}{\partial \mathbf{P}_{k}}    
    \bigg]
    +
    \frac{m^2 j_{l} j_{k}}{4}
    \bigg[ 
    R (i f_0^{*} - i f_0) \frac{ \partial R }{\partial \mathbf{P}_{k} } 
    \bigg]
    \nonumber\\
    &=
    \frac{m^2 j_{l} j_{k}}{4}
    \bigg[ 
        \frac{\partial[R^2 (i f_0^{*} - i f_0)]}{\partial \mathbf{P}_{k}}    
    \bigg]
    =
    \frac{m^2 j_{l} j_{k}}{2}
        \frac{\partial}{\partial \mathbf{P}_{k}}  
        \left[  \frac{(R)^2 \gamma}{\dot{\Phi}^2 + \gamma^2}\right]
\end{align}
\end{subequations}
\end{widetext} 
Combining Eqs.~\eqref{eq:sim-Pldot-1} and \eqref{eq:sim-Pldot-2}, we obtain Eq.~\eqref{eq:Pldot-Plprimedot-1}.
We can use the same steps presented above to derive Eq.~\eqref{eq:Pldot-Plprimedot-2}.
\section{Derivation of the Bessel function relationship}\label{appendix:bessel-function-derivation}
The modified Bessel function of the second kind is defined by 
\begin{align}
    K_0 (\omega) = \frac{1}{2} \int_{-\infty}^{\infty} \frac{e^{i \omega t}}{\sqrt{1+t^2}} dt \,, \quad \mathrm{Re} (\omega) >0 \,.
\end{align}
Let now $|1-\alpha| \equiv \epsilon$, so that we can use Eq.~\eqref{eq:relativistic-Laplace-Coefficients} to write
\begin{align}\label{eq:ak-coeff-v1-appendix}
    b^{j}_{1/2}(\alpha,a,b) &= \frac{1}{\pi}\int_{-\pi}^{\pi} d \Psi \frac{\cos( j \Psi)}{\sqrt{a \epsilon^2 + b \Psi^2}} \,,  \nonumber\\
    &=
    \frac{1}{\pi \sqrt{a} \epsilon }\int_{-\pi}^{\pi} d \Psi \frac{\cos( j \Psi)}{\sqrt{1 + \frac{b}{a \epsilon^2} \Psi^2}} \,, \nonumber \\
    &=
    \frac{1}{\pi \sqrt{b} }\int_{-\frac{\pi}{\epsilon}\sqrt{\frac{b}{a}}}^{\frac{\pi}{\epsilon}\sqrt{\frac{b}{a}}} dt \frac{\cos\left( j t \epsilon\sqrt{\frac{a}{b}}\right) }{\sqrt{1 + t^2}} dt \,,\nonumber \\
    &\approx
     \frac{1}{\pi \sqrt{b} }\int_{-\infty}^{\infty} dt \frac{\cos\left( j t \epsilon\sqrt{\frac{a}{b}}\right) }{\sqrt{1 + t^2}} dt  \,,\nonumber \\
    &= \frac{2}{\pi \sqrt{b} } K_0 \left(\sqrt{\frac{a}{b}} j \epsilon \right)\,.
\end{align}
\section{Secular evolution: Relativistic expressions for corotation resonance}\label{appendix:secular-evol-resonances}
In this appendix, we provide the expressions for the secular evolution of the eccentricity due to corotation resonances. 
We also provide these expressions in the supplementary \texttt{MATHEMATICA} notebook.

The expressions for $\left< \dot{e}' \right>_{\mathrm{Cr,k}}^{(0)}$ is given by
\begin{widetext}
\begin{align}
    &\left< \dot{e}' \right>_{\mathrm{Cr,k}}^{(0)}
    =
    -\frac{e'{} k \left(8 (p'{}-3)^{9/2} (p'{}-9)\right) q}{(p'{}-6)^{7/2} (p'{}-2) \sqrt{p'{}}}\Sigma(p')
    \left[K_0\left(\frac{\kappa }{2 | A| }\right)+\frac{\sqrt{p'{}-6} }{2 \sqrt{p'{}-3}}K_1\left(\frac{\kappa }{2 | A| }\right)\right]^2
    \nonumber\\
    &
    -\frac{e'{} q \left(16 k \sqrt{p'{}} (p'{}-3)^{11/2}\right)}{3 (p'{}-6)^{5/2} (p'{}-2)^2}
    \Sigma'
    \left[K_0\left(\frac{\kappa }{2 | A| }\right)+\frac{\sqrt{p'{}-6} }{2 \sqrt{p'{}-3}}K_1\left(\frac{\kappa }{2 | A| }\right)\right]^2
    \,,
\end{align}
\end{widetext}
where
\begin{align}
    \frac{\kappa}{2|A|} = \frac{2 \sqrt{(p'{}-6) (p'{}-3)}}{3 (p'{}-2)}
    \,.
\end{align}
The expression for $\left< \dot{e}' \right>_{\mathrm{Cr,k}}^{(1)}$ is too long to be displayed here and we present it in the supplementary \texttt{MATHEMATICA} file.
\section{Components of the metric in comoving Fermi normal coordinates}\label{appendix:comoving-coordinates}
We list the components of the metric in co-moving Fermi normal coordinates below:
\begin{widetext}
\begin{subequations}
\begin{align}
    g_{\tau \tau} &= -1
    +
    \frac{3 (\tilde{p}-2) x_{c}^2 {}}{M^2 (\tilde{p}-3) \tilde{p}^3}-\frac{z_c^2 {}}{M^2 (\tilde{p}-3) \tilde{p}^2}
    \,,\\
    g_{\tau x_{c}} &= 
    -\frac{2 \sqrt{\tilde{p}-2} x_{c} y_{c} {}}{M^2 (\tilde{p}-3) \tilde{p}^3}-\frac{y_{c} {}}{M \tilde{p}^{3/2}}
    \,,\\
    g_{\tau y_{c}} &= 
    \frac{2 \sqrt{\tilde{p}-2} x_{c}^2 {}}{M^2 (\tilde{p}-3) \tilde{p}^3}-\frac{2 \sqrt{\tilde{p}-2} z_c^2 {}}{M^2 (\tilde{p}-3) \tilde{p}^3}+\frac{x_{c} {}}{M \tilde{p}^{3/2}}
    \,,\\
    g_{\tau z_{c}} &= 
    \frac{2 \sqrt{\tilde{p}-2} y_{c} z_c {}}{M^2 (\tilde{p}-3) \tilde{p}^3}
    \,,\\
    g_{x_{c} x_{c}} &= 1
    +
    \frac{z_c^2 {}}{3 M^2 \tilde{p}^3}+\frac{y_{c}^2 {}}{3 M^2 (\tilde{p}-3) \tilde{p}^2}
    \,,\\
    g_{x_{c} y_{c}} &= 
    -\frac{x_{c} y_{c} {}}{3 M^2 (\tilde{p}-3) \tilde{p}^2}
    \,,\\
    g_{x_{c} z_{c}} &= 
    -\frac{x_{c} z_c {}}{3 M^2 \tilde{p}^3}
    \,,\\
    g_{y_{c} y_{c}} &= 
    1
    -\frac{(2 \tilde{p}-3) z_c^2 {}}{3 M^2 (\tilde{p}-3) \tilde{p}^3}+\frac{x_{c}^2 {}}{3 M^2 (\tilde{p}-3) \tilde{p}^2}
    \,,\\
    g_{y_{c} z_{c}} &= 
    \frac{(2 \tilde{p}-3) y_{c} z_c {}}{3 M^2 (\tilde{p}-3) \tilde{p}^3}
    \,,\\
    g_{z_{c}, z_{c}} &= 
    1
    +
    \frac{x_{c}^2 {}}{3 M^2 \tilde{p}^3}-\frac{(2 \tilde{p}-3) y_{c}^2 {}}{3 M^2 (\tilde{p}-3) \tilde{p}^3}
    \,.
\end{align}
\end{subequations}
\end{widetext}
\bibliography{ref}

@ARTICLE{Blaes:2014,
       author = {{Blaes}, Omer},
        title = "{General Overview of Black Hole Accretion Theory}",
      journal = {\ssr},
         year = 2014,
        month = sep,
       volume = {183},
       number = {1-4},
        pages = {21-41},
          doi = {10.1007/s11214-013-9985-6},
archivePrefix = {arXiv},
       eprint = {1304.4879},
 primaryClass = {astro-ph.HE},
       adsurl = {https://ui.adsabs.harvard.edu/abs/2014SSRv..183...21B}
}

@article{Duque:2025yfm,
    author = "Duque, Francisco and Sberna, Laura and Spiers, Andrew and Vicente, Rodrigo",
    title = "{Extreme-mass-ratio inspirals in relativistic accretion discs}",
    eprint = "2510.02433",
    archivePrefix = "arXiv",
    primaryClass = "gr-qc",
    month = "10",
    year = "2025"
}

@ARTICLE{yuan_2014,
       author = {{Yuan}, Feng and {Narayan}, Ramesh},
        title = "{Hot Accretion Flows Around Black Holes}",
      journal = {\araa},
         year = 2014,
        month = aug,
       volume = {52},
        pages = {529-588},
          doi = {10.1146/annurev-astro-082812-141003},
archivePrefix = {arXiv},
       eprint = {1401.0586},
 primaryClass = {astro-ph.HE},
       adsurl = {https://ui.adsabs.harvard.edu/abs/2014ARA&A..52..529Y}
}

@BOOK{Binney-Tremaine-Book,
       author = {{Binney}, James and {Tremaine}, Scott},
        title = "{Galactic Dynamics: Second Edition}",
         year = 2008,
       adsurl = {https://ui.adsabs.harvard.edu/abs/2008gady.book.....B}
}

@ARTICLE{jiang_2019_super,
       author = {{Jiang}, Yan-Fei and {Stone}, James M. and {Davis}, Shane W.},
        title = "{Super-Eddington Accretion Disks around Supermassive Black Holes}",
      journal = {\apj},
         year = 2019,
        month = aug,
       volume = {880},
       number = {2},
          eid = {67},
        pages = {67},
          doi = {10.3847/1538-4357/ab29ff},
archivePrefix = {arXiv},
       eprint = {1709.02845},
 primaryClass = {astro-ph.HE},
       adsurl = {https://ui.adsabs.harvard.edu/abs/2019ApJ...880...67J}
}

@ARTICLE{aird_2013,
       author = {{Aird}, James and {Coil}, Alison L. and {Moustakas}, John and {Diamond-Stanic}, Aleksandar M. and {Blanton}, Michael R. and {Cool}, Richard J. and {Eisenstein}, Daniel J. and {Wong}, Kenneth C. and {Zhu}, Guangtun},
        title = "{PRIMUS: An Observationally Motivated Model to Connect the Evolution of the Active Galactic Nucleus and Galaxy Populations out to z \raisebox{-0.5ex}\textasciitilde 1}",
      journal = {\apj},
         year = 2013,
        month = sep,
       volume = {775},
       number = {1},
          eid = {41},
        pages = {41},
          doi = {10.1088/0004-637X/775/1/41},
archivePrefix = {arXiv},
       eprint = {1301.1689},
 primaryClass = {astro-ph.CO},
       adsurl = {https://ui.adsabs.harvard.edu/abs/2013ApJ...775...41A}
}

@ARTICLE{turner_2004,
       author = {{Turner}, N.~J.},
        title = "{On the Vertical Structure of Radiation-dominated Accretion Disks}",
      journal = {\apjl},
         year = 2004,
        month = apr,
       volume = {605},
       number = {1},
        pages = {L45-L48},
          doi = {10.1086/386545},
archivePrefix = {arXiv},
       eprint = {astro-ph/0402539},
 primaryClass = {astro-ph},
       adsurl = {https://ui.adsabs.harvard.edu/abs/2004ApJ...605L..45T}
}

@ARTICLE{balbus_2003,
       author = {{Balbus}, Steven A.},
        title = "{Enhanced Angular Momentum Transport in Accretion Disks}",
      journal = {\araa},
         year = 2003,
        month = jan,
       volume = {41},
        pages = {555-597},
          doi = {10.1146/annurev.astro.41.081401.155207},
archivePrefix = {arXiv},
       eprint = {astro-ph/0306208},
 primaryClass = {astro-ph},
       adsurl = {https://ui.adsabs.harvard.edu/abs/2003ARA&A..41..555B}
}

@article{White:2023,
   title={An Extension of the Athena++ Code Framework for Radiation-magnetohydrodynamics in General Relativity Using a Finite-solid-angle Discretization},
   volume={949},
   ISSN={1538-4357},
   url={http://dx.doi.org/10.3847/1538-4357/acc8cf},
   DOI={10.3847/1538-4357/acc8cf},
   number={2},
   journal={The Astrophysical Journal},
   publisher={American Astronomical Society},
   author={White, Christopher J. and Mullen, Patrick D. and Jiang, Yan-Fei and Davis, Shane W. and Stone, James M. and Morozova, Viktoriya and Zhang , Lizhong},
   year={2023},
   month=jun, pages={103} }

@ARTICLE{Davis:2020,
       author = {{Davis}, Shane W. and {Tchekhovskoy}, Alexander},
        title = "{Magnetohydrodynamics Simulations of Active Galactic Nucleus Disks and Jets}",
      journal = {\araa},
         year = 2020,
        month = aug,
       volume = {58},
        pages = {407-439},
          doi = {10.1146/annurev-astro-081817-051905},
archivePrefix = {arXiv},
       eprint = {2101.08839},
 primaryClass = {astro-ph.HE},
       adsurl = {https://ui.adsabs.harvard.edu/abs/2020ARA&A..58..407D}
}

@ARTICLE{Ward-1988,
       author = {{Ward}, W.~R.},
        title = "{On disk-planet interactions and orbital eccentricities}",
      journal = {\icarus},
         year = 1988,
        month = feb,
       volume = {73},
       number = {2},
        pages = {330-348},
          doi = {10.1016/0019-1035(88)90103-0},
       adsurl = {https://ui.adsabs.harvard.edu/abs/1988Icar...73..330W}
}

@article{Hirata_2011,
   title={Lindblad resonance torques in relativistic discs - I. Basic equations: Lindblad resonances - I},
   volume={414},
   ISSN={0035-8711},
   url={http://dx.doi.org/10.1111/j.1365-2966.2011.18617.x},
   DOI={10.1111/j.1365-2966.2011.18617.x},
   number={4},
   journal={Monthly Notices of the Royal Astronomical Society},
   publisher={Oxford University Press (OUP)},
   author={Hirata, Christopher M.},
   year={2011},
   month=may, pages={3198–3211} }

@article{Hirata_2011_II,
   title={Lindblad resonance torques in relativistic discs - II. Computation of resonance strengths: Lindblad resonances - II},
   volume={414},
   ISSN={0035-8711},
   url={http://dx.doi.org/10.1111/j.1365-2966.2011.18619.x},
   DOI={10.1111/j.1365-2966.2011.18619.x},
   number={4},
   journal={Monthly Notices of the Royal Astronomical Society},
   publisher={Oxford University Press (OUP)},
   author={Hirata, Christopher M.},
   year={2011},
   month=may, pages={3212–3230} }

@ARTICLE{GT-disc-satellite-interaction,
       author = {{Goldreich}, P. and {Tremaine}, S.},
        title = "{Disk-satellite interactions.}",
      journal = {APJ},
         year = 1980,
        month = oct,
       volume = {241},
        pages = {425-441},
          doi = {10.1086/158356},
       adsurl = {https://ui.adsabs.harvard.edu/abs/1980ApJ...241..425G}
}

@article{MEYERVERNET1987157,
title = {On the physics of resonant disk-satellite interaction},
journal = {Icarus},
volume = {69},
number = {1},
pages = {157-175},
year = {1987},
issn = {0019-1035},
doi = {https://doi.org/10.1016/0019-1035(87)90011-X},
url = {https://www.sciencedirect.com/science/article/pii/001910358790011X},
author = {N. Meyer-Vernet and B. Sicardy}
}

@article{Warburton_2017,
   title={Evolution of small-mass-ratio binaries with a spinning secondary},
   volume={96},
   ISSN={2470-0029},
   url={http://dx.doi.org/10.1103/PhysRevD.96.084057},
   DOI={10.1103/physrevd.96.084057},
   number={8},
   journal={Physical Review D},
   publisher={American Physical Society (APS)},
   author={Warburton, Niels and Osburn, Thomas and Evans, Charles. R.},
   year={2017},
   month=oct }

@article{Heffernan_2012,
   title={High-order expansions of the Detweiler-Whiting singular field in Schwarzschild spacetime},
   volume={86},
   ISSN={1550-2368},
   url={http://dx.doi.org/10.1103/PhysRevD.86.104023},
   DOI={10.1103/physrevd.86.104023},
   number={10},
   journal={Physical Review D},
   publisher={American Physical Society (APS)},
   author={Heffernan, Anna and Ottewill, Adrian and Wardell, Barry},
   year={2012},
   month=nov }

@article{Pound_2014,
   title={Practical, covariant puncture for second-order self-force calculations},
   volume={89},
   ISSN={1550-2368},
   url={http://dx.doi.org/10.1103/PhysRevD.89.104020},
   DOI={10.1103/physrevd.89.104020},
   number={10},
   journal={Physical Review D},
   publisher={American Physical Society (APS)},
   author={Pound, Adam and Miller, Jeremy},
   year={2014},
   month=may }

@ARTICLE{GT-Uranus-Ring,
       author = {{Goldreich}, P. and {Tremaine}, S.},
        title = "{The origin of the eccentricities of the rings of Uranus}",
      journal = {APJ},
         year = 1981,
        month = feb,
       volume = {243},
        pages = {1062-1075},
          doi = {10.1086/158671},
       adsurl = {https://ui.adsabs.harvard.edu/abs/1981ApJ...243.1062G}
}

@article{Gammie_2004,
   title={The Magnetorotational Instability in the Kerr Metric},
   volume={614},
   ISSN={1538-4357},
   url={http://dx.doi.org/10.1086/423443},
   DOI={10.1086/423443},
   number={1},
   journal={The Astrophysical Journal},
   publisher={American Astronomical Society},
   author={Gammie, Charles F.},
   year={2004},
   month=oct, pages={309–313} }

@INPROCEEDINGS{Novikov-Thorne,
       author = {{Novikov}, I.~D. and {Thorne}, K.~S.},
        title = "{Astrophysics of black holes.}",
    booktitle = {Black Holes (Les Astres Occlus)},
         year = 1973,
       editor = {{Dewitt}, C. and {Dewitt}, B.~S.},
        month = jan,
        pages = {343-450},
       adsurl = {https://ui.adsabs.harvard.edu/abs/1973blho.conf..343N}
}

@ARTICLE{ogilvie-gordon-2007,
       author = {{Ogilvie}, Gordon I.},
        title = "{Mean-motion resonances in satellite-disc interactions}",
      journal = {MNRAS},
         year = 2007,
        month = jan,
       volume = {374},
       number = {1},
        pages = {131-149},
          doi = {10.1111/j.1365-2966.2006.11141.x},
archivePrefix = {arXiv},
       eprint = {astro-ph/0610082},
 primaryClass = {astro-ph},
       adsurl = {https://ui.adsabs.harvard.edu/abs/2007MNRAS.374..131O}
}

@BOOK{Goldstein-Safko-book,
       author = {{Goldstein}, H. and {Poole}, C. and {Safko}, J.},
        title = "{Classical mechanics}",
         year = 2002,
       adsurl = {https://ui.adsabs.harvard.edu/abs/2002clme.book.....G}
}

@ARTICLE{Ogilvie_2007,
       author = {{Ogilvie}, Gordon I.},
        title = "{Mean-motion resonances in satellite-disc interactions}",
      journal = {MNRAS},
         year = 2007,
        month = jan,
       volume = {374},
       number = {1},
        pages = {131-149},
          doi = {10.1111/j.1365-2966.2006.11141.x},
archivePrefix = {arXiv},
       eprint = {astro-ph/0610082},
 primaryClass = {astro-ph},
       adsurl = {https://ui.adsabs.harvard.edu/abs/2007MNRAS.374..131O}
}

@article{Poisson_2011,
   title={The Motion of Point Particles in Curved Spacetime},
   volume={14},
   ISSN={1433-8351},
   url={http://dx.doi.org/10.12942/lrr-2011-7},
   DOI={10.12942/lrr-2011-7},
   number={1},
   journal={Living Reviews in Relativity},
   publisher={Springer Science and Business Media LLC},
   author={Poisson, Eric and Pound, Adam and Vega, Ian},
   year={2011},
   month=sep }

@BOOK{Murray-Dermott-Book,
       author = {{Murray}, Carl D. and {Dermott}, Stanley F.},
        title = "{Solar System Dynamics}",
         year = 1999,
          doi = {10.1017/CBO9781139174817},
       adsurl = {https://ui.adsabs.harvard.edu/abs/1999ssd..book.....M}
}

@ARTICLE{Ward-1997,
       author = {{Ward}, William R.},
        title = "{Protoplanet Migration by Nebula Tides}",
      journal = {\icarus},
         year = 1997,
        month = apr,
       volume = {126},
       number = {2},
        pages = {261-281},
          doi = {10.1006/icar.1996.5647},
       adsurl = {https://ui.adsabs.harvard.edu/abs/1997Icar..126..261W}
}

@misc{katz2024efficientgpuacceleratedmultisourceglobal,
      title={An efficient GPU-accelerated multi-source global fit pipeline for LISA data analysis}, 
      author={Michael L. Katz and Nikolaos Karnesis and Natalia Korsakova and Jonathan R. Gair and Nikolaos Stergioulas},
      year={2024},
      eprint={2405.04690},
      archivePrefix={arXiv},
      primaryClass={gr-qc},
      url={https://arxiv.org/abs/2405.04690}, 
}

@article{Bender:1997hs,
    author = "Bender, P. L. and Hils, D.",
    title = "{Confusion noise level due to galactic and extragalactic binaries}",
    doi = "10.1088/0264-9381/14/6/008",
    journal = "Class. Quant. Grav.",
    volume = "14",
    pages = "1439--1444",
    year = "1997"
}

@misc{strub2024globalanalysislisadata,
      title={Global Analysis of LISA Data with Galactic Binaries and Massive Black Hole Binaries}, 
      author={Stefan H. Strub and Luigi Ferraioli and Cédric Schmelzbach and Simon C. Stähler and Domenico Giardini},
      year={2024},
      eprint={2403.15318},
      archivePrefix={arXiv},
      primaryClass={gr-qc},
      url={https://arxiv.org/abs/2403.15318}, 
}

@article{Silva:2025lkl,
    author = "Silva, Makana and Blake-Goszyk, Harrison G. and Hirata, Christopher M.",
    title = "{Resonant interactions from dynamical perturbers on generic orbits around an extreme mass ratio inspiral}",
    eprint = "2507.22260",
    archivePrefix = "arXiv",
    primaryClass = "gr-qc",
    month = "7",
    year = "2025"
}

@article{Yunes:2010sm,
    author = "Yunes, Nicolas and Coleman Miller, M. and Thornburg, Jonathan",
    title = "{The Effect of Massive Perturbers on Extreme Mass-Ratio Inspiral Waveforms}",
    eprint = "1010.1721",
    archivePrefix = "arXiv",
    primaryClass = "astro-ph.GA",
    doi = "10.1103/PhysRevD.83.044030",
    journal = "Phys. Rev. D",
    volume = "83",
    pages = "044030",
    year = "2011"
}

@article{Naoz_2013,
   title={RESONANT POST-NEWTONIAN ECCENTRICITY EXCITATION IN HIERARCHICAL THREE-BODY SYSTEMS},
   volume={773},
   ISSN={1538-4357},
   url={http://dx.doi.org/10.1088/0004-637X/773/2/187},
   DOI={10.1088/0004-637x/773/2/187},
   number={2},
   journal={The Astrophysical Journal},
   publisher={American Astronomical Society},
   author={Naoz, Smadar and Kocsis, Bence and Loeb, Abraham and Yunes, Nicolás},
   year={2013},
   month=aug, pages={187} }

@article{Silva:2022blb,
    author = "Silva, Makana and Hirata, Christopher",
    title = "{Dynamical perturbations around an extreme mass ratio inspiral near resonance}",
    eprint = "2207.07733",
    archivePrefix = "arXiv",
    primaryClass = "gr-qc",
    doi = "10.1103/PhysRevD.106.084058",
    journal = "Phys. Rev. D",
    volume = "106",
    number = "8",
    pages = "084058",
    year = "2022"
}

@ARTICLE{2017CQGra..34x4002R,
       author = {{Robson}, Travis and {Cornish}, Neil},
        title = "{Impact of galactic foreground characterization on a global analysis for the LISA gravitational wave observatory}",
      journal = {Classical and Quantum Gravity},
         year = 2017,
        month = dec,
       volume = {34},
       number = {24},
          eid = {244002},
        pages = {244002},
          doi = {10.1088/1361-6382/aa9601},
archivePrefix = {arXiv},
       eprint = {1705.09421},
 primaryClass = {gr-qc},
       adsurl = {https://ui.adsabs.harvard.edu/abs/2017CQGra..34x4002R}
}

@inbook{C_rdenas_Avenda_o_2024,
   title={Testing Gravity with Extreme-Mass-Ratio Inspirals},
   ISBN={9789819728718},
   ISSN={2731-7358},
   url={http://dx.doi.org/10.1007/978-981-97-2871-8_8},
   DOI={10.1007/978-981-97-2871-8_8},
   booktitle={Recent Progress on Gravity Tests},
   publisher={Springer Nature Singapore},
   author={Cárdenas-Avendaño, Alejandro and Sopuerta, Carlos F.},
   year={2024},
   pages={275–359} }

@article{Amaro_Seoane_2018,
   title={Relativistic dynamics and extreme mass ratio inspirals},
   volume={21},
   ISSN={1433-8351},
   url={http://dx.doi.org/10.1007/s41114-018-0013-8},
   DOI={10.1007/s41114-018-0013-8},
   number={1},
   journal={Living Reviews in Relativity},
   publisher={Springer Science and Business Media LLC},
   author={Amaro-Seoane, Pau},
   year={2018},
   month=may }

@misc{witzany2022actionanglecoordinatesblackholegeodesics,
      title={Action-angle coordinates for black-hole geodesics I: Spherically symmetric and Schwarzschild}, 
      author={Vojtěch Witzany},
      year={2022},
      eprint={2203.11952},
      archivePrefix={arXiv},
      primaryClass={gr-qc},
      url={https://arxiv.org/abs/2203.11952}, 
}

@article{Wardell_2023,
   title={Gravitational Waveforms for Compact Binaries from Second-Order Self-Force Theory},
   volume={130},
   ISSN={1079-7114},
   url={http://dx.doi.org/10.1103/PhysRevLett.130.241402},
   DOI={10.1103/physrevlett.130.241402},
   number={24},
   journal={Physical Review Letters},
   publisher={American Physical Society (APS)},
   author={Wardell, Barry and Pound, Adam and Warburton, Niels and Miller, Jeremy and Durkan, Leanne and Le Tiec, Alexandre},
   year={2023},
   month=jun }

@article{Miranda_2019,
   title={Multiple Spiral Arms in Protoplanetary Disks: Linear Theory},
   volume={875},
   ISSN={1538-4357},
   url={http://dx.doi.org/10.3847/1538-4357/ab0f9e},
   DOI={10.3847/1538-4357/ab0f9e},
   number={1},
   journal={The Astrophysical Journal},
   publisher={American Astronomical Society},
   author={Miranda, Ryan and Rafikov, Roman R.},
   year={2019},
   month=apr, pages={37} }

@article{Ogilvie-Lubow-wake,
    author = {Ogilvie, G. I. and Lubow, S. H.},
    title = {On the wake generated by a planet in a disc},
    journal = {Monthly Notices of the Royal Astronomical Society},
    volume = {330},
    number = {4},
    pages = {950-954},
    year = {2002},
    month = {03},
    issn = {0035-8711},
    doi = {10.1046/j.1365-8711.2002.05148.x},
    url = {https://doi.org/10.1046/j.1365-8711.2002.05148.x},
    eprint = {https://academic.oup.com/mnras/article-pdf/330/4/950/3296251/330-4-950.pdf},
}

@book{Poisson-Will, 
place={Cambridge}, 
title={Gravity: Newtonian, Post-Newtonian, Relativistic}, 
DOI={10.1017/CBO9781139507486}, 
publisher={Cambridge University Press}, 
author={Poisson, Eric and Will, Clifford M.}, 
year={2014}}

@article{Fujita_2009,
   title={Analytical solutions of bound timelike geodesic orbits in Kerr spacetime},
   volume={26},
   ISSN={1361-6382},
   url={http://dx.doi.org/10.1088/0264-9381/26/13/135002},
   DOI={10.1088/0264-9381/26/13/135002},
   number={13},
   journal={Classical and Quantum Gravity},
   publisher={IOP Publishing},
   author={Fujita, Ryuichi and Hikida, Wataru},
   year={2009},
   month=jun, pages={135002} }

@article{Pound_2008,
   title={Osculating orbits in Schwarzschild spacetime, with an application to extreme mass-ratio inspirals},
   volume={77},
   ISSN={1550-2368},
   url={http://dx.doi.org/10.1103/PhysRevD.77.044013},
   DOI={10.1103/physrevd.77.044013},
   number={4},
   journal={Physical Review D},
   publisher={American Physical Society (APS)},
   author={Pound, Adam and Poisson, Eric},
   year={2008},
   month=feb }

@BOOK{Chandrasekar-BH-Book,
       author = {{Chandrasekhar}, S.},
        title = "{The mathematical theory of black holes}",
         year = 1983,
       adsurl = {https://ui.adsabs.harvard.edu/abs/1983mtbh.book.....C}
}

@book{K-Cole-perturbation-theory,
    author = {Kevorkian, J. and Cole, J. D.} ,
    title = "Multiple Scale and Singular Perturbation Methods",
    publisher = "Springer New York, NY",
    year = "1996",
    doi = {https://doi.org/10.1007/978-1-4612-3968-0}
}

@INPROCEEDINGS{2014prpl.conf..667B,
       author = {{Baruteau}, C. and {Crida}, A. and {Paardekooper}, S. -J. and {Masset}, F. and {Guilet}, J. and {Bitsch}, B. and {Nelson}, R. and {Kley}, W. and {Papaloizou}, J.},
        title = "{Planet-Disk Interactions and Early Evolution of Planetary Systems}",
    booktitle = {Protostars and Planets VI},
         year = 2014,
       editor = {{Beuther}, Henrik and {Klessen}, Ralf S. and {Dullemond}, Cornelis P. and {Henning}, Thomas},
        month = jan,
        pages = {667-689},
          doi = {10.2458/azu_uapress_9780816531240-ch029},
archivePrefix = {arXiv},
       eprint = {1312.4293},
 primaryClass = {astro-ph.EP},
       adsurl = {https://ui.adsabs.harvard.edu/abs/2014prpl.conf..667B}
}

@ARTICLE{Ward-1986,
       author = {{Ward}, W.~R.},
        title = "{Density waves in the solar nebula: Diffential Lindblad torque}",
      journal = {ICARUS},
         year = 1986,
        month = jul,
       volume = {67},
       number = {1},
        pages = {164-180},
          doi = {10.1016/0019-1035(86)90182-X},
       adsurl = {https://ui.adsabs.harvard.edu/abs/1986Icar...67..164W}
}

@misc{duque2025constrainingaccretionphysicsgravitational,
      title={Constraining accretion physics with gravitational waves from eccentric extreme-mass-ratio inspirals}, 
      author={Francisco Duque and Shubham Kejriwal and Laura Sberna and Lorenzo Speri and Jonathan Gair},
      year={2025},
      eprint={2411.03436},
      archivePrefix={arXiv},
      primaryClass={gr-qc},
      url={https://arxiv.org/abs/2411.03436}, 
}

@article{Kocsis_2011,
   title={Observable signatures of extreme mass-ratio inspiral black hole binaries embedded in thin accretion disks},
   volume={84},
   ISSN={1550-2368},
   url={http://dx.doi.org/10.1103/PhysRevD.84.024032},
   DOI={10.1103/physrevd.84.024032},
   number={2},
   journal={Physical Review D},
   publisher={American Physical Society (APS)},
   author={Kocsis, Bence and Yunes, Nicolás and Loeb, Abraham},
   year={2011},
   month=jul }

@article{Fujita_2012,
   title={Gravitational Waves from a Particle in Circular Orbits around a Schwarzschild Black Hole to the 22nd Post-Newtonian Order},
   volume={128},
   ISSN={1347-4081},
   url={http://dx.doi.org/10.1143/PTP.128.971},
   DOI={10.1143/ptp.128.971},
   number={5},
   journal={Progress of Theoretical Physics},
   publisher={Oxford University Press (OUP)},
   author={Fujita, R.},
   year={2012},
   month=nov, pages={971–992} }

@inbook{Pound_2021,
   title={Black Hole Perturbation Theory and Gravitational Self-Force},
   ISBN={9789811547027},
   url={http://dx.doi.org/10.1007/978-981-15-4702-7_38-1},
   DOI={10.1007/978-981-15-4702-7_38-1},
   booktitle={Handbook of Gravitational Wave Astronomy},
   publisher={Springer Singapore},
   author={Pound, Adam and Wardell, Barry},
   year={2021},
   pages={1–119} }

@BOOK{2002mcma.book.....M,
       author = {{Morbidelli}, Alessandro},
        title = "{Modern celestial mechanics : aspects of solar system dynamics}",
         year = 2002,
       adsurl = {https://ui.adsabs.harvard.edu/abs/2002mcma.book.....M}
}

@misc{fairbairn2025pushinglimitseccentricityplanetdisc,
      title={Pushing the limits of eccentricity in planet-disc interactions}, 
      author={Callum W. Fairbairn and Alexander J. Dittmann},
      year={2025},
      eprint={2506.19917},
      archivePrefix={arXiv},
      primaryClass={astro-ph.EP},
      url={https://arxiv.org/abs/2506.19917}, 
}

@ARTICLE{2002ApJ...565.1257T,
       author = {{Tanaka}, Hidekazu and {Takeuchi}, Taku and {Ward}, William R.},
        title = "{Three-Dimensional Interaction between a Planet and an Isothermal Gaseous Disk. I. Corotation and Lindblad Torques and Planet Migration}",
      journal = {\apj},
         year = 2002,
        month = feb,
       volume = {565},
       number = {2},
        pages = {1257-1274},
          doi = {10.1086/324713},
       adsurl = {https://ui.adsabs.harvard.edu/abs/2002ApJ...565.1257T}
}

@ARTICLE{1993Icar..102..150K,
       author = {{Korycansky}, D.~G. and {Pollack}, J.~B.},
        title = "{Numerical Calculations of the Linear Response of a Gaseous Disk to a Protoplanet}",
      journal = {ICARUS},
         year = 1993,
        month = mar,
       volume = {102},
       number = {1},
        pages = {150-165},
          doi = {10.1006/icar.1993.1039},
       adsurl = {https://ui.adsabs.harvard.edu/abs/1993Icar..102..150K}
}

@misc{BHPToolkit,
  title = {{Black Hole Perturbation Toolkit}},
  howpublished = {(\href{http://bhptoolkit.org/}{bhptoolkit.org})},
}

@misc{colpi2024lisadefinitionstudyreport,
      title={LISA Definition Study Report}, 
      author={Monica Colpi and others},
      year={2024},
      eprint={2402.07571},
      archivePrefix={arXiv},
      primaryClass={astro-ph.CO},
      url={https://arxiv.org/abs/2402.07571}, 
}

@misc{amaroseoane2017laserinterferometerspaceantenna,
      title={Laser Interferometer Space Antenna}, 
      author={Pau Amaro-Seoane and others},
      year={2017},
      eprint={1702.00786},
      archivePrefix={arXiv},
      primaryClass={astro-ph.IM},
      url={https://arxiv.org/abs/1702.00786}, 
}

@ARTICLE{2000ApJ...536..663N,
       author = {{Narayan}, Ramesh},
        title = "{Hydrodynamic Drag on a Compact Star Orbiting a Supermassive Black Hole}",
      journal = {\apj},
         year = 2000,
        month = jun,
       volume = {536},
       number = {2},
        pages = {663-667},
          doi = {10.1086/308956},
archivePrefix = {arXiv},
       eprint = {astro-ph/9907328},
 primaryClass = {astro-ph},
       adsurl = {https://ui.adsabs.harvard.edu/abs/2000ApJ...536..663N}
}

@article{Babak:2017tow,
    author = "Babak, Stanislav and Gair, Jonathan and Sesana, Alberto and Barausse, Enrico and Sopuerta, Carlos F. and Berry, Christopher P. L. and Berti, Emanuele and Amaro-Seoane, Pau and Petiteau, Antoine and Klein, Antoine",
    title = "{Science with the space-based interferometer LISA. V: Extreme mass-ratio inspirals}",
    eprint = "1703.09722",
    archivePrefix = "arXiv",
    primaryClass = "gr-qc",
    doi = "10.1103/PhysRevD.95.103012",
    journal = "Phys. Rev. D",
    volume = "95",
    number = "10",
    pages = "103012",
    year = "2017"
}

@article{Speri_2023,
   title={Probing Accretion Physics with Gravitational Waves},
   volume={13},
   ISSN={2160-3308},
   url={http://dx.doi.org/10.1103/PhysRevX.13.021035},
   DOI={10.1103/physrevx.13.021035},
   number={2},
   journal={Physical Review X},
   publisher={American Physical Society (APS)},
   author={Speri, Lorenzo and Antonelli, Andrea and Sberna, Laura and Babak, Stanislav and Barausse, Enrico and Gair, Jonathan R. and Katz, Michael L.},
   year={2023},
   month=jun }

@article{Yunes:2011ws,
    author = "Yunes, Nicolas and Kocsis, Bence and Loeb, Abraham and Haiman, Zoltan",
    title = "{Imprint of Accretion Disk-Induced Migration on Gravitational Waves from Extreme Mass Ratio Inspirals}",
    eprint = "1103.4609",
    archivePrefix = "arXiv",
    primaryClass = "astro-ph.CO",
    doi = "10.1103/PhysRevLett.107.171103",
    journal = "Phys. Rev. Lett.",
    volume = "107",
    pages = "171103",
    year = "2011"
}

@article{Derdzinski:2018qzv,
    author = "Derdzinski, A. M. and D'Orazio, D. and Duffell, P. and Haiman, Z. and MacFadyen, A.",
    title = "{Probing gas disc physics with LISA: simulations of an intermediate mass ratio inspiral in an accretion disc}",
    eprint = "1810.03623",
    archivePrefix = "arXiv",
    primaryClass = "astro-ph.HE",
    doi = "10.1093/mnras/stz1026",
    journal = "Mon. Not. Roy. Astron. Soc.",
    volume = "486",
    number = "2",
    pages = "2754--2765",
    year = "2019",
    note = "[Erratum: Mon.Not.Roy.Astron.Soc. 489, 4860--4861 (2019)]"
}

@misc{li2025extrememassratioinspiralultralight,
      title={Extreme mass-ratio inspiral within an ultralight scalar cloud I. Scalar radiation}, 
      author={Dongjun Li and Colin Weller and Patrick Bourg and Michael LaHaye and Nicolás Yunes and Huan Yang},
      year={2025},
      eprint={2507.02045},
      archivePrefix={arXiv},
      primaryClass={gr-qc},
      url={https://arxiv.org/abs/2507.02045}, 
}

@misc{dyson2025environmentaleffectsextrememass,
      title={Environmental effects in extreme mass ratio inspirals: perturbations to the environment in Kerr}, 
      author={Conor Dyson and Thomas F. M. Spieksma and Richard Brito and Maarten van de Meent and Sam Dolan},
      year={2025},
      eprint={2501.09806},
      archivePrefix={arXiv},
      primaryClass={gr-qc},
      url={https://arxiv.org/abs/2501.09806}, 
}

@article{Baumann_2022,
   title={Sharp Signals of Boson Clouds in Black Hole Binary Inspirals},
   volume={128},
   ISSN={1079-7114},
   url={http://dx.doi.org/10.1103/PhysRevLett.128.221102},
   DOI={10.1103/physrevlett.128.221102},
   number={22},
   journal={Physical Review Letters},
   publisher={American Physical Society (APS)},
   author={Baumann, Daniel and Bertone, Gianfranco and Stout, John and Tomaselli, Giovanni Maria},
   year={2022},
   month=jun }

@article{Derdzinski:2020wlw,
    author = "Derdzinski, A. and D'Orazio, D. and Duffell, P. and Haiman, Z. and MacFadyen, A.",
    title = "{Evolution of gas disc{\textendash}embedded intermediate mass ratio inspirals in the $LISA$ band}",
    eprint = "2005.11333",
    archivePrefix = "arXiv",
    primaryClass = "astro-ph.HE",
    doi = "10.1093/mnras/staa3976",
    journal = "Mon. Not. Roy. Astron. Soc.",
    volume = "501",
    number = "3",
    pages = "3540--3557",
    year = "2021"
}

@article{Copparoni_2025,
   title={Implications of stochastic gas torques for asymmetric binaries in the LISA band},
   volume={111},
   ISSN={2470-0029},
   url={http://dx.doi.org/10.1103/PhysRevD.111.104079},
   DOI={10.1103/physrevd.111.104079},
   number={10},
   journal={Physical Review D},
   publisher={American Physical Society (APS)},
   author={Copparoni, Lorenzo and Barausse, Enrico and Speri, Lorenzo and Sberna, Laura and Derdzinski, Andrea},
   year={2025},
   month=may }

@article{Sberna:2022qbn,
    author = "Sberna, Laura and others",
    title = "{Observing GW190521-like binary black holes and their environment with LISA}",
    eprint = "2205.08550",
    archivePrefix = "arXiv",
    primaryClass = "gr-qc",
    doi = "10.1103/PhysRevD.106.064056",
    journal = "Phys. Rev. D",
    volume = "106",
    number = "6",
    pages = "064056",
    year = "2022"
}

@article{Sasaki_2003,
   title={Analytic Black Hole Perturbation Approach to Gravitational Radiation},
   volume={6},
   ISSN={1433-8351},
   url={http://dx.doi.org/10.12942/lrr-2003-6},
   DOI={10.12942/lrr-2003-6},
   number={1},
   journal={Living Reviews in Relativity},
   publisher={Springer Science and Business Media LLC},
   author={Sasaki, Misao and Tagoshi, Hideyuki},
   year={2003},
   month=nov }

@article{Wardell:2021fyy,
    author = "Wardell, Barry and Pound, Adam and Warburton, Niels and Miller, Jeremy and Durkan, Leanne and Le Tiec, Alexandre",
    title = "{Gravitational Waveforms for Compact Binaries from Second-Order Self-Force Theory}",
    eprint = "2112.12265",
    archivePrefix = "arXiv",
    primaryClass = "gr-qc",
    doi = "10.1103/PhysRevLett.130.241402",
    journal = "Phys. Rev. Lett.",
    volume = "130",
    number = "24",
    pages = "241402",
    year = "2023"
}

@article{Albertini:2022rfe,
    author = "Albertini, Angelica and Nagar, Alessandro and Pound, Adam and Warburton, Niels and Wardell, Barry and Durkan, Leanne and Miller, Jeremy",
    title = "{Comparing second-order gravitational self-force, numerical relativity, and effective one body waveforms from inspiralling, quasicircular, and nonspinning black hole binaries}",
    eprint = "2208.01049",
    archivePrefix = "arXiv",
    primaryClass = "gr-qc",
    doi = "10.1103/PhysRevD.106.084061",
    journal = "Phys. Rev. D",
    volume = "106",
    number = "8",
    pages = "084061",
    year = "2022"
}

@article{Arvanitaki_2011,
   title={Exploring the string axiverse with precision black hole physics},
   volume={83},
   ISSN={1550-2368},
   url={http://dx.doi.org/10.1103/PhysRevD.83.044026},
   DOI={10.1103/physrevd.83.044026},
   number={4},
   journal={Physical Review D},
   publisher={American Physical Society (APS)},
   author={Arvanitaki, Asimina and Dubovsky, Sergei},
   year={2011},
   month=feb }

@article{Zhang_2020,
   title={Dynamic signatures of black hole binaries with superradiant clouds},
   volume={101},
   ISSN={2470-0029},
   url={http://dx.doi.org/10.1103/PhysRevD.101.043020},
   DOI={10.1103/physrevd.101.043020},
   number={4},
   journal={Physical Review D},
   publisher={American Physical Society (APS)},
   author={Zhang, Jun and Yang, Huan},
   year={2020},
   month=feb }

@article{LISAConsortiumWaveformWorkingGroup:2023arg,
    author = "Afshordi, Niayesh and others",
    collaboration = "LISA Consortium Waveform Working Group",
    title = "{Waveform Modelling for the Laser Interferometer Space Antenna}",
    eprint = "2311.01300",
    archivePrefix = "arXiv",
    primaryClass = "gr-qc",
    month = "11",
    year = "2023"
}

@article{Maselli_2020,
   title={Detecting Scalar Fields with Extreme Mass Ratio Inspirals},
   volume={125},
   ISSN={1079-7114},
   url={http://dx.doi.org/10.1103/PhysRevLett.125.141101},
   DOI={10.1103/physrevlett.125.141101},
   number={14},
   journal={Physical Review Letters},
   publisher={American Physical Society (APS)},
   author={Maselli, Andrea and Franchini, Nicola and Gualtieri, Leonardo and Sotiriou, Thomas P.},
   year={2020},
   month=sep }

@article{Vicente_2022,
   title={Dynamical friction of black holes in ultralight dark matter},
   volume={105},
   ISSN={2470-0029},
   url={http://dx.doi.org/10.1103/PhysRevD.105.083008},
   DOI={10.1103/physrevd.105.083008},
   number={8},
   journal={Physical Review D},
   publisher={American Physical Society (APS)},
   author={Vicente, Rodrigo and Cardoso, Vitor},
   year={2022},
   month=apr }

@article{Hughes_2021,
   title={Adiabatic waveforms for extreme mass-ratio inspirals via multivoice decomposition in time and frequency},
   volume={103},
   ISSN={2470-0029},
   url={http://dx.doi.org/10.1103/PhysRevD.103.104014},
   DOI={10.1103/physrevd.103.104014},
   number={10},
   journal={Physical Review D},
   publisher={American Physical Society (APS)},
   author={Hughes, Scott A. and Warburton, Niels and Khanna, Gaurav and Chua, Alvin J.K. and Katz, Michael L.},
   year={2021},
   month=may }

@article{Bourg:2024vre,
    author = "Bourg, Patrick and Leather, Benjamin and Casals, Marc and Pound, Adam and Wardell, Barry",
    title = "{Implementation of a Green-Hollands-Zimmerman-Teukolsky puncture scheme for gravitational self-force calculations}",
    eprint = "2403.12634",
    archivePrefix = "arXiv",
    primaryClass = "gr-qc",
    doi = "10.1103/PhysRevD.110.044007",
    journal = "Phys. Rev. D",
    volume = "110",
    number = "4",
    pages = "044007",
    year = "2024"
}

@article{Shah_2014,
   title={Gravitational-wave flux for a particle orbiting a Kerr black hole to 20th post-Newtonian order: A numerical approach},
   volume={90},
   ISSN={1550-2368},
   url={http://dx.doi.org/10.1103/PhysRevD.90.044025},
   DOI={10.1103/physrevd.90.044025},
   number={4},
   journal={Physical Review D},
   publisher={American Physical Society (APS)},
   author={Shah, Abhay G.},
   year={2014},
   month=aug }

@article{Warburton:2024xnr,
    author = "Warburton, Niels and Wardell, Barry and Trestini, David and Henry, Quentin and Pound, Adam and Blanchet, Luc and Durkan, Leanne and Faye, Guillaume and Miller, Jeremy",
    title = "{Comparison of 4.5PN and 2SF gravitational energy fluxes from quasicircular compact binaries}",
    eprint = "2407.00366",
    archivePrefix = "arXiv",
    primaryClass = "gr-qc",
    month = "6",
    year = "2024"
}

@ARTICLE{1978ApJ...222..850G,
       author = {{Goldreich}, P. and {Tremaine}, S.},
        title = "{The excitation and evolution of density waves.}",
      journal = {\apj},
         year = 1978,
        month = jun,
       volume = {222},
        pages = {850-858},
          doi = {10.1086/156203},
       adsurl = {https://ui.adsabs.harvard.edu/abs/1978ApJ...222..850G}
}

@ARTICLE{1979ApJ...233..857G,
       author = {{Goldreich}, P. and {Tremaine}, S.},
        title = "{The excitation of density waves at the Lindblad and corotation resonances by an external potential.}",
      journal = {\apj},
         year = 1979,
        month = nov,
       volume = {233},
        pages = {857-871},
          doi = {10.1086/157448},
       adsurl = {https://ui.adsabs.harvard.edu/abs/1979ApJ...233..857G}
}

@INPROCEEDINGS{Paardekooper:2023,
       author = {{Paardekooper}, S. and {Dong}, R. and {Duffell}, P. and {Fung}, J. and {Masset}, F.~S. and {Ogilvie}, G. and {Tanaka}, H.},
        title = "{Planet-Disk Interactions and Orbital Evolution}",
    booktitle = {Protostars and Planets VII},
         year = 2023,
       editor = {{Inutsuka}, S. and {Aikawa}, Y. and {Muto}, T. and {Tomida}, K. and {Tamura}, M.},
       series = {Astronomical Society of the Pacific Conference Series},
       volume = {534},
        month = jul,
        pages = {685},
          doi = {10.48550/arXiv.2203.09595},
archivePrefix = {arXiv},
       eprint = {2203.09595},
 primaryClass = {astro-ph.EP},
       adsurl = {https://ui.adsabs.harvard.edu/abs/2023ASPC..534..685P}
}

@article{Franchini_2023,
   title={Quasi-periodic eruptions from impacts between the secondary and a rigidly precessing accretion disc in an extreme mass-ratio inspiral system},
   volume={675},
   ISSN={1432-0746},
   url={http://dx.doi.org/10.1051/0004-6361/202346565},
   DOI={10.1051/0004-6361/202346565},
   journal={Astronomy and amp; Astrophysics},
   publisher={EDP Sciences},
   author={Franchini, Alessia and Bonetti, Matteo and Lupi, Alessandro and Miniutti, Giovanni and Bortolas, Elisa and Giustini, Margherita and Dotti, Massimo and Sesana, Alberto and Arcodia, Riccardo and Ryu, Taeho},
   year={2023},
   month=jul, pages={A100} }

@article{Barausse:2007dy,
    author = "Barausse, Enrico and Rezzolla, Luciano",
    title = "{The Influence of the hydrodynamic drag from an accretion torus on extreme mass-ratio inspirals}",
    eprint = "0711.4558",
    archivePrefix = "arXiv",
    primaryClass = "gr-qc",
    doi = "10.1103/PhysRevD.77.104027",
    journal = "Phys. Rev. D",
    volume = "77",
    pages = "104027",
    year = "2008"
}

@article{Berti:2018vdi,
    author = "Berti, Emanuele and Yagi, Kent and Yang, Huan and Yunes, Nicol{\'a}s",
    title = "{Extreme Gravity Tests with Gravitational Waves from Compact Binary Coalescences: (II) Ringdown}",
    eprint = "1801.03587",
    archivePrefix = "arXiv",
    primaryClass = "gr-qc",
    doi = "10.1007/s10714-018-2372-6",
    journal = "Gen. Rel. Grav.",
    volume = "50",
    number = "5",
    pages = "49",
    year = "2018"
}

@article{Caprini:2018mtu,
    author = "Caprini, Chiara and Figueroa, Daniel G.",
    title = "{Cosmological Backgrounds of Gravitational Waves}",
    eprint = "1801.04268",
    archivePrefix = "arXiv",
    primaryClass = "astro-ph.CO",
    doi = "10.1088/1361-6382/aac608",
    journal = "Class. Quant. Grav.",
    volume = "35",
    number = "16",
    pages = "163001",
    year = "2018"
}

@article{LISA:2022yao,
    author = "Seoane, Pau Amaro and others",
    collaboration = "LISA",
    title = "{Astrophysics with the Laser Interferometer Space Antenna}",
    eprint = "2203.06016",
    archivePrefix = "arXiv",
    primaryClass = "gr-qc",
    doi = "10.1007/s41114-022-00041-y",
    journal = "Living Rev. Rel.",
    volume = "26",
    number = "1",
    pages = "2",
    year = "2023"
}

@misc{blanchet2024postnewtoniantheorygravitationalwaves,
      title={Post-Newtonian Theory for Gravitational Waves}, 
      author={Luc Blanchet},
      year={2024},
      eprint={1310.1528},
      archivePrefix={arXiv},
      primaryClass={gr-qc},
      url={https://arxiv.org/abs/1310.1528}, 
}

@misc{zhang2025radiationgrmhdmodelsaccretion,
      title={Radiation GRMHD Models of Accretion onto Stellar-Mass Black Holes: I. Survey of Eddington Ratios}, 
      author={Lizhong Zhang and James M. Stone and Patrick D. Mullen and Shane W. Davis and Yan-Fei Jiang and Christopher J. White},
      year={2025},
      eprint={2506.02289},
      archivePrefix={arXiv},
      primaryClass={astro-ph.HE},
      url={https://arxiv.org/abs/2506.02289}, 
}

@inbook{Baruteau_2014,
   title={Planet-Disk Interactions and Early Evolution of Planetary Systems},
   url={http://dx.doi.org/10.2458/azu_uapress_9780816531240-ch029},
   DOI={10.2458/azu_uapress_9780816531240-ch029},
   booktitle={Protostars and Planets VI},
   publisher={University of Arizona Press},
   author={Baruteau, C. and Crida, A. and Paardekooper, S.-J. and Masset, F. and Guilet, J. and Bitsch, B. and Nelson, R. and Kley, W. and Papaloizou, J.},
   year={2014} }

@ARTICLE{1993ApJ...419..155A,
       author = {{Artymowicz}, Pawel},
        title = "{On the Wave Excitation and a Generalized Torque Formula for Lindblad Resonances Excited by External Potential}",
      journal = {\apj},
         year = 1993,
        month = dec,
       volume = {419},
        pages = {155},
          doi = {10.1086/173469},
       adsurl = {https://ui.adsabs.harvard.edu/abs/1993ApJ...419..155A}
}

@article{Abramowicz_2013,
   title={Foundations of Black Hole Accretion Disk Theory},
   volume={16},
   ISSN={1433-8351},
   url={http://dx.doi.org/10.12942/lrr-2013-1},
   DOI={10.12942/lrr-2013-1},
   number={1},
   journal={Living Reviews in Relativity},
   publisher={Springer Science and Business Media LLC},
   author={Abramowicz, Marek A. and Fragile, P. Chris},
   year={2013},
   month=jan }

@BOOK{MTW,
       author = {{Misner}, Charles W. and {Thorne}, Kip S. and {Wheeler}, John Archibald},
        title = "{Gravitation}",
         year = 1973,
       adsurl = {https://ui.adsabs.harvard.edu/abs/1973grav.book.....M}
}

@article{Ni-Zimmermann,
  title = {Inertial and gravitational effects in the proper reference frame of an accelerated, rotating observer},
  author = {Ni, Wei-Tou and Zimmermann, Mark},
  journal = {Phys. Rev. D},
  volume = {17},
  issue = {6},
  pages = {1473--1476},
  numpages = {0},
  year = {1978},
  month = {Mar},
  publisher = {American Physical Society},
  doi = {10.1103/PhysRevD.17.1473},
  url = {https://link.aps.org/doi/10.1103/PhysRevD.17.1473}
}

@article{Parker-Luis,
  title = {Gravitational perturbation of the hydrogen spectrum},
  author = {Parker, Leonard and Pimentel, Luis O.},
  journal = {Phys. Rev. D},
  volume = {25},
  issue = {12},
  pages = {3180--3190},
  numpages = {0},
  year = {1982},
  month = {Jun},
  publisher = {American Physical Society},
  doi = {10.1103/PhysRevD.25.3180},
  url = {https://link.aps.org/doi/10.1103/PhysRevD.25.3180}
}

@INPROCEEDINGS{2023ASPC..534..685P,
       author = {{Paardekooper}, S. and {Dong}, R. and {Duffell}, P. and {Fung}, J. and {Masset}, F.~S. and {Ogilvie}, G. and {Tanaka}, H.},
        title = "{Planet-Disk Interactions and Orbital Evolution}",
    booktitle = {Protostars and Planets VII},
         year = 2023,
       editor = {{Inutsuka}, S. and {Aikawa}, Y. and {Muto}, T. and {Tomida}, K. and {Tamura}, M.},
       series = {Astronomical Society of the Pacific Conference Series},
       volume = {534},
        month = jul,
        pages = {685},
          doi = {10.48550/arXiv.2203.09595},
archivePrefix = {arXiv},
       eprint = {2203.09595},
 primaryClass = {astro-ph.EP},
       adsurl = {https://ui.adsabs.harvard.edu/abs/2023ASPC..534..685P}
}
\end{document}